\pdfoutput=1
\documentclass[useAMS,usenatbib, usegraphicx]{mn2e}

\usepackage{verbatim} 
\usepackage[usenames,dvipsnames]{color}
\usepackage{amssymb, amsmath}
\usepackage{multicol}
\usepackage{longtable}
\usepackage{rotating}
\usepackage{pdflscape}
\usepackage{url}
\usepackage{times}
\usepackage{multirow}
\usepackage{lscape}

\usepackage{graphicx}
\usepackage{epstopdf}
\usepackage[labelfont=bf,labelsep=space]{caption}

\usepackage{enumitem}
\usepackage[justification=centering]{caption}
\usepackage[]{subfigure}

\usepackage{natbib}
\bibpunct{(}{)}{;}{a}{}{,}

\usepackage{array}
\newcolumntype{x}[1]{>{\centering\arraybackslash\hspace{0pt}}m{#1}}

\makeatletter
\def\url@leostyle{%
  \@ifundefined{selectfont}{\def\UrlFont{\sf}}{\def\UrlFont{\small\ttfamily}}}
\makeatother
\usepackage{footnote}
\usepackage{tablefootnote}

\usepackage{newtxtext,newtxmath}

\DeclareRobustCommand{\ion}[2]{%
\relax\ifmmode
\ifx\testbx\f@series
{\mathbf{#1\,\mathsc{#2}}}\else
{\mathrm{#1\,\mathsc{#2}}}\fi
\else\textup{#1\,{\mdseries\textsc{#2}}}%
\fi}

\title[GAMA: Enhanced SFR and clustering of SF galaxies]{\color{black}Galaxy And Mass Assembly (GAMA): the signatures of galaxy interactions as viewed from small scale galaxy clustering }

\color{black} 
\author[Gunawardhana \it{et.\,al}]{M.\,L.\,P.\,Gunawardhana$^{1, 2, 3}$\thanks{Marie Sk\l{}odowska-Curie Fellow, FONDECYT fellow 2016--2017}\thanks{ E-mail: gunawardhana@strw.leidenuniv.nl}, P.\,Norberg$^{1}$\thanks{E-mail: peder.norberg@durham.ac.uk}, I.\,Zehavi$^{4}$, D.\,J.\,Farrow$^{5}$, J.\,Loveday$^{6}$, 
\newauthor
 A.\,M.\,Hopkins$^{7}$, L.\,J.\,M.\,Davies$^{8}$, L.\,Wang$^{9, 10}$, M.\,Alpaslan$^{11}$, J.\,Bland-Hawthorn$^{12}$,
 \newauthor
 S.\,Brough$^{13}$, B.\,W.\,Holwerda$^{14}$, M.\,S.\,Owers$^{7,15}$, A.\,H.\,Wright$^{16}$\\
$^{1}$ICC $\&$ CEA Department of Physics, Durham University, South Road, Durham, DH1 3LE, UK\\
$^{2}$Instituto de Astrof\'isica and Centro de Astroingenier\'ia, Facultad de F\'isica, Pontificia Universidad Cat\'olica de Chile, \\ Vicu\~na Mackenna 4860, 7820436 Macul, Santiago, Chile \\
$^{3}$Leiden Observatory, University of Leiden, Niels Bohrweg 2, NL-2333 CA Leiden, The Netherlands\\
$^{4}$Department of Astronomy and Department of Physics, Case Western Reserve University, Cleveland, OH 44106, USA \\
$^{5}$Max-Planck-Institut f{\"u}r extraterrestrische Physik, Postfach 1312 Giessenbachstrasse, 85741 Garching, Germany\\
$^{6}$Astronomy Centre, University of Sussex, Falmer, Brighton, BN1 9QH, UK\\
$^{7}$The Australian Astronomical Observatory, PO Box 915, North Ryde, NSW, 1670, Australia \\
$^{8}$ICRAR, The University of Western Australia, 35 Stirling Highway, Crawley, WA 6009, Australia \\
$^{9}$SRON Netherlands Institute for Space Research, Landleven 12, 9747 AD, Groningen, The Netherlands\\
$^{10}$Kapteyn Astronomical Institute, University of Groningen, Postbus 800, 9700 AV, Groningen, The Netherlands\\
$^{11}$NASA Ames Research Center, N244-30, Moffett Field, Mountain View, CA 94035, USA\\
$^{12}$Sydney Institute for Astronomy, School of Physics A28, University of Sydney, NSW 2006, Australia\\
$^{13}$School of Physics, University of New South Wales, NSW 2052, Australia\\
$^{14}$Department of Physics and Astronomy, University of Louisville, Louisville KY 40292, USA\\
$^{15}$Department of Physics and Astronomy, Macquarie University, NSW, 2109, Australia\\
$^{16}$Argelander-Institut f{\"u}r Astronomie, Universit{\"a}t Bonn, D-53121 Bonn, Germany\\
}

\begin{document}
\color{black}
\date{Accepted date. Received date; in original form date}

\pagerange{\pageref{firstpage}--\pageref{lastpage}} \pubyear{2002}

\maketitle

\label{firstpage}

\begin{abstract}
Statistical studies of galaxy-galaxy interactions often utilise net change in physical properties of progenitors as a function of the separation between their nuclei to trace both the strength and the observable timescale of their interaction.
In this study, we use two-point auto, cross and mark correlation functions to investigate the extent to which small-scale clustering properties of star forming galaxies can be used to gain physical insight into galaxy-galaxy interactions between galaxies of similar optical brightness and stellar mass. The H$\alpha$ star formers, drawn from the highly spatially complete Galaxy And Mass Assembly (GAMA) survey, show an increase in clustering on small separations. Moreover, the clustering strength shows a strong dependence on optical brightness and stellar mass, where (1) the clustering amplitude of optically brighter galaxies at a given separation is larger than that of optically fainter systems, (2) the small scale clustering properties (e.g.\,the strength, the scale at which the signal relative to the fiducial power law plateaus) of star forming galaxies appear to differ as a function of increasing optical brightness of galaxies.  According to cross and mark correlation analyses, the former result is largely driven by the increased dust content in optically bright star forming galaxies. The latter could be interpreted as evidence of a correlation between interaction-scale and optical brightness of galaxies, where physical evidence of interactions between optically bright star formers, likely hosted within relatively massive halos, persist over larger separations than those between optically faint star formers. 

\end{abstract}

\begin{keywords}
galaxies: interactions -- galaxies: starburst -- galaxies: star formation -- galaxies: haloes -- galaxies: statistics -- galaxies: distances and redshifts
\end{keywords}

\section{Introduction}\label{sec:intro}

Historically, the field of galaxy interactions dates as far back as the 1940s, however, it was not until 1970s that the concept of tidal forces being the underlying drivers of morphological distortions in galaxies was fully accepted. It was the pioneering works by \cite{Toomre1972} on numerically generating "galactic bridges and tails" from galaxy interactions, and by \cite{Larson1978} on broadband optical observations of discrepancies in "star formation rates in normal and peculiar galaxies" that essentially solidified this concept. Since then, the progress that followed revealed that interacting galaxies often show enhancements in H$\alpha$ emission \citep[e.g.][]{Keel1985, Kennicutt1987}, infrared emission \citep[e.g.][]{Lonsdale1984, Soifer1984, Sanders1986, Solomon1988}, in radio continuum emission \citep[e.g.][]{Condon1982}, and in molecular (CO) emission \citep[e.g.][]{Young1996} compared to isolated galaxies.

{{Over the past decade or so, numerous studies based on large sky survey datasets have provided ubiquitous evidence for, and signatures of tidal interactions. The enhancement of star formation is perhaps the most important and direct signature of a gravitational interaction \citep{Kennicutt1998, Wong2011}, however, not all starbursts are interaction driven, and not all interactions trigger starbursts. Starbursts, by definition, are short-lived intense periods of concentrated star formation confined within the galaxy and are expected to be triggered only by the increase in molecular gas surface density in the inner regions over a short timescale. The tidal torques generated during the interactions of gas-rich galaxies are, therefore, one of the most efficient ways of funnelling gas to the centre of a galaxy \citep{Smith2007, DiMatteo2007, Cox2008}. In the absence of an interaction, however, bars of galaxies, which are prominent in spiral galaxies, can effectively facilitate both gas inflows and outflows \citep{Regan2004, Owers2007, Ellison2011b, Martel2013}, and trigger starbursts. Nuclear starbursts appear to be a common occurrence of interactions and mergers, however, there are cases where starbursts have been observed to occur, for example, in the overlapping regions between two galaxies \citep[e.g.\,the Antennae galaxies;][]{Snijders2007}.

In the local Universe, most interacting galaxies have been observed to have higher than average central star formation \citep[e.g.][]{Lambas2003, Smith2007, Ellison2008, Xu2010, Scott2014, Robotham2014, Knapen2015}, though in a handful of cases, depending on the nature of the progenitors, moderate \citep[e.g.][]{Rogers2009, Darg2010, Knapen2015} to no enhancements \citep[e.g.][]{Bergvall2003, Lambas2003} have also been reported. Likewise, interactions have been observed to impact circumnuclear gas-phase metallicities. In most cases, interactions appear to dilute nuclear gas-phase metallicities \citep[e.g.][]{Kewley2006, Scudder2012, Ellison2013} and flatten metallicity gradients \citep[e.g.][]{Kewley2006b, Ellison2008}. There are also cases where an enhancement in central gas-phase metallicities \citep[e.g.][]{Barrera-Ballesteros2015} has also been observed. The other observational signatures of galaxy-galaxy interactions include enhancements in optical colours, with enhancements in bluer colours \citep[e.g.][]{DePropris2005, Darg2010, Patton2011} observed to be tied to gas-rich and redder colours to gas-poor interactions \citep[e.g.][]{Rogers2009, Darg2010}, increased Active Galactic Nuclei activities \citep[AGNs,\,e.g.;][]{Rogers2009, Ellison2011, Kaviraj2015, Sabater2015} and substantially distorted galaxy morphologies \citep[e.g.][]{Casteels2013}.       

The strength and the duration of a physical change triggered in an interaction can potentially shed light on to the nature of that interaction, progenitors and the roles of their galaxy- and halo-scale environments in driving and sustaining that change. In this regard, the projected separation between galaxies, R$_p$, can essentially be used as a clock for dating an interaction, measuring either the time elapsed since or time to the pericentric passage.

One of the more widely used approaches to understanding the effects of galaxy-galaxy interactions involves directly quantifying net enhancement or decrement of a physical property as a function of R$_p$. 
For example, the strongest enhancements in SFR have typically been observed over $<30\,h^{-1}_{70}$ kpc \citep[e.g.][]{Ellison2008, Li2008, Wong2011, Scudder2012, Patton2013}. The lower-level enhancements, on the other hand, have been observed to persist for relatively longer timescales. \cite{Ellison2008} report a net enhancement in SFRs and a decrement in metallicity of $\sim0.05-0.1$ dex out to separations of $\sim30-40\,h^{-1}_{70}$ kpc, and an enhancement in SFR out to wider separations for galaxy pairs of equal mass. \cite{Wong2011} report observations of SFR enhancements out to $\sim50$ h$_{70}^{-1}$ kpc based galaxy pair sample drawn from PRIMUS, \cite{Scudder2012} find that net changes in both SFR and metallicity persist out to at least $\sim80\,h^{-1}_{70}$ kpc, \cite{Patton2013} find a clear enhancement in SFR out to $\sim150$ kpc with no net enhancement beyond, while \cite{Patton2011} report enhancement in colours out to $\sim80$ h$_{70}^{-1}$ kpc, and \cite{Nikolic2004} report an enhancement in SFR out to $\sim300$ kpc for their sample of actively star forming late-type galaxy pairs. 

Even though the direct measure of a net change is advantageous as it can provide insight into dissipation rates and observable timescales of interaction-driven alterations \citep{Lotz2011, Robotham2014}, as highlighted above, the reported values of R$_p$ out to which a given change persists often varies. The strength and the scale out to which a physical change is observable is expected to be influenced by orbital parameters and properties of progenitors \citep{Nikolic2004, Owers2007, Ellison2010, Patton2011}, as well as by the differences in dynamical timescales associated with short and long duration star formation events \citep{Davies2015}. Furthermore, galaxy-galaxy interactions do not always lead to observable changes. In particular, the subtle physical changes on R$_{\rm{p}}$s at which progenitors are just starting to experience the effects of an interaction can be too weak to be observed. A further caveat is that this method fails to provide any physical insights into potential causes for the observed changes, i.e.\,whether the change is a result of the first pericentric passage, second or environment. 

Another approach to studying the effects of galaxy-galaxy interactions involves two-point and higher order correlation statistics. The correlation statistics are often used in the interpretation of clustering properties of galaxies within one- and two-halo terms, and  can to be utilised with or without incorporating physical information of galaxies. In this study, we aim to investigate whether large-scale environment plays any role in driving and sustaining interaction-driven changes in star forming galaxies with the aid of two-point correlation statistics. 

In the local Universe, correlation functions have been ubiquitously used to study the clustering strength of galaxies with respect to galaxy properties like stellar mass, galaxy luminosities, and optical colours. \cite{Norberg2002} and \cite{Madgwick2003}, for example, find clustering strength to be dependent strongly upon galaxy luminosity. \cite{Zehavi2005b, Li2006, Li2009, Zehavi2011, Ross2014, Favole2016} and \cite{Loh2010} report that galaxies with optically redder colours, which tend to be characterised with bulge dominated morphologies and higher surface brightnesses, correlate strongest with the strength of clustering than those residing in the green valley or in the blue cloud.   

Even though much work has been done in this area, very few of those studies have focussed on investigating clustering of galaxies with respect to their star forming properties such as star formation rate (SFR), specific SFR (sSFR) and dust. 
The Sloan Digital Sky Survey (SDSS) based analysis of \cite{Li2008} reports a strong dependence of the amplitude of the correlation function on specific star formation rate (sSFR) of galaxies on R$_{\rm{p}}\lesssim100$ kpc. They find a dependence between clustering amplitude and sSFR, where the amplitude is observed to increase smoothly with increasing sSFR such that galaxies with high specific SFRs are clustered more strongly than those with low specific SFRs. The strongest enhancements in amplitude are found to be associated with the lowest mass galaxies and over the smallest R$_p$. They interpret this behaviour as being due to tidal interactions. Using \textit{GALEX} imaging data of SDSS galaxies, \cite{Heinis2009} investigate the clustering dependence with respect to both ($NUV-r$) and sSFR. Over $0.01<$R$_{p}$[h$^{-1}$Mpc]$<10$, they find a smooth transition in clustering strength from weak-to-strong as a function of the blue-to-red change in ($NUV-r$) and the low-to-high change in sSFR. It must be noted, however, that on the smallest scales the clustering of the bluest ($NUV-r$) galaxies shows an enhancement. 

\cite{Coil2016} use the PRIMUS and DEEP2 galaxy surveys spanning $0.2<z<1.2$ to measure the stellar mass and sSFR dependence of the clustering of galaxies. They find that clustering dependence is as strong of a function of sSFR as of stellar mass, such that clustering smoothly increases with increasing stellar mass and decreasing sSFR, and find no significant dependence on stellar mass a fixed sSFR. This same trend is also found within the quiescent population. The DEEP2 survey based study of \cite{Mostek2013} too finds that within the star forming population the clustering amplitude increases as a function of increasing SFR and decreasing sSFR. Their analysis of small scale clustering of both star forming and quiescent populations, however, shows a clustering excess for high sSFR galaxies, which they attribute to galaxy-galaxy interactions.    

The spatial and redshift completenesses of a galaxy survey largely determine the smallest R$_p$ that can be reliably probed by two-point correlation statistics, thus the ability to trace galaxy-galaxy interactions reliably. The lack of sufficient overlap between pointings to ensure the full coverage of all sources can significantly impact the spatial completeness of a fibre-based spectroscopic survey. The resulting spatial incompleteness can considerably decrease the clustering signal on R$_p\lesssim0.2$ [Mpc], especially for non-projected statistics \citep{Yoon2008}, and can have non-negligible effects even on larger scales \citep{Zehavi2005b}. Therefore many of the aforementioned studies are generally limited to probing clustering on R$_p\gtrsim0.1$ [Mpc h$^{-1}$]. 

For this study, we draw a star forming sample of galaxies from the Galaxy And Mass Assembly (GAMA) survey \citep{Driver2011, Liske2015}, which has very high spatial and redshift completenesses ($>98.5\%$). The GAMA achieves this very high spatial completeness  by surveying the same field over and over ($\sim8-10$ times) until all targets have been observed \citep[][see the subsequent section for a discussion on the characteristics of the survey]{Robotham2010}. Galaxy surveys like SDSS are limited both by the finite size of individual fibre heads as well as by the number of overlaps ($\sim1.3$ times). Therefore GAMA survey is ideal for a study, such as ours, that investigates the small-scale clustering properties of star forming galaxies as a function of the star forming properties. } }   

This paper is structured as follows. In \S\,\ref{sec:gama}, we describe the characteristics of the GAMA survey and the different GAMA catalogues that have been used in this study. This section also details the spectroscopic completeness of the GAMA survey, the selection of a reliable star forming galaxy sample from GAMA and the construction of galaxy samples for the clustering analyses. The different clustering techniques and definitions used in this analyses, as well as the modelling of the selection function associated with random galaxies, are described in \S\,\ref{sec:clustering_methods}. Subsequently, in \S\,\ref{sec:SF_signatures}, we present the trends of star forming galaxies with respect to different potential indicators of galaxy-galaxy interactions, and the correlation functions of star forming based on auto, cross and mark correlation statistics. Finally, in \S\,\ref{sec:discuss} and \ref{sec:conc}, we discuss and compare the results of this study with the results reported in other published studies of star forming galaxies in the local Universe. This paper also includes four appendices, which are structured as follows. A discussion on sample selection and systematics is given in Appendix A. In Appendices B and C, we present a volume limited analysis involving auto and cross correlation functions, and further correlation results involving different galaxy samples introduced in \S\,\ref{sec:gama}. Finally, in Appendix\,\ref{app:mark_CFs}, we present the mark correlation analyses as we chose  to show only the rank ordered mark correlation analysis in the main paper.

The assumed cosmological parameters are H$_{\circ}=70$ km s$^{-1}$Mpc$^{-1}$, $\Omega_{\rm{M}}=0.3$ and $\Omega_\Lambda=0.7$. All magnitudes are presented in the AB system, and a \cite{Chabrier2003} IMF is assumed throughout.

\section{Galaxy And Mass Assembly (GAMA) Survey} \label{sec:gama}

We utilise the GAMA \citep[][]{Driver2011, Liske2015} survey data for the analysis presented in this paper. In the subsequent sections, we briefly describe the characteristics of the GAMA survey and the workings of the GAMA spectroscopic pipeline. 

\subsection{GAMA survey characteristics}
\subsubsection{GAMA imaging}

GAMA is a comprehensive multi-wavelength photometric and spectroscopic survey of the nearby Universe. GAMA brings together several independent imaging campaigns to provide a near-complete sampling of the UV to far-IR (0.15--500$\mu$m) wavelength range, through 21 broad-band filters; FUV, NUV \citep[\textit{GALEX};][]{Martin2005}, $ugriz$ \citep[Sloan Digital Sky Survey data release 7, i.e.\,SDSS DR7;][]{Fukugita1996, Gunn1998, Abazajian2009}, Z, Y, J, H, K \citep[VIsta Kilo-degree INfrared Galaxy survey, i.e.\,VIKING;][]{Edge2013}, W1, W2, W3, W4 \citep[Wide-field Infrared Survey Explorer, i.e.\,\textit{WISE};][]{Wright2010}, 100$\mu$m, 160$\mu$m, 250$\mu$m, 350$\mu$m, and 500$\mu$m \citep[\textit{Herschel}-ATLAS;][]{Eales2010}. A complete analysis of the multi-wavelength successes of GAMA is presented in the end of survey report of \cite{Liske2015} and in the panchromatic data release of \cite{Driver2015}.

\subsubsection{GAMA redshifts}

GAMA's independent spectroscopic campaign was primarily conducted with the 2dF/AAOmega multi-object instrument \citep{Sharp2006} on the 3.9m Anglo-Australian Telescope (AAT). Between 2008 and 2014, GAMA has surveyed a total sky area of $\sim286$ deg$^{2}$ split into five independent regions; three equatorial (called GAMA-09hr or G09, G12, and G15) and two southern (G02 and G23) fields of $12\times5$ deg$^{2}$ each. The GAMA equatorial targets are drawn primarily from SDSS DR7 \citep[][]{Abazajian2009}. We refer the readers to the paper by \cite{Baldry2010} for detailed discussions on target selection strategies and input catalogues. The equatorial fields have been surveyed to an extinction corrected Petrosian $r$-band magnitude depth of $19.8$. A key strength of GAMA is its high spatial completeness, both in terms of the overall completeness and completeness on small spatial scales. This is also advantageous for the present study aimed at investigating SFR enhancement due to galaxy interactions via small scale galaxy clustering. The tiling and observing strategies of the survey are discussed in detail in \cite{Robotham2010} and \cite{Driver2011}. At the conclusion of the spectroscopic survey, GAMA has achieved a high redshift completeness of $98.5\%$ for the equatorial regions, and we discuss in detail the spectroscopic completeness of the survey in \S\,\ref{subsec:selection}.

\subsubsection{GAMA spectroscopic pipeline}

A detailed summary of the GAMA redshift assignment, re-assignment, and quality control procedure is given in \cite{Liske2015}, according to which galaxy redshifts with normalised redshift qualities (NQ) $\geq3$ are secure redshifts. GAMA does not re-observe galaxies with high-quality spectra originating from other surveys, such that the GAMA spectroscopic catalogues comprise spectra from a number of other sources, e.g.\,SDSS, 2-degree Field Galaxy Redshift Survey \citep[2dFGRS;][]{Colless2001}, Millennium Galaxy Catalogue \citep[MGC;][]{Driver2007}, see \S\,\ref{subsec:selection} for a discussion on the contribution of non-GAMA spectral measures to our analysis. Finally, given the exceptionally high redshift completeness of the GAMA equatorial fields, we restrict our analysis to the equatorial data. 

The GAMA spectroscopic analysis procedure, including data reduction, flux calibration, and spectral line measurements, is presented in \cite{Hopkins2013}. The GAMA emission line catalogue ({\tt{SpecLineSFR}}) provides line fluxes and equivalent width measurements for all strong emission line measurements. A more detailed description of the spectral line measurement procedure and {\tt{SpecLineSFR}} catalogue, in general, can be found in \cite{Gordon2017}. Additionally, the strength of the ${\rm{\lambda}}4000$-\AA\,break (D$_{\rm{4000}}$) is measured over the D$_{\rm{4000}}$ bandpasses (i.e.\,$3850-3950$\AA\, and $4000-4100$\AA) defined in \cite{Balogh1999} following the method of \cite{Cardiel1998}.  {\tt{SpecLineSFR}} also provides a continuum ($6383-6538$\AA) signal-to-noise per pixel measurement, which is representative of the red-end of the spectrum.

\subsection{Galaxy properties}

The two main intrinsic galaxy properties used in this investigation are H$\alpha$ SFRs and galaxy stellar masses. Below, we briefly overview the derivation of these properties and discuss their uncertainties. 

\subsubsection{H$\alpha$ Star Formation Rates} \label{subsubsec:HaSFR}

The GAMA intrinsic H$\alpha$ SFRs are derived following the prescription of \cite{Hopkins2003}, using the Balmer emission line fluxes provided in {\color{black}{\tt{SpecLineSFR}}}. The spectroscopic redshifts used in the calculation are corrected for the effects of local and large-scale flows using the parametric multi-attractor model of \cite{Tonry2000} as described in \cite{Baldry2012}, and the application of stellar absorption, dust obscuration and fibre aperture corrections to SFRs is described in detail in \cite{Gunawardhana2013}. 

\begin{figure}
\begin{center}
\includegraphics[scale=0.4]{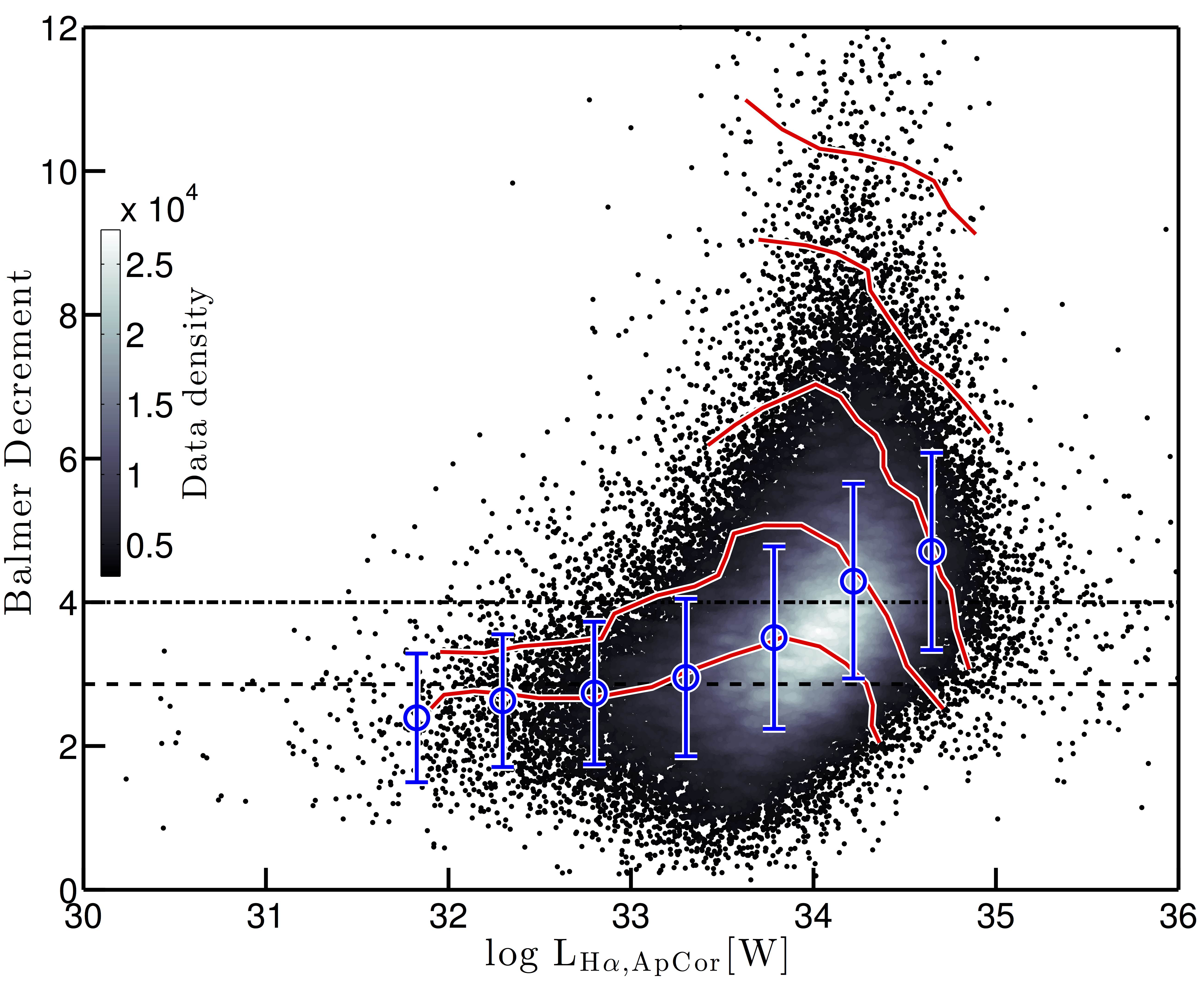}
\caption{\color{black}The distribution of Balmer decrement in aperture corrected H$\alpha$ luminosity (L$_{\rm{H}\alpha, \rm{ApCor}}$, i.e.\,H$\alpha$ luminosity before correcting for dust obscuration) illustrating the luminosity dependence of dust obscuration. The grey colour scale shows the data density distribution of all star forming galaxies. The black dashed and dot-dashed lines indicate the theoretical Case B recombination ratio of 2.86, and the Balmer decrement corresponding to the assumption of one magnitude extinction at the wavelength of H$\alpha$. The blue points denote the mean variation and one-sigma error in dust obscuration as a function of L$_{\rm{H}\alpha, \rm{ApCor}}$. The constant log sSFR contours, shown in red, are defined in steps of 0.3 dex, where $\log$ sSFR increases from $-10.2$[yr$^{-1}$] at low Balmer decrements to $-9$[yr$^{-1}$] at high Balmer decrements.}
\label{fig:BD_vs_lum}
\end{center}
\end{figure}
\begin{figure*}
\begin{minipage}{\textwidth}
\begin{center}
\includegraphics[trim={1.2cm 1.0cm 1.5cm 0.8cm},clip,width=0.49\textwidth]{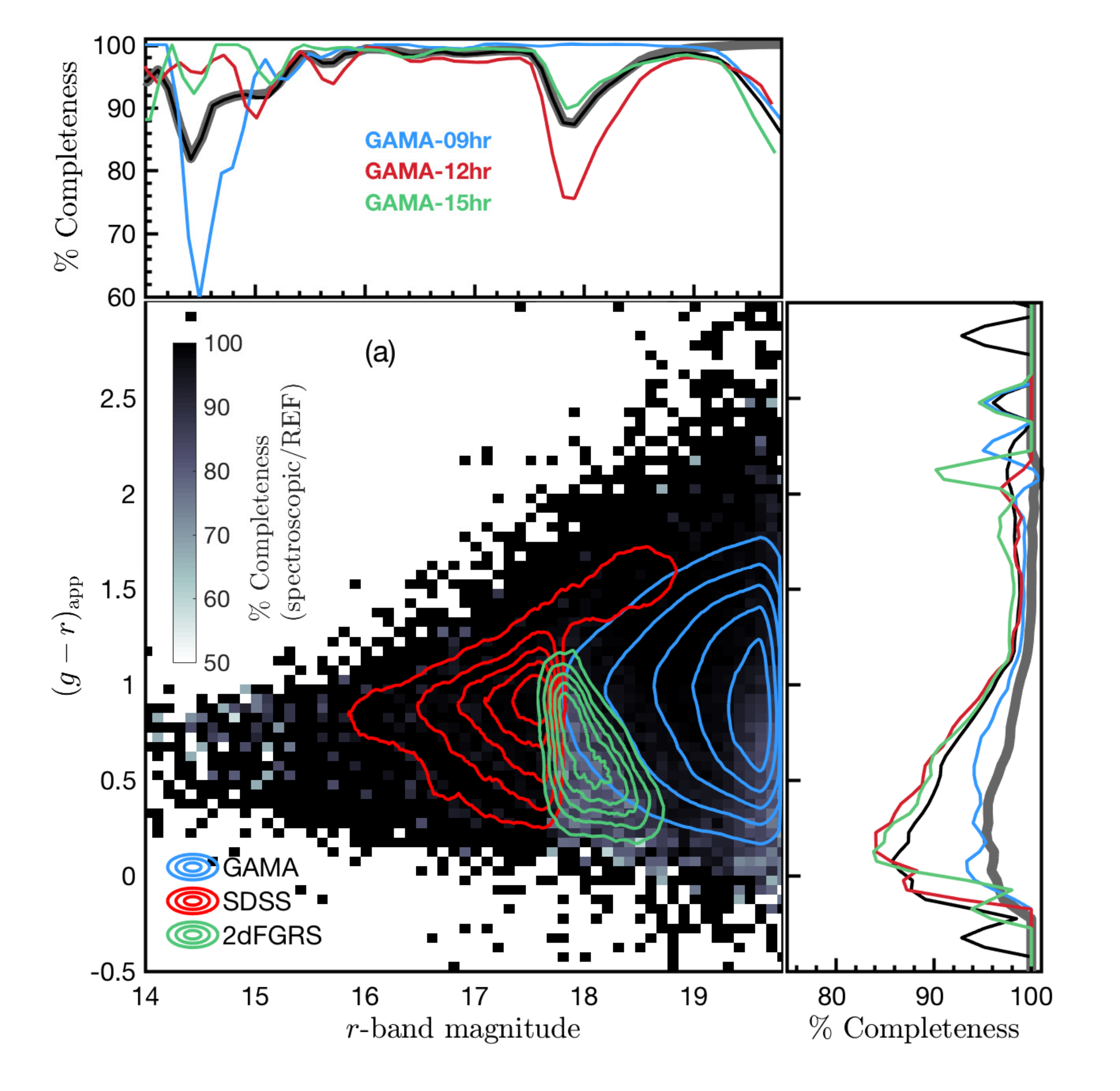}
\includegraphics[trim={0.2cm 0.5cm 0.9cm 0.4cm},clip,width=0.5\textwidth]{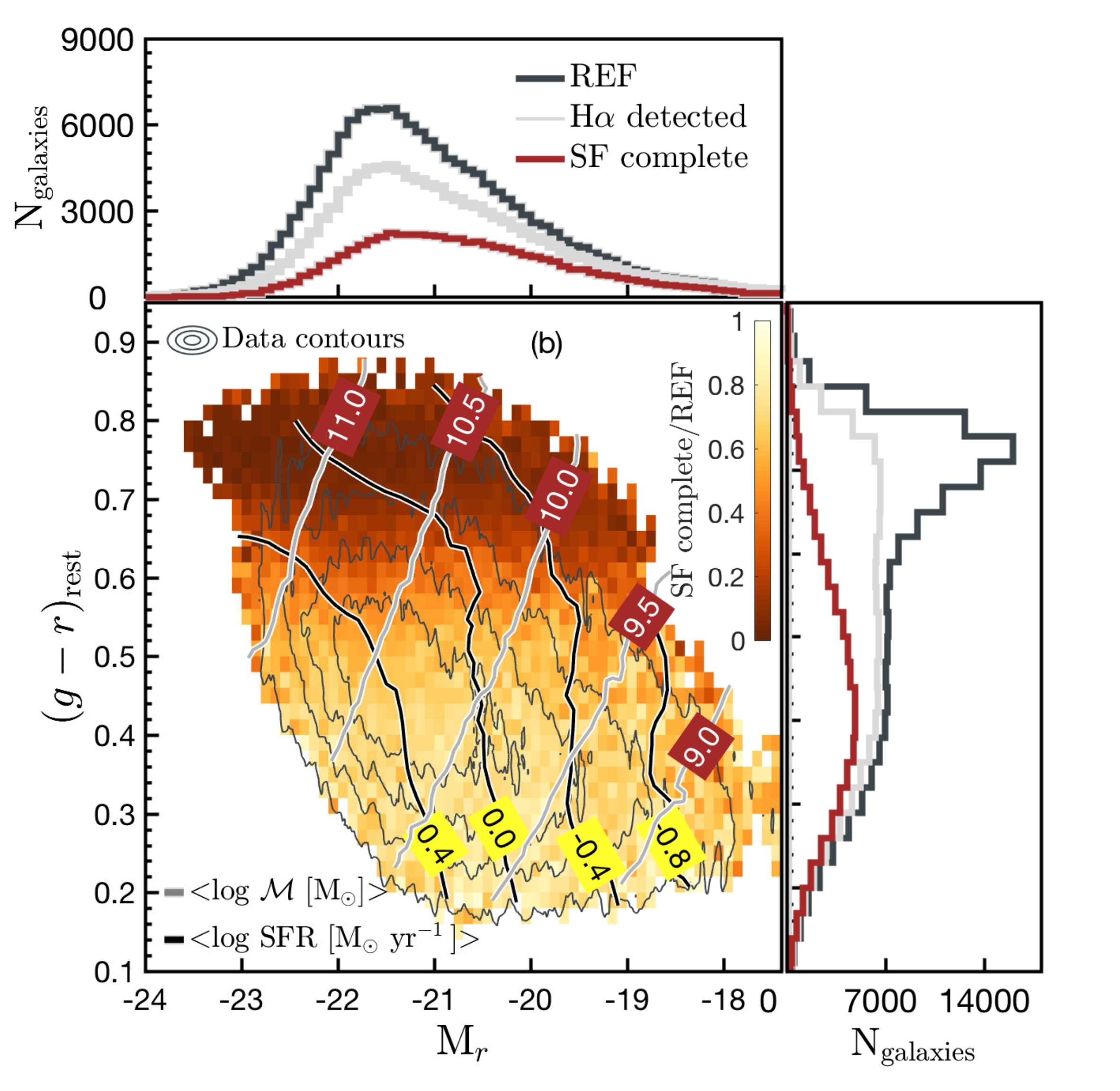}
\caption{\color{black}(a) The apparent $g-r$ colour, ($g-r$)$_{\rm{app}}$, and $r$-band Petrosian magnitude distributions of the ratios of spectroscopic-to-REF galaxies. The colour code corresponds to the percentage completeness with lighter colours indicating the deviation of the ratios from unity. The coloured contours show the approximate distribution of galaxies in our sample originating from GAMA, SDSS, 2dFGRS surveys. The top and side panels show completeness as a function of $r$-band Petrosian magnitude and ($g-r$)$_{\rm{app}}$, respectively, with black and thick grey lines showing the overall completeness across the three equatorial fields with (black) and without (grey) a spectral signal-to-noise cut, and the coloured lines showing the completenesses for individual GAMA fields. (b) The ($g-r$)$_{\rm{rest}}$ and M$_r$ distribution of the ratio of SF complete-to-REF galaxies. The closed contours from inwards-to-outwards enclose  $\sim25$, $50$ ,$75$ and $90\%$ of the SF complete data. Also shown are the constant mean log stellar mass ($<\log \mathcal{M}/$M$_\odot>$) and mean log SFR ($<\log$ SFR [M$_\odot$ yr$^{-1}$]$>$) contours corresponding to SF complete galaxies. The top and side panels show the univariate M$_r$ and ($g-r$)$_{\rm{rest}}$ distributions of REF (black) and SF complete (brown) galaxies, as well as the distribution all SF galaxies with reliably measured H$\alpha$ emission line fluxes (grey).  }
\label{fig:completeness}
\end{center}
\end{minipage}
\end{figure*}

The luminosity (or SFR) dependent dust obscuration, reflecting that massive star forming galaxies also contain large amounts of dust relative to their low-SFR counterparts, is observationally well established in the local Universe \citep[e.g.][]{Hopkins2003, Brinchmann2004, Garn2010, Ly2012, Zahid2013, Jimmy2016}. The mean variation in Balmer decrement with aperture corrected H$\alpha$ luminosity for our sample is shown as blue points in Figure\,\ref{fig:BD_vs_lum}, with red contours indicating the dependence of Balmer decrement on specific SFR. The dot-dashed line denotes the Balmer decrement approximately corresponding to the assumption of an extinction of one magnitude at the wavelength of H$\alpha$ for all galaxy luminosities \citep{Kennicutt1992}. In this study, for galaxies without reliable H$\beta$ flux measurements, we approximate a Balmer decrement based on the relation shown in blue in Figure\,\ref{fig:BD_vs_lum}.

\subsubsection{Stellar masses}

The GAMA stellar masses and absolute magnitudes\footnote{The rest-frame colours used in this analysis are based on these absolute magnitudes.} provided in \textsc{StellarMassesv16} \citep{Taylor2011, Kelvin2012} catalogue are used for this study. A Bayesian approach is used in the derivation of the stellar masses, and are based on $u,g,r,i,z$ spectral distributions and \cite{Bruzual2003} population synthesis models. Furthermore, the derivation assumes a \cite{Chabrier2003} stellar IMF and \cite{Calzetti2000} dust law. The stellar mass uncertainties, modulo any uncertainties associated with stellar population synthesis models, are determined to be $\sim0.1\,$dex. A detailed discussion on the estimation of GAMA stellar masses and the associated uncertainties can be found in \cite{Taylor2011}.

\subsection{Sample selection and spectroscopic completeness}\label{subsec:selection}

We select a \textit{reference} sample of galaxies, henceforth REF, consisting only of equatorial objects that satisfy both the GAMA main survey selection criteria \citep{Baldry2010}, and have spectroscopic redshifts, $z_{\rm{spec}}$, in the range $0.002\leq z_{\rm{spec}}<0.35$, representing the $z$ window over which the H$\alpha$ spectral feature is observable in the GAMA spectra \citep{Driver2011}. The REF sample consists of {157\,079} objects in total. 

Out of the REF galaxies, those observed either as a part of GAMA and/or SDSS spectroscopic surveys with spectral signal-to-noise $>$3 form the \textit{spectroscopic} sample. Objects with other survey spectra (e.g.\,2dFGRS, MGC) are excluded as they lack the necessary information needed to reliably flux calibrate their spectra, and the objects with duplicate spectra\footnote{In cases where an object has an independent GAMA and a SDSS spectrum, the SDSS spectrum is generally found to have the highest spectral signal-to-noise, and is selected to be part of the sample.} are removed on the basis of their spectral signal-to-noise, leaving 148\,834 galaxies in the spectroscopic sample.    

We assess the spectroscopic completeness of the survey by comparing the bivariate colour-magnitude distributions of REF and spectroscopic samples. Figure\,\ref{fig:completeness}(a) shows the colour-magnitude distribution of the ratio of spectroscopic--to--REF galaxies in a given $r$-band magnitude and apparent $g-r$ colour, hereafter ($g-r$)$_{\rm{app}}$, cell, and the top and right-side panels show the completeness as a function of the $r$-band magnitude and ($g-r$)$_{\rm{app}}$. The exclusion of 2dFGRS spectra, in particular, leads to an overall incompleteness of $\sim20\%$ across the three equatorial regions over the magnitude range probed by the 2dFGRS (green contours in Figure\,\ref{fig:completeness}(a) highlight the colour and magnitude range corresponding to the 2dFGRS galaxy distribution). The incompleteness present in each field, however, varies considerably, with G12 being the most incomplete (i.e.\,relatively a larger number of 2dFGRS galaxies reside in this region) and G09 being the most complete (i.e.\,no 2dFGRS galaxies reside in this region) as shown in the top panel of Figure\,\ref{fig:completeness}(a). Additionally, recall that GAMA spectral signal-to-noise measures are representative of the red end of the spectrum, therefore, the application of a signal-to-noise cut results in the incompleteness evident at fainter magnitudes and bluer colours in the same figure. The implication being that the spectroscopic sample is biased against optically faint bluer galaxies (the thin and thick black lines shown in the side panels of Figure\,\ref{fig:completeness}(a) clearly demonstrate this bias). Note that the variations in completeness seen at optically redder colours is largely driven by small number statistics. See \S\,\ref{subsec:systematics} for discussion on the impact of spectroscopic incompleteness on the results and conclusions of this study. 

Out of the galaxies with detected H$\alpha$ emission in the spectroscopic sample, those dominated by active galactic nuclei (AGN) emission are removed using the standard optical emission line ([\ion{N}{ii}]~$\lambda6584$/H$\alpha$ and [\ion{O}{iii}] $\lambda5007$/H$\beta$) diagnostics \citep[BPT;][]{Baldwin1981} and the \cite{Kauffmann2003} pure star forming (SF) and AGN discrimination prescription. If all four emission lines needed for a BPT diagnostic are not detected for a given galaxy, then the two line diagnostics based on the \cite{Kauffmann2003} method (e.g.\,$\log$ [\ion{N}{ii}] $\lambda6584$/H$\alpha>0.2$ and $\log$ [\ion{O}{iii}] $\lambda5007$/H$\beta>1.0$) are used for the classification. The galaxies that were unable to be classified this way are retained in our sample as a galaxy with measured H$\alpha$ flux but without an [\ion{N}{ii}] $\lambda6584$ or [\ion{O}{iii}] $\lambda5007$ measurement are more likely to be SF galaxies than AGNs \citep{CidFernandes2011}. Overall, $\sim$16$\%$ of objects are classified either as an AGN or as an AGN--SF composite and are removed from the sample, and the $\sim$28$\%$  unable to be classified are retained in the sample. 

As a consequence of the bivariate magnitude and H$\alpha$ flux selection that is applied to our sample, our sample is biased against optically faint SF galaxies. This is a bias that not only affects any star forming galaxy sample drawn a broadband magnitude survey, but it becomes progressively more significant with increasing $z$ \citep{Gunawardhana2015}. Therefore to select an approximately complete SF galaxy sample, henceforth \textit{SF complete}, we impose an additional flux cut of $1\times10^{-18}$\,Wm$^{-2}$, which roughly corresponds to the turn-over in the observed H$\alpha$ flux distribution of GAMA H$\alpha$ detected galaxies \citep{Gunawardhana2013}.

A comparison between the SF complete sample and REF galaxies in rest-frame $g-r$ colour, hereafter ($g-r$)$_{\rm{rest}}$, and M$_r$ space is shown in Figure\,\ref{fig:completeness}(b). The closed contours denote the fraction of the data enclosed, while the open black and grey contours denote constant $\langle\log$ SFR [M$_\odot$ yr$^{-1}$]$\rangle$ and $\langle\log \mathcal{M}/$M$_\odot\rangle$ lines, respectively.  Even though the SF complete galaxies are dominated by optically bluer systems, a significant fraction of galaxies with optically redder colours have reliably measured H$\alpha$ SFRs, indicating on-going star formation, albeit at lower rates. Also shown are the univariate M$_r$ and ($g-r$)$_{\rm{rest}}$ distributions of REF (black), SF complete (brown), and of galaxies with reliable H$\alpha$ emission detections that are classified as SF following the removal of AGNs (grey) to illustrate how the H$\alpha$ flux cut of $1\times10^{-18}$ Wm$^{-2}$ act to largely exclude optically redder systems from our sample. 

\subsection{REF and SF complete samples for clustering analysis}\label{subsec:clustering_samples}
\begin{table*}
\caption{The key characteristics of the three disjoint luminosity selected sub-samples (M$_b$: $-23.5\leq$ M$_r$<$-21.5$; M$_*$: $-21.5\leq$ M$_r$<$-20.5$; M$_f$: $-20.5\leq$ M$_r$<$-19.5$) drawn from the SF complete and REF samples are given. For each sample, we provide the size of the sample, the average redshift and central $\sim50\%$ redshift range, median log sSFR [yr$^{-1}$], ($g-r$)$_{\rm{rest}}$ and $\log\,\mathcal{M}$ [M$_{\odot}$] along with their central $\sim50\%$ ranges. We define two redshift samples for each M$_b$, M$_*$ and M$_f$, where one sample covers the full redshift range over which the H$\alpha$ feature is visible in GAMA spectra (i.e.\,$0.001<z<0.34$), and the second covers a narrower $0.001<z\leq0.24$ range (see \S\,\ref{subsec:cross_corr}). {Using both the $r$--band magnitude selection of the GAMA survey and the H$\alpha$ flux selection of our sample, we estimate a completeness for each disjoint luminosity selected sub-sample, which is shown within brackets under N$_{\rm{galaxies}}$.}}
\begin{minipage}{1\textwidth}
\begin{tabular}{clcccccccc}
\hline
subset  & N$_{\rm{galaxies}}$ & $\langle z\rangle$ & $z$ & ${\log}$ sSFR & log sSFR & $\langle$($g-r$)$_{\rm{rest}}\rangle$ & ($g-r$)$_{\rm{rest}}$ & $\langle$log $\mathcal{M}\rangle$ & log $\mathcal{M}$  \\
  & & & $_{\sigma=25\%, 75\%}$ & [yr$^{-1}$] & $_{\sigma=25\%, 75\%}$ & & $_{\sigma=25\%, 75\%}$ & [M$_{\odot}$] & $_{\sigma=25\%, 75\%}$ \\
\hline
\hline
\multicolumn{10}{c}{{\textbf{SF complete}} } \\
\hline
M$_b$ &  8\,100 (53\%)\footnote{The sample completeness}   & 0.24 & (0.19, 0.29) & -10.28 & (-10.70, -9.87) & 0.55 & (0.47, 0.63) & 10.9 & (10.8, 11.1) \\
	   &  3\,749 (68\%)  & 0.17  & (0.13, 0.21)  & -10.67  & (-11.08, -10.13)  & 0.60  & (0.51, 0.69)  & 10.8  &  (10.68, 10.99)\\
M$_*$  & 20\,976  (12\%) & 0.21 & (0.18, 0.27) & -9.90 & (-10.20, -9.61) & 0.48 & (0.39, 0.56) & 10.46 & (10.31, 10.65) \\
  	    & 12\,308 (62\%)  & 0.17  & (0.13, 0.21)  & -10.11  & (-10.52, -9.79) & 0.50  & (0.41, 0.59)  & 10.32 &  (10.15, 10.50)\\
M$_f$  & 14\,000 ($<1\%$)  & 0.14 & (0.11, 0.18) & -9.84 & (-10.14, -9.54) & 0.42 & (0.32, 0.51) & 9.98 & (9.81, 10.16) \\
	    & 13\,650 ($<1\%$)  & 0.14  &  (0.11, 0.18) & -9.94 & (-10.24, -9.64)  & 0.42  & (0.33, 0.51)  &  9.83 &  (9.66, 10.02)\\
\hline 
\multicolumn{10}{c}{{\textbf{REF}} } \\
\hline
M$_b$  & 33\,406 & 0.25 & (0.20, 0.30) & -- & -- & 0.67 & (0.59, 0.75) & 10.95 & (10.83, 11.09)  \\
M$_*$  & 64\,618  & 0.22 & (0.18, 0.27) & -- & --  & 0.59 & (0.48, 0.72) & 10.50 & (10.34, 10.69)  \\
M$_f$  & 34\,868  & 0.15 & (0.13, 0.19) & --  & --  & 0.51 & (0.37, 0.67) & 9.98 & (9.76, 10.20)  \\
\hline
\end{tabular}
\end{minipage}
\label{table:stats1}
\end{table*}%

\begin{table*}
\begin{minipage}{1\textwidth}
\caption{The key characteristics of the three disjoint stellar mass selected sub-samples ($\mathcal{M}_{\mathcal{H}}$: $10.5\leq\log\mathcal{M}/$M$_{\odot}\leq11.0$; $\mathcal{M}_{\mathcal{I}}$: $10.0\leq\log\mathcal{M}/$M$_{\odot}\leq10.5$; $\mathcal{M}_{\mathcal{L}}$: $9.5\leq\log\mathcal{M}/$M$_{\odot}\leq10.0$) drawn from the SF complete and REF samples are given. For each sample, we provide the size of the sample, the average redshift and central $\sim50\%$ range, median log sSFR, ($g-r$)$_{\rm{rest}}$ and M$_r$ along with their central $\sim50\%$ ranges. As described in the caption of Table\,\ref{table:stats1} above, we define two redshift samples for each $\mathcal{M}_{\mathcal{H}}$, $\mathcal{M}_{\mathcal{I}}$ and $\mathcal{M}_{\mathcal{L}}$. The completeness of each sample due to the dual $r$-band magnitude and H$\alpha$ flux is indicated within brackets in the second column (after N$_{\rm{galaxies}}$), which is approximately the fraction of galaxies seen over the full volume. This value does not take into account the maximum volume out to which a galaxy of a given stellar mass would be detected.}
\begin{tabular}{clcccccccc}
\hline
subset  & N$_{\rm{galaxies}}$ & $\langle z\rangle$ & $z$ & ${\log}$ sSFR & log sSFR & $\langle$($g-r$)$_{\rm{rest}}\rangle$ & ($g-r$)$_{\rm{rest}}$ & $\langle$M$_r\rangle$ & M$_r$  \\
  & & & $_{\sigma=25\%, 75\%}$ & [yr$^{-1}$] & $_{\sigma=25\%, 75\%}$ & & $_{\sigma=25\%, 75\%}$ &  & $_{\sigma=25\%, 75\%}$ \\
\hline
\hline
\multicolumn{10}{c}{{\textbf{SF complete}} } \\
\hline
$\mathcal{M}_{\mathcal{H}}$ &  11\,600 (36\%)   & 0.23 & (0.18, 0.30) & -10.35 & (-10.72, -9.98) & 0.57 & (0.50, 0.64) & -21.53 & (-21.78, -21.29) \\
	   				    &  5\,597 (61\%) & 0.16 & (0.12, 0.21)  & -10.57 & (-10.99, -10.17) & 0.61 & (0.54, 0.68) & -21.46 & (-21.72, -21.20) \\
$\mathcal{M}_{\mathcal{I}}$  & 18\,103  (11\%) & 0.20 & (0.14, 0.26) & -10.01 & (-10.29, -9.71) & 0.47 & (0.40, 0.54) & -20.82 & (-21.10, -20.55) \\
  	    				     & 12\,135 (47\%)  & 0.16 & (0.12, 0.21) & -10.12 & (-10.43, -9.81) & 0.51 & (0.43, 0.58) & -20.69 & (-20.96, -20.43) \\
$\mathcal{M}_{\mathcal{L}}$  & 12\,647 ($<1\%$)          & 0.15 & (0.11, 0.19)      & -9.86 & (-10.16, -9.57) & 0.39 & (0.31, 0.45) & -20.01 & (-20.34, -19.69) \\
	    				      & 11\,648 ($\sim14\%$)    & 0.14  & (0.10, 0.18)  & -9.90 & (-10.18, -9.62) & 0.40 & (0.32, 0.46) & -19.95  & (-20.27, -19.66) \\
\hline 
\multicolumn{10}{c}{{\textbf{REF}} } \\
\hline
$\mathcal{M}_{\mathcal{H}}$  & 54\,681 & 0.24 & (0.19, 0.29) & -- & -- & 0.67 & (0.60, 0.74) & -21.36 & (-21.61, -21.10)  \\
$\mathcal{M}_{\mathcal{I}}$  & 44\,146  & 0.19 & (0.15, 0.24) & -- & --  & 0.55 & (0.44, 0.67) & -21.64 & (-20.95, -20.33)  \\
$\mathcal{M}_{\mathcal{L}}$  & 23\,615  & 0.15 & (0.11, 0.18) & --  & --  & 0.42 & (0.33, 0.50) & -19.91 & (-20.26, -19.57)  \\
\hline
\end{tabular}
\label{table:stats3}
\end{minipage}
\end{table*}%

In order to investigate the clustering properties of star forming galaxies with respect to optical luminosity and stellar mass (\S\,\ref{subsec:auto_corr} to \S\,\ref{subsec:mark_cfs}), we use REF and SF complete samples to further define three disjoint luminosity selected, three disjoint stellar mass selected, and several volume limited samples, for which all selection effects are carefully modelled. 

{{The three disjoint luminosity selected samples, called M$_f$, M$_*$ and M$_b$, together cover the $-23.5\leq$M$_r<-19.5$ range, and the three disjoint stellar mass selected samples, called $\mathcal{M}_{\mathcal{L}}$, $\mathcal{M}_{\mathcal{I}}$ and $\mathcal{M}_{\mathcal{H}}$, together span the $9.5\leq \log \mathcal{M}$/M$_{\odot}<11$. See Tables\,\ref{table:stats1} and \ref{table:stats3} for individual magnitude and stellar mass coverages of each luminosity and stellar mass selected sample, as well as for a description of their key characteristics. We also define two redshift samples for each M$_b$, M$_*$, and M$_f$, and for each $\mathcal{M}_{\mathcal{H}}$, $\mathcal{M}_{\mathcal{I}}$ and $\mathcal{M}_{\mathcal{L}}$, where one set covers the full redshift range of the SF complete galaxies, and the second spans only the $0.001\leq z\leq0.24$ range.}} 

Out of the two redshift samples mentioned above, the former (i.e.\,the samples covering the full redshift range) is used for the auto correlation analysis, and the latter for the cross and mark correlation analyses (\S\,\ref{subsec:cross_corr} and \ref{subsec:mark_cfs}).
The main reason for restricting the redshift coverage of galaxy samples in the latter case is to overcome the effects of the equivalent width bias\footnote{{{Emission line samples drawn from a broadband survey, like GAMA, can be biased against low SFR and weak-line systems. This can become significant with increasing redshift and apparent magnitude, and the differences in clustering results obtained from different clustering estimators can be used to quantify the significance of such biases.}}} \citep[][see also Appendix\,\ref{app:modelling}]{Liang2004, Groves2012}. In this study, we find that the cross correlation functions, hereafter CCFs, of low sSFR galaxies spanning $0.24\leq z<0.34$ in redshift computed using two different clustering estimators, the \cite{LS1993} and \cite{Hamilton1993} estimators, differ systematically from each other, suggesting a failure in the modelling of the selection function of low sSFR galaxies over the $0.24\leq z<0.34$ range. The respective results for the low sSFR galaxies in the $0.01\leq z\leq0.24$ range, on the other hand, are consistent with each other. Therefore we limit the redshift range of all galaxy samples used for the cross and mark correlation analyses to $0.01\leq z\leq0.24$.        

\begin{figure}
\begin{center}
\includegraphics[width=0.48\textwidth]{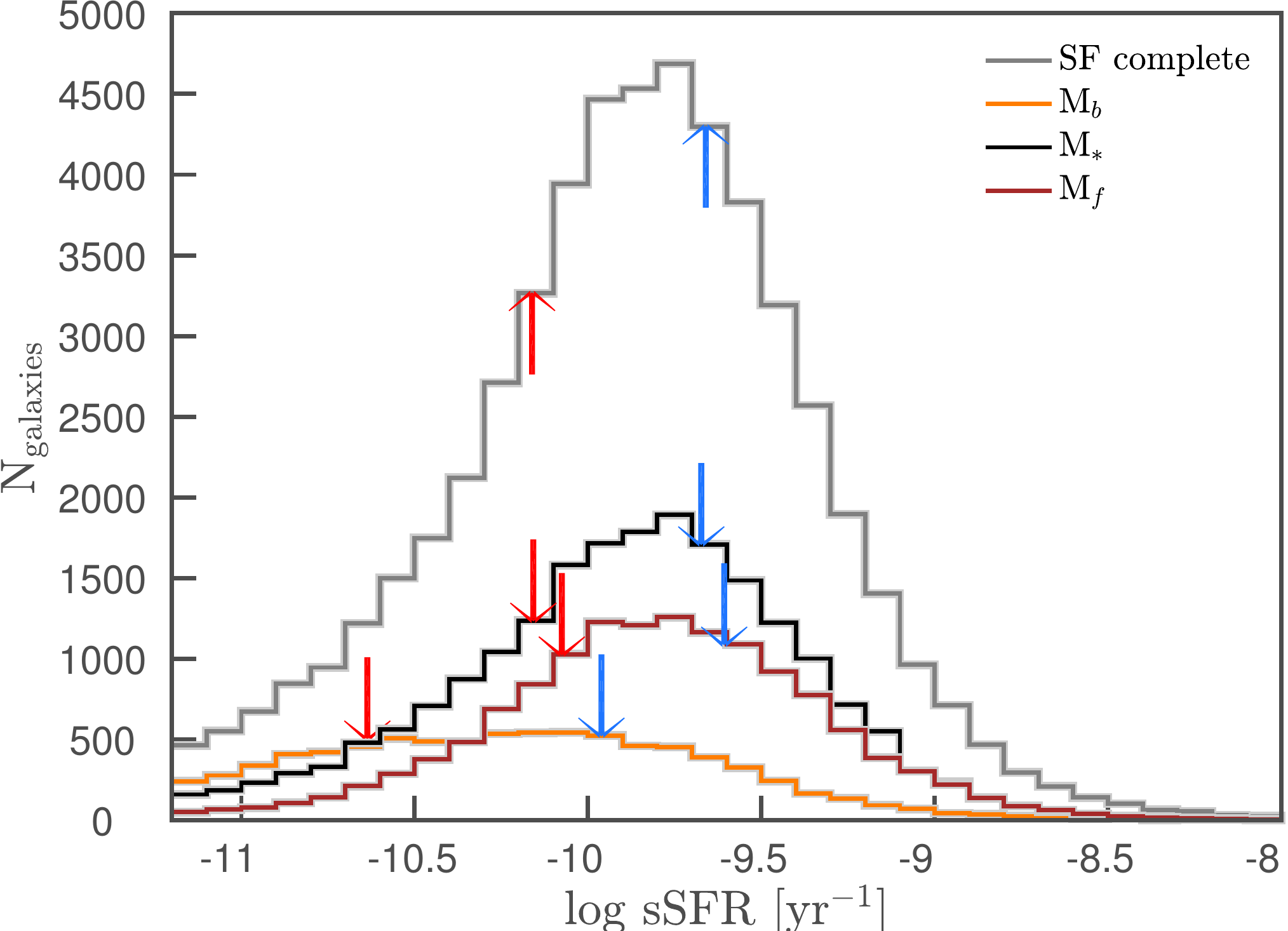}
\caption{The $\log$ sSFR distributions of all SF complete galaxies (grey), as well as M$_b$, M$_*$, and M$_f$ galaxies of SF complete sample. The redshift range considered is $0.001<z\leq0.24$, and the arrows indicate the sSFR cuts used to select the $30\%$ highest (blue arrows) and the $30\%$ lowest (red arrows) sSFR galaxies from each distribution.}
\label{fig:colour_ssfr_cuts2}
\end{center}
\end{figure}
\begin{figure}
\begin{center}
\includegraphics[width=0.48\textwidth]{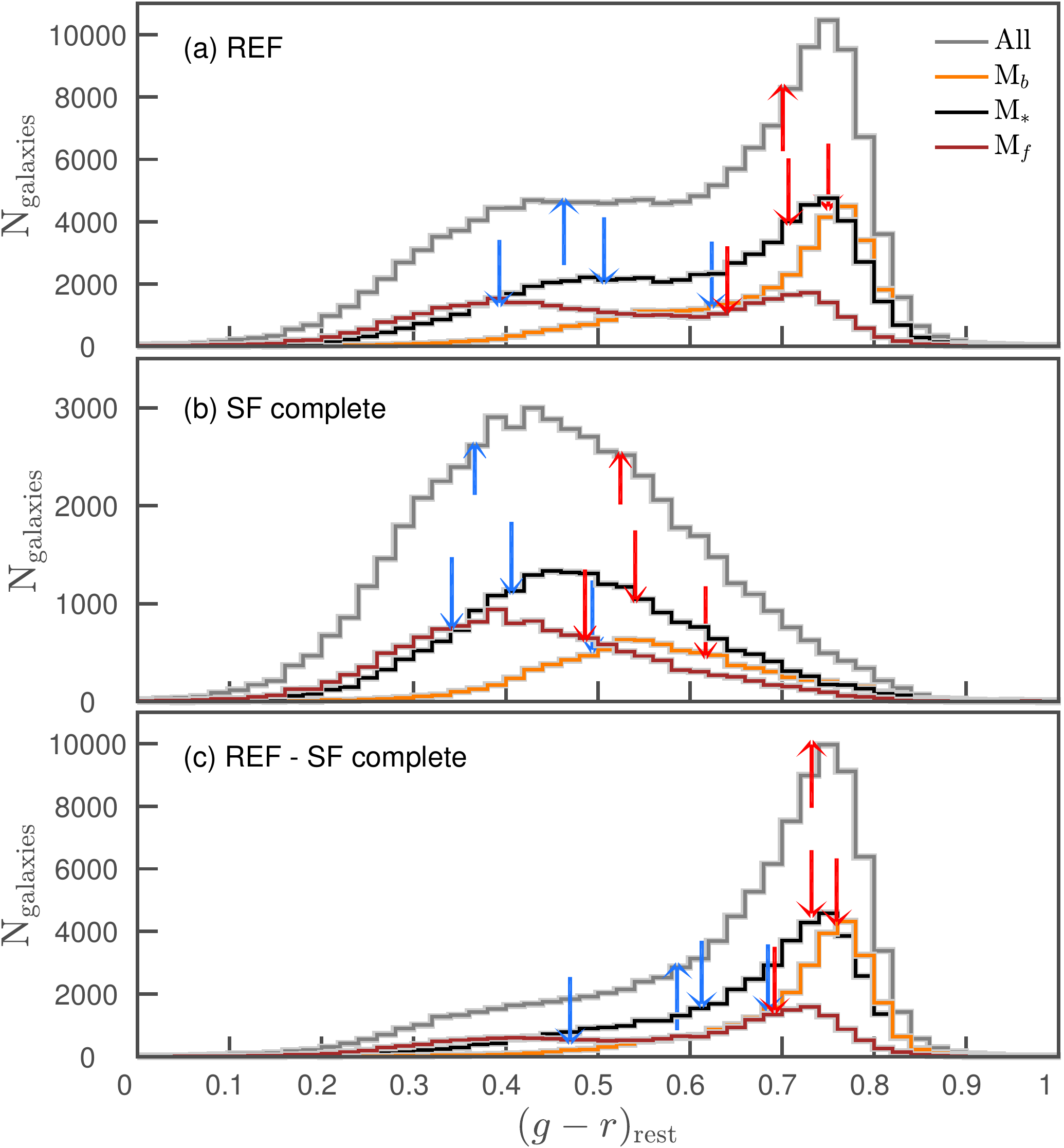}
\caption{The ($g-r$)$_{\rm{rest}}$ distributions of (a) all REF and (b) all SF complete galaxies, as well as the distributions of their respective M$_b$, M$_*$ and M$_f$ sub-samples. For completeness, we also show in panel (c) the distributions of REF - SF complete galaxies. The redshift range considered is $0.001<z\leq0.24$, and the arrows indicate the colour cuts used to select the $30\%$ bluest (blue arrows) and the $30\%$ reddest (red arrows) colour galaxies from each distribution. The arrows show a clear change in position with luminosity (i.e.\,arrows move towards redder colours with increasing optical brightness), that is not seen with log sSFR (Figure\,\ref{fig:colour_ssfr_cuts2}).}
\label{fig:colour_ssfr_cuts1}
\end{center}
\end{figure}

The log sSFR and ($g-r$)$_{\rm{rest}}$ distributions of the three disjoint luminosity selected samples are shown in Figures\,\ref{fig:colour_ssfr_cuts2} and \ref{fig:colour_ssfr_cuts1}. In Figure\,\ref{fig:colour_ssfr_cuts2}, with increasing optical luminosity, the peak of the distribution of log sSFRs moves progressively towards lower sSFRs. The notably broader peak of the M$_b$ distribution arises as a result of the bimodality present in the bivariate SFR (or sSFR) and $\mathcal{M}$ distribution (see, for example, Figure\,\ref{fig:ssfr_mass_dist}). Similarly, the ($g-r$)$_{\rm{rest}}$ distributions show a progressive shift towards redder colours with increasing optical luminosity. 
From each disjoint luminosity (stellar mass) selected sample, we select the 30$\%$ highest and lowest sSFR (SFR), ($g-r$)$_{\rm{rest}}$, Balmer decrement and {D$_{\rm{4000}}$ \citep[i.e.\,the strength of the $4000$\,\AA\,break, ][]{Kauffmann2003b}} galaxies to be used in the cross correlation analysis (\S\,\ref{subsec:cross_corr}). The red and blue arrows in Figures\,\ref{fig:colour_ssfr_cuts2} and \ref{fig:colour_ssfr_cuts1} show these $30\%$ selections. 
   
As none of the samples defined so far are truly volume limited, we define a series of volume limited luminosity and stellar mass samples, which are described in Table\,\ref{table:stats2}. The volume limited SF complete samples are defined to be at least $95\%$ complete\footnote{This completeness is achieved through excluding very low-SFR sources as they can significantly limit the redshift coverage of a volume limited sample, resulting in samples with small number statistics.} with respect to the bivariate $r$--band magnitude and H$\alpha$ flux selections. While this implies, by definition, that each volume limited luminosity sample is at least $95\%$ volume limited, the same cannot be said about the volume limited stellar mass samples. To achieve a $95\%$ completeness in volume limited stellar mass samples would require the additional consideration of the detectability of a galaxy of a given stellar mass within the survey volume. It is, however, reasonable to assume that the "volume limited stellar mass" samples are close to $95\%$ volume limited given the strong correlation between stellar mass and optical luminosity. {For our sample, the 1$\sigma$ scatter in stellar mass--luminosity correlation is $\sim0.4$ dex}. The volume limited REF samples have the same redshift coverage as their SF counterparts, and as such, they are $100\%$ complete with respect to their univariate magnitude selection. 

\section{Clustering Methods} \label{sec:clustering_methods}

In this section, we describe the modelling of the galaxy selection function using GAMA random galaxy catalogues, and introduce two-point galaxy correlation function estimators used in the analysis. \vspace{1.5cm}

\subsection{Modelling of the selection function}\label{subsec:modelling_randoms}

To model the selection function, we use the GAMA random galaxy catalogues ({\color{black}Random DMU}) introduced in \cite{Farrow2015}. Briefly, \cite{Farrow2015} employ the method of \cite{Cole2011} to generate clones of observed galaxies, where the number of clones generated per galaxy is proportional to the ratio of the maximum volume out to which that galaxy is visible given the magnitude constraints of the survey (V$_{\rm{max} , r}$) to the same volume weighted by the number density with redshift, {taking into account targeting and redshift incompletenesses}.  
\begin{figure*}
\begin{center}
\includegraphics[width=0.353\textwidth]{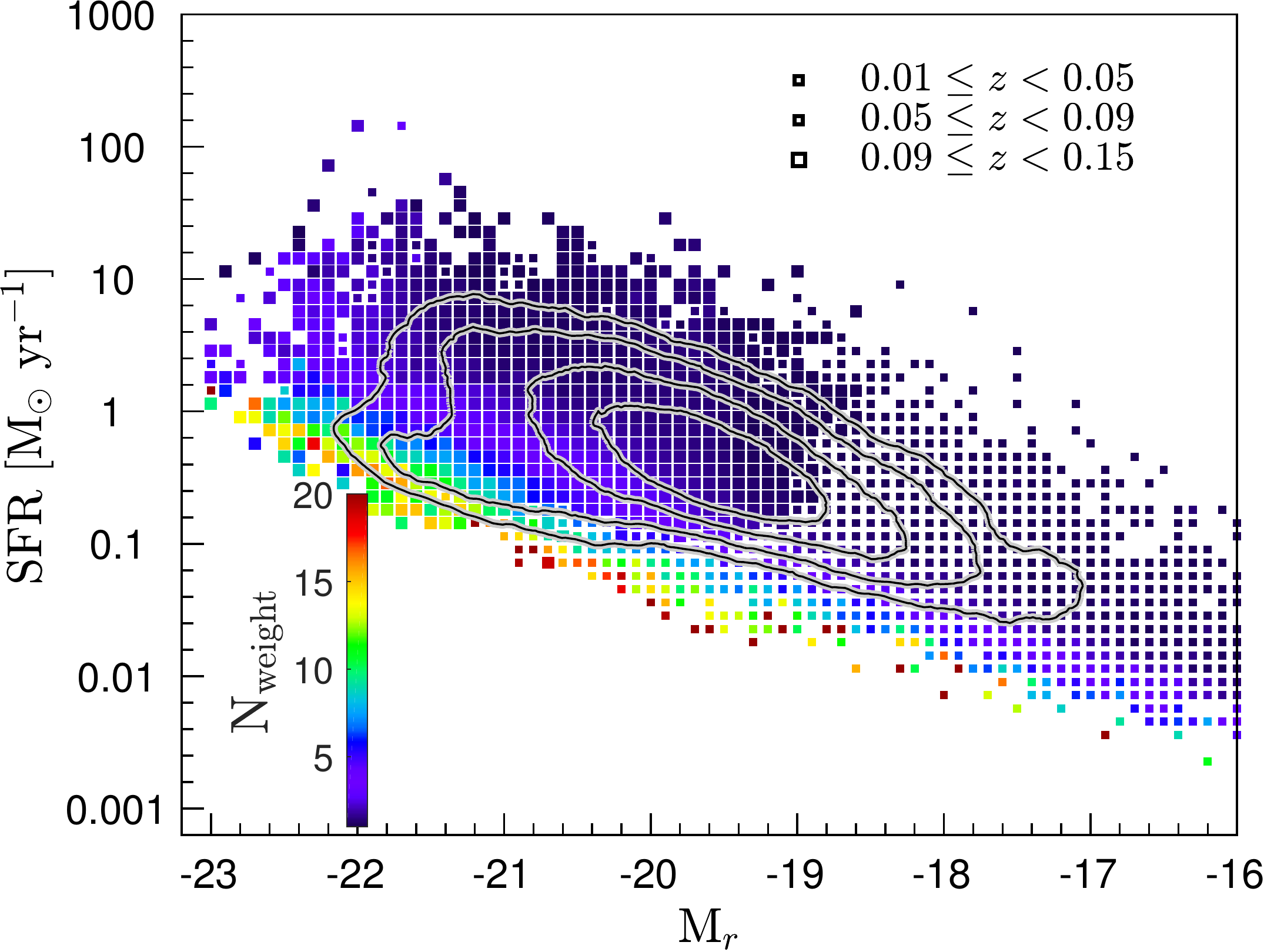}
\includegraphics[width=0.319\textwidth]{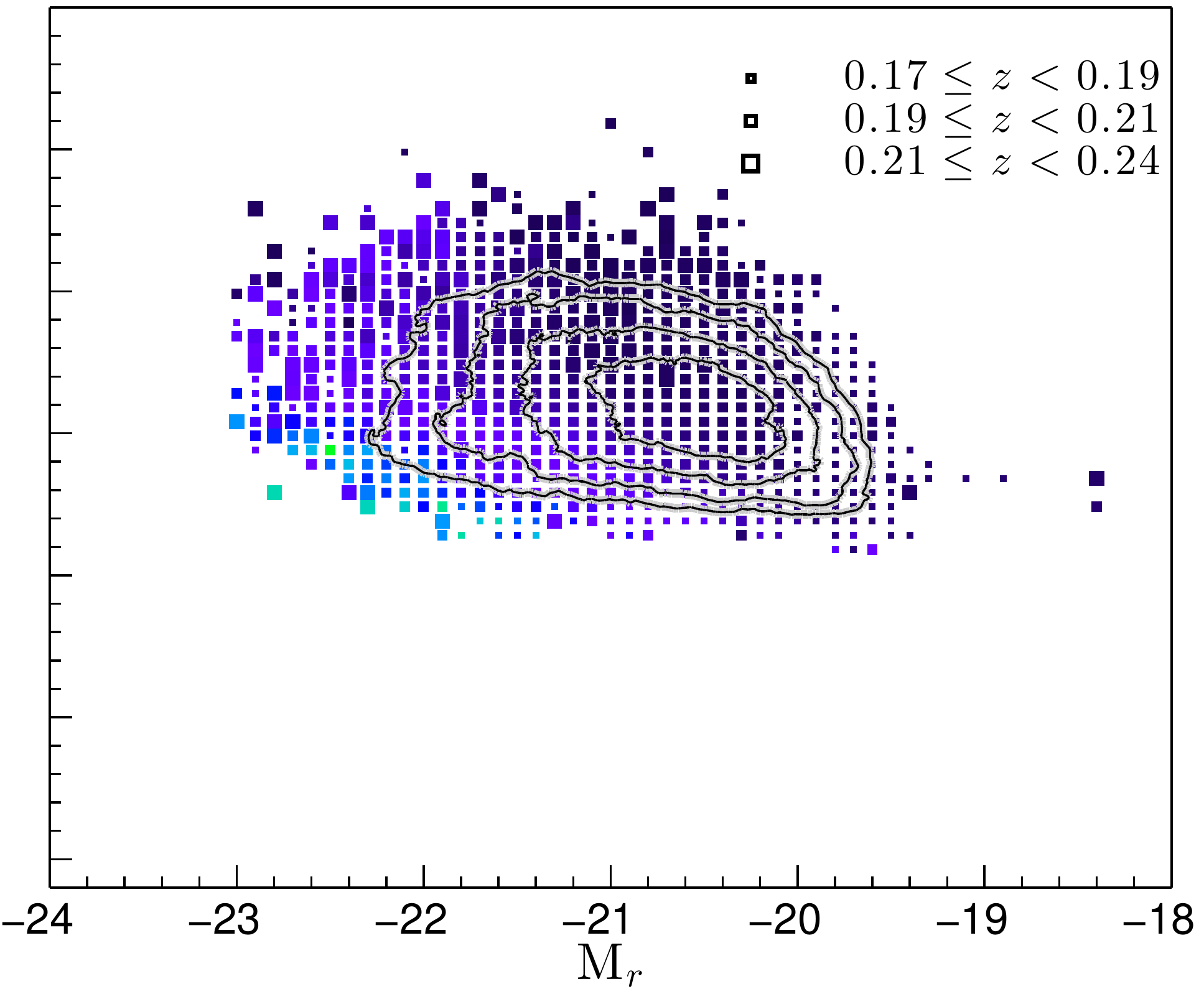}
\includegraphics[width=0.319\textwidth]{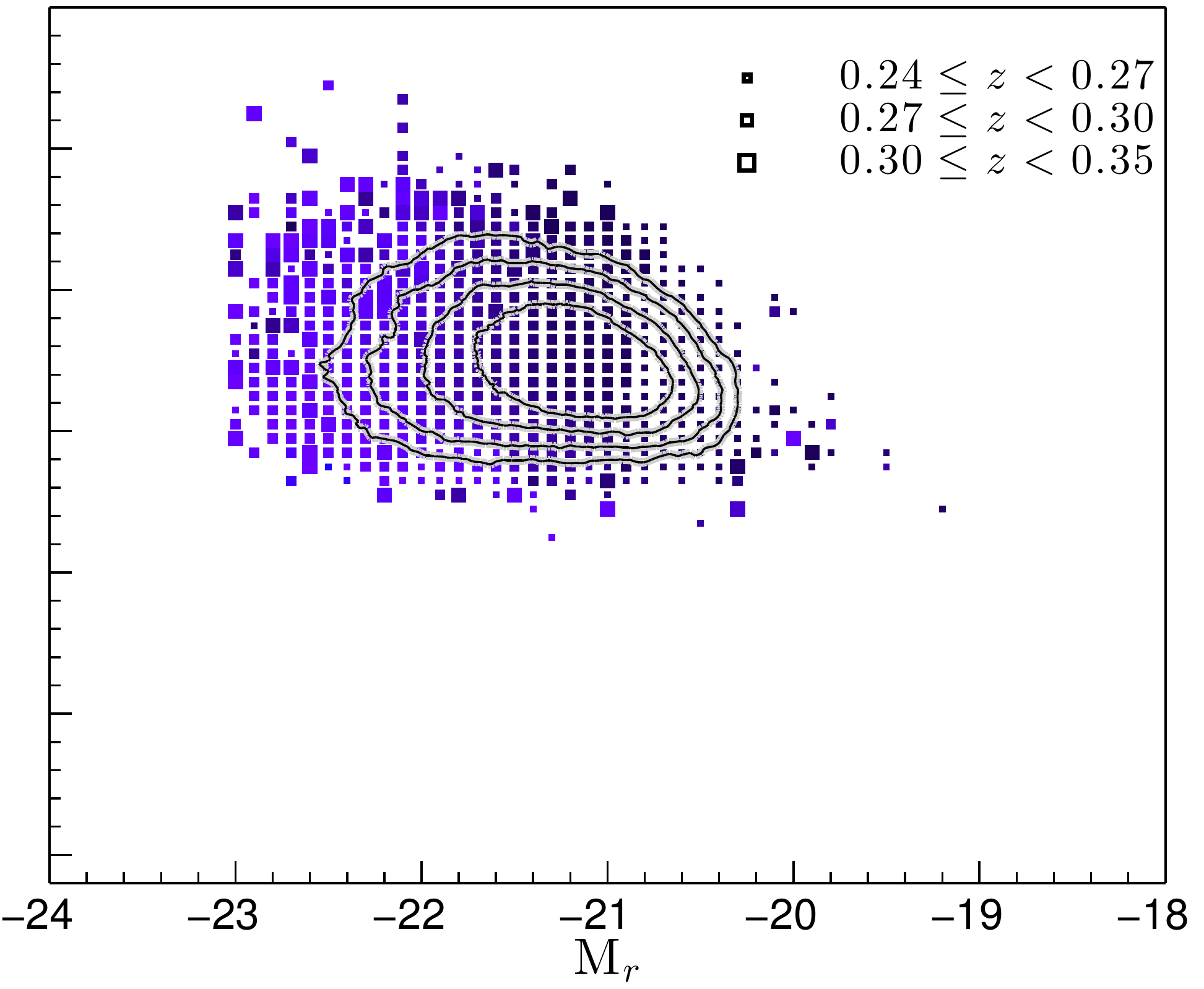}
\caption{\color{black}Mean weight applied to the random SF complete sample as a function of their intrinsic SFR and M$_r$. The size of the markers indicates the mean redshift of GAMA SF complete galaxies with a given SFR and M$_r$. The closed contours from inwards-to-outwards enclose 25, 50, 75 and 90$\%$ of the data in the $0.01\lesssim z<0.15$, $0.17<z<0.24$ and $0.24\lesssim z<0.35$ ranges (left-to-right panels). Only the lowest-redshift sample (left panel) contains galaxies with large N$_{\rm{weight}}$ measures.}
\label{fig:random_weights}
\end{center}
\end{figure*}

In effect, {\color{black}\tt{Random DMU}} provides N$_{\rm{r}}$, with $\langle$N$_{\rm{r}}\rangle\approx400$, clones per GAMA galaxy in {\color{black}\tt{TilingCatv43}}. The clones share all intrinsic physical properties (e.g.\,SFR, stellar mass, etc.) as well as the unique galaxy identification (i.e.\,{\tt{CATAID}}) of the parent GAMA galaxy, and are randomly distributed within the parent's V$_{\rm{max}, r}$, while ensuring that the angular selection function of the clones matches that of GAMA. Therefore for any galaxy sample drawn from {\color{black}\tt{TilingCatv43}} based on galaxy intrinsic properties, an equivalent sample of randomly distributed clones can be selected from {\color{black}\tt{Random DMU}} by applying the same selection. If however, a selection involves observed properties, then the clones need to be tagged with "observed" properties before applying the same selection. 

\begin{figure}
\begin{center}	
\includegraphics[width=0.5\textwidth]{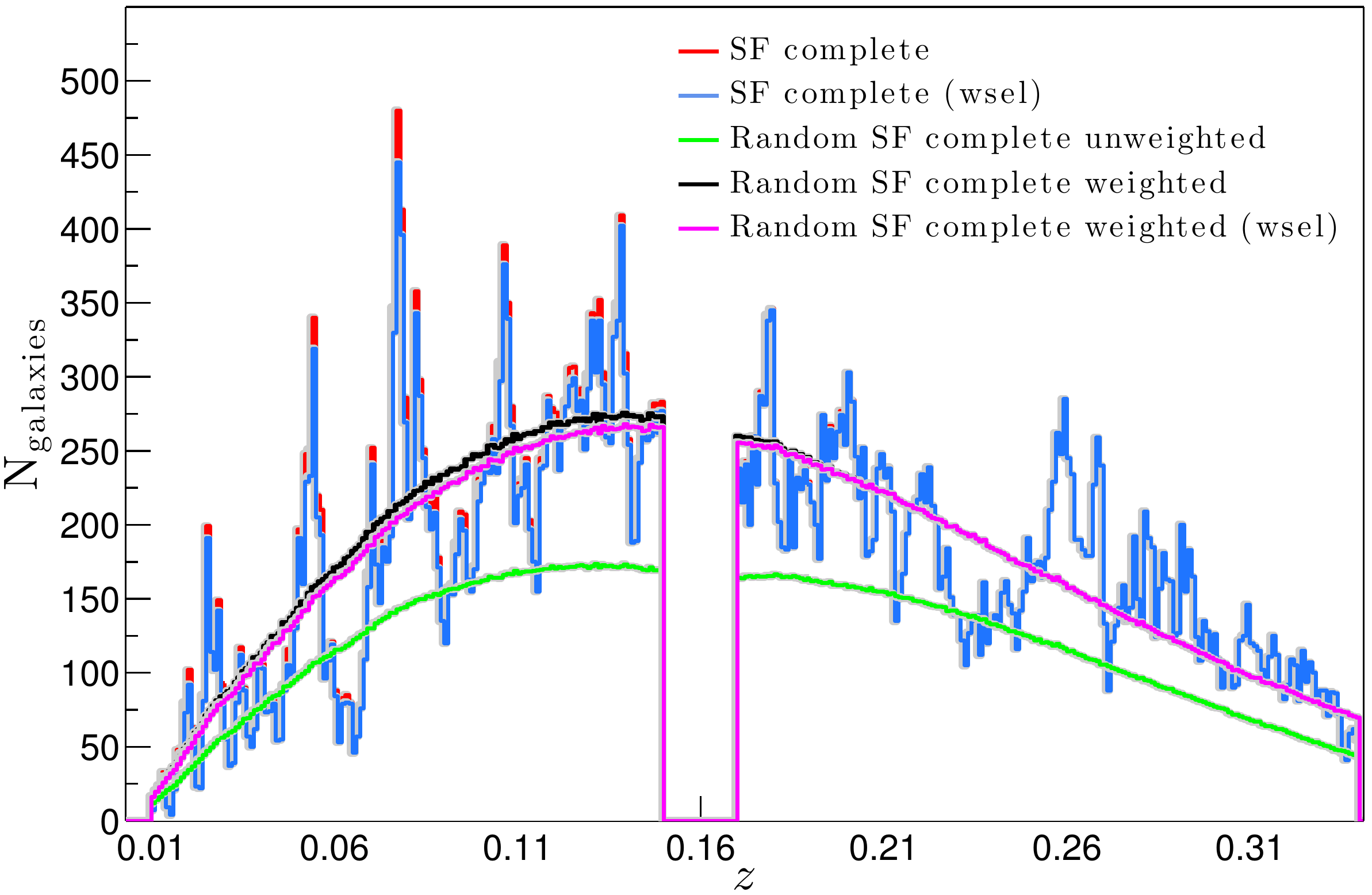}
\caption{\color{black}The redshift ($0.01\lesssim z<0.35$) distribution of the SF complete sample in comparison to the weighted (black and magenta lines) and not-weighted (green) distributions of the random SF complete sample. The weights are determined according to Eq.\,\ref{eq:weights_def}, and the gap in the distributions centred around $z\sim0.16$ indicates the redshift range where the redshifted H$\alpha$ line overlaps with the atmospheric Oxygen-A band. The galaxies, both GAMA and random, with redshifts in this range are excluded from the analysis as described in \S\,\ref{subsec:selection}. Shown also are the weight-selected (wsel) distributions of the SF complete sample and the equivalent weighted random SF complete sample. These distributions exclude all galaxies (and their random clones) with $N_{\rm{weights}}>10$.}
\label{fig:redshift_distributions_all}
\end{center}
\end{figure}

In order to select a sample of clones representative of galaxies in SF complete sample, firstly we exclude the clones of GAMA galaxies not part of SF complete sample. Secondly, each clone is assigned an "observed" H$\alpha$ flux based on their redshift and their parent's intrinsic H$\alpha$ luminosity. Finally, the clones with H$\alpha$ fluxes $>1\times10^{-18}$ W/m$^2$ and with redshifts outside the wavelength range dominated by the O$_2$ atmospheric band but within the detection range of H$\alpha$ (i.e.\,SF complete selection criteria) are selected for the analysis. The redshift distribution of the selected clones,  hereafter \textit{random SF complete}, normalised by the approximate number of replications (i.e.\,$\langle$N$_{\rm{r}}\rangle$) is shown in Figure\,\ref{fig:redshift_distributions_all} (green line). Also shown for reference is the redshift distribution of the GAMA SF complete sample (red line). The clear disagreement between the two distributions is a result of the differences in the selections. Recall that only the $r$--band selection of the survey is considered in the generation of clones, i.e.\,the clones are distributed within their parent's V$_{\rm{max}, r}$, whereas we also impose an H$\alpha$ flux cut to select the SF complete sample. In essence, we require the clones to be distributed within their parent's $min$(V$_{\rm{max}, r}$, V$_{\rm{max}, H\alpha}$), where V$_{\rm{max}, H\alpha}$ is the maximum volume given the H$\alpha$ flux limit, in order to resolve the disagreement between the two distributions.   

Instead of regenerating the random DMU with a bivariate selection, we adopt a weighting scheme for the clones, where the original distribution of clones within a given parent's V$_{\rm{max}, r}$ is altered to a distribution within min(V$_{\rm{max} ,\,r}$, V$_{\rm{max} , \,H\alpha}$, V$_{\rm{zlim}}$), where V$_{\rm{zlim}}$ is the volume out to the detection limit of the H$\alpha$ spectral line in GAMA spectra. The weight of a galaxy, $i$, is defined as 
\begin{equation}
N^{i}_{\rm{weight}} = \frac{N^{i}_{{V_{\rm{max, \,r}}}}}{N^{i}_{{\min(V_{\rm{max, \,H\alpha}}, \, V_{\rm{max}, \,r}, \, V_{\rm{zlim}})}}},
\label{eq:weights_def}
\end{equation}
where $N^{i}_{{V_{\rm{max, \,r}}}}\equiv$\,N$_{\rm{r}}$ is the total number of clones originally generated for the galaxy $i$ and distributed within its V$_{\rm{max}, \,r}$, and $N^{i}_{{\min(V_{\rm{max, \,H\alpha}}, \, V_{\rm{max}, \,r}, \, V_{\rm{zlim}})}}$ is the number of clones within min(V$_{\rm{max} , \,H\alpha}$, V$_{\rm{zlim}}$) of the $i^{\rm{th}}$ galaxy.  

We show the mean variation of N$_{\rm{weight}}$ in SFR and M$_r$ space in Figure\,\ref{fig:random_weights} for three different redshift bins. At a fixed M$_r$, N$_{\rm{weight}}$ declines with increasing SFR and redshift, and at a fixed SFR, N$_{\rm{weight}}$ decreases with increasing optical brightness and decreasing redshift. The implication being that the maximum volume out to which a high-SFR galaxy would be detectable is limited only by the $r$--band magnitude selection of the survey (i.e.\,no weighting is required), and vice versa. For example, a (low SFR) galaxy with N$_{\rm{weight}}\approx20$ has $\sim20$ clones out of $\sim400$ within its V$_{\rm{max, \,H\alpha}}$. While low SFR galaxies can have larger values of N$_{\rm{weight}}$, we demonstrate in Figure\,\ref{fig:redshift_distributions_all} that the modelling of the redshift distribution is only very marginally affected by cutting the sample on N$_{\rm{weight}}$. Moreover, in Appendix\,\ref{app:modelling}, we show that the differences between the redshift distributions of clones weighted by N$_{\rm{weight}}$ with and without removing large values of N$_{\rm{weight}}$ are minimal. The differences are largely confined to lower redshifts, where most low-SFR systems reside. The impact of galaxies with large values of N$_{\rm{weight}}$ on the clustering results is, again, minimal, and is not surprising as most of the low-SFR systems with large N$_{\rm{weight}}$ lie outside the $90\%$ data contour (Figure\,\ref{fig:random_weights}). 

A comparison between the redshift distribution of the clones weighted by N$_{\rm{weight}}$, called {\textit{random SF complete weighted}}, and the distributions of the unweighted clones and GAMA star forming galaxies is presented in Figure\,\ref{fig:redshift_distributions_all}. We also illustrate the relatively small effect on the weighted distribution if objects with N$_{\rm{weight}}>10$ (i.e.\,wsel selection in Figure\,\ref{fig:redshift_distributions_all}) are removed from the analysis. Consequently, the impact on the results of the correlation analyses are also minimal as demonstrated in Appendix\,\ref{app:randoms_weights}.   

Alternatively, N$_{\rm{weight}}$ can also be calculated in redshift slices. We refer readers to Appendix\,\ref{app:modelling} for a discussion on the resulting redshift distributions, mean N$_{\rm{weight}}$ variations with respect to SFR, M$_r$, and redshift, as well as on the clustering analysis. The main caveat in calculating N$_{\rm{weight}}$ in (smaller) redshift slices is that a relatively higher fraction of clones will require larger weights as V$_{\rm{zlim}}$ now defines the volume of a given redshift slice. For this reason we choose to use N$_{\rm{weight}}$ calculated assuming a V$_{\rm{zlim}}$ defined by the detection limit of H$\alpha$ spectral line in GAMA spectra as described above for the clustering analysis presented in subsequent sections. 

In summary, in this section, we presented a technique with which the available random clones of GAMA galaxies can be used, without the need to recompute them to take into account any additional constraints resulting from star formation selections.

\subsection{Two-point galaxy correlation function}\label{subsec:two_point_corr}

The spatial two-point correlation function, $\xi$($r$), is defined as the \textit{excess} probability $dP$, relative to that expected for a random distribution, of finding a galaxy in a volume element $dV$ at a distance $r$ from another galaxy \citep{Peebles1980}, i.e., 
\begin{equation}
dP = n[1 + \xi(r)]\,dV,
\end{equation}
where $n$ is the galaxy number density determined from a given galaxy catalogue.

To disentangle the effects of redshift space distortions from intrinsic spatial clustering, the galaxy CF is often estimated in a two--dimensional grid of pair separations parallel ($\pi$) and perpendicular ($r_p$) to the line of sight, where $r=\sqrt{\pi^2 + r_p^2}$. Using the notation of \cite{Fisher1994}, for a pair of galaxies with redshift positions \textbf{v$_1$} and \textbf{v$_2$},  we define the redshift separation vector  \textbf{s $\equiv$ v$_1$-v$_2$} and the line of sight vector \textbf{$\boldsymbol\ell$ $\equiv$ $\frac{1}{2}$(v$_1$+v$_2$)}. The parallel and perpendicular separations are then,
\begin{equation}
\pi \equiv \mid \textbf{s . $\boldsymbol\ell$} \mid / \mid  \textbf{$\boldsymbol\ell$} \mid \text{\hspace{0.5cm} and    \hspace{0.5cm}}  r_p^2 \equiv \textbf{s . s} - \pi^2.
\end{equation}

The projected two point CF, $\omega_p(r_p)$, obtained by integrating the two-point CF over the line-of-sight ($\pi$) direction, then allows the real space $\xi(r)$ to be recovered devoid of  redshift distortion effects \citep{Davis1983}. The $\omega_p(r_p)$ is defined as,  
\begin{equation}
\begin{split}
\omega_p(r_p) & = 2 \int\limits_0^{\pi_{\rm{max}}} \xi(r_p, \pi) d\pi = 2 \sum\limits_i \xi(r_p, \pi_i) \,\Delta \pi_i. 
\end{split}
\label{eq:omegaP}
\end{equation}

We integrate to $\pi_{\rm{max}}\approx40\,h^{-1}$ Mpc, which is determined to be large enough to include all the correlated pairs, and suppress the noise in the estimator \citep{Skibba2009, Farrow2015}.  

The statistical errors on clustering measures are generally estimated using jackknife resampling \citep[e.g.][]{Zehavi2005, Zehavi2011}, using several spatially contiguous subsets of the full sample omitting each of the subsets in turn. The uncertainties are estimated from the error covariance matrix,  
\begin{equation}
\begin{split}
C_{ij} & = \frac{N_{\rm{JK}}}{N_{\rm{JK}}-1} \times \\ 
& \sum\limits_{n=1}^{N_{\rm{JK}}}[\omega_{p}^n(r_{p_i})-{\omega}_{p}(r_{p_i})][\omega_{p}^n(r_{p_j})-{\omega}_{p}(r_{p_j})], 
\end{split}
\label{eq:error_cov}
\end{equation}
where $N_{\rm{JK}}$ is the number of jackknife samples used. We use $18$ spatially contiguous subsets (i.e.\,N$_{\rm{JK}}=18$), each covering $16$\,deg$^{2}$ of the full area, {and the results are robust to the number of samples considered (e.g.\,from 12 to 24).}

There are several two-point galaxy CF estimators widely used in the literature \citep[e.g.][]{Hamilton1993, LS1993, Davis1983, Peebles1974}. Here we adopt the \cite{LS1993} estimator to perform; (i) two-point auto correlation, (ii) two-point cross correlation, and (iii) mark two-point cross correlation analyses, as explained in the subsequent subsections.  In Appendix\,\ref{app:modelling}, we compare the results of  \cite{LS1993} with that obtained from the \cite{Hamilton1993} estimator to check whether our results are in fact independent of the estimator used.   

\subsubsection{Two-point auto correlation function} \label{subsubsec:auto_function}

The two-point auto CF, $\xi_a$, estimator by \cite{LS1993} is,
\begin{equation}
\xi_a(r_p, \pi)_{_{\rm{LS}}} = \frac{DD(r_p, \pi)}{RR(r_p, \pi)} - 2\frac{{DR(r_p, \pi)}}{RR(r_p, \pi)} + 1.  
\label{eq:auto_ls}
\end{equation}

The DD($r_{\rm{p}}, \pi$), RR($r_{\rm{p}}, \pi$) and DR($r_{\rm{p}}, \pi$) are normalised data-data, random--random and galaxy--random pair counts, {and randoms are weighted by N$_{\rm{weight}}$ (Eq.\,\ref{eq:weights_def})}. 

\subsubsection{Two-point cross correlation function}

The estimators given in Eq.\,\ref{eq:auto_ls} and \ref{eq:auto_hamilton} are adapted for the two-point galaxy cross CF, $\xi_c$, respectively, as follows; 
\begin{equation}
\xi_c(r_p, \pi)_{_{\rm{LS}}} = \frac{D_1D_2(r_p, \pi) - D_1R_2(r_p, \pi) - D_2R_1(r_p, \pi)}{R_1R_2(r_p, \pi)} + 1, 
\label{eq:cross_ls}
\end{equation}

The D$_{1}$D$_{2}$($r_{\rm{p}}, \pi$) is the normalised galaxy--galaxy pair count between data samples 1 and 2, and R$_{1}$R$_{2}$($r_{\rm{p}}, \pi$) is the normalised random--random pair count between random clone samples 1 and 2, {and the randoms are weighted by N$_{\rm{weight}}$ as defined in Eq.\,\ref{eq:weights_def}}.  

The projected cross CFs and their uncertainties are estimated following the same principles as the auto CFs (\S\,\ref{subsubsec:auto_function}).

Finally, in most cases below, we present GAMA auto and cross correlation functional results relative to the \cite{Zehavi2011} power law fit to their $-21\leq$M$_r^{0.1}-5\log\,$h$\leq-20$ sample, hereafter $\omega_p^{\rm{Z11}}$, given by, 
\begin{equation}
\omega_p^{Z11} = \frac{5.33}{r_p}^{\gamma}  \Gamma(0.5)\Gamma[0.5(\gamma-1)]\Gamma(0.5\gamma),
\label{eq:zehavi}
\end{equation}
where $\gamma=1.81$.

\subsubsection{Two-point mark correlation function}

Over the last few decades, numerous clustering studies based on auto and cross correlation techniques have quantitatively characterised the galaxy clustering dependence on galaxy properties in the low-to-moderate redshift Universe. While these studies use the physical information to define galaxy samples for auto and cross correlation analyses, that specific information is not considered in the analysis itself. In other words, galaxies are weighted as "ones" or "zeros" regardless of their physical properties, leading to a potential loss of valuable information. The \textit{mark} clustering statistics, on the other hand, allow physical properties or "marks" of galaxies to be used in the clustering estimation.  

The two-point mark CF relates the conventional galaxy clustering to clustering in which each galaxy in a pair is weighted by its mark, therefore, allowing not only clustering as a function of galaxy properties to be measured, but also the spatial distribution of galaxy properties themselves and their correlation with the environment to be efficiently quantified \citep{Sheth2005}. As it is the difference between weighted to unweighted clustering at a particular scale that is considered, the mark CF has serval advantages over conventional clustering statistics; (1) it essentially quantifies the degree to which a galaxy mark is correlated with the environment at that scale, and (2) it is less affected by issues related survey/sample selection and incompleteness than conventional methods \citep{Skibba2009}.  
The two-point mark CF is defined as,
\begin{equation}
M(r_p, \pi) = \frac{1+ W(r_p, \pi)}{1 + \xi(r_p, \pi)}, 
\label{eq:mark_xcrr}
\end{equation}
where $\xi(r_p, \pi)$ is the galaxy two-point CF defined above, and $W(r_p, \pi)$ is the weighted CF in which the product of the weights of each galaxy pair taken into account. 

For the galaxy pair weighting, we adopt a multiplicative scheme, i.e.,
\begin{equation}
DD(r_p, \pi) = \sum_{ij} \omega_i \times \omega_j, 
\label{eq:weight_scheme}
\end{equation}
where $\omega_i$ is the weight of the $i^{\rm{th}}$ galaxy given by the ratio of its mark to the mean mark across the whole sample. Thus $\frac{1}{\rm{N_D}}\sum\omega_d^i = 1$ by construction. 

The projected mark two-point CF is defined in a similar fashion: 
\begin{equation}
E_{m}(r_p) = \frac{1+ W_p(r_p)/r_p}{1 + \omega_p(r_p)/r_p}. 
\label{eq:mark_projected}
\end{equation}
On large scales, $M(r)$ and $E_m(r_p)$ approach unity \citep{Skibba2009}.

Again, we adopt the \cite{LS1993} and \cite{Hamilton1993} clustering estimators for this analysis. 

\begin{figure}
\begin{center}
\includegraphics[scale=0.35]{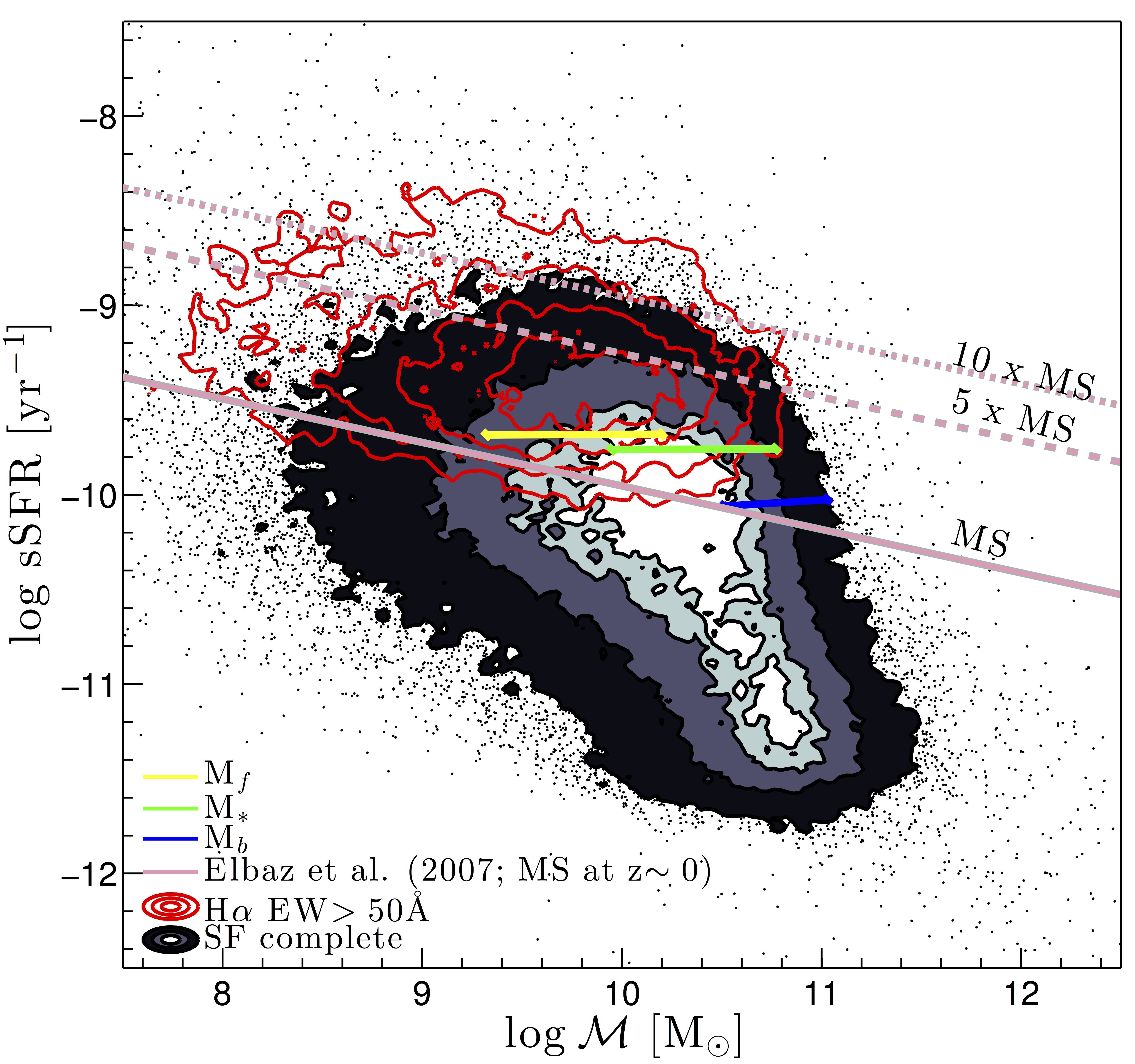}
\caption{The sSFR and $\mathcal{M}$ distribution of SF complete galaxies. The filled-in and red contours enclose 25, 50, 75, and 94$\%$ of SF complete galaxies and SF complete galaxies with H$\alpha$ EW$>50$\AA\,\citep[i.e.\,the "starburst" definition of][]{Rodighiero2011}, respectively. The dark pink lines denote the $z\sim0$ star formation main sequence \citep[solid line,][]{Elbaz2007}, and two starburst selections, 5$\times$ (dashed line) and 10$\times$ the main sequence (dotted line), generally used in the literature \citep[e.g.][]{Rodighiero2011, Silverman2015}. The rest of the lines (yellow, green and blue) show the $30\%$ highest sSFR selections applied to the three disjoint luminosity selected galaxy samples used in this analysis (see Table\,\ref{table:stats1}).}
\label{fig:sfr_mass}
\end{center}
\end{figure}

{{
\section{Signatures of Interaction driven star formation}\label{sec:SF_signatures}

In this study, we consider several different physical properties of galaxies, such as sSFR, colour, dust obscuration and {the strength of the $4000$\,\AA\,break (D$_{\rm{4000}}$)}, that are most likely to be altered in a galaxy-galaxy interaction. A discussion of these properties is given in \S\,\ref{subsec:gamaSF}, followed by the results of the auto and cross correlation analyses in \S\,\ref{subsec:auto_corr} and \ref{subsec:cross_corr}, respectively. Finally, in \S\,\ref{subsec:mark_cfs}, we present the results of the mark correlation analysis, where sSFRs and ($g-r$)$_{\rm{rest}}$ of galaxies are used as marks to investigate the spatial correlations of star forming galaxies. 

\subsection{Characteristics of GAMA star forming galaxies}\label{subsec:gamaSF}
\begin{figure*}
\centering
\includegraphics[width=0.335\textwidth, height=0.5\textwidth, trim={0.6cm 0.4cm 0 0.3cm}, clip]{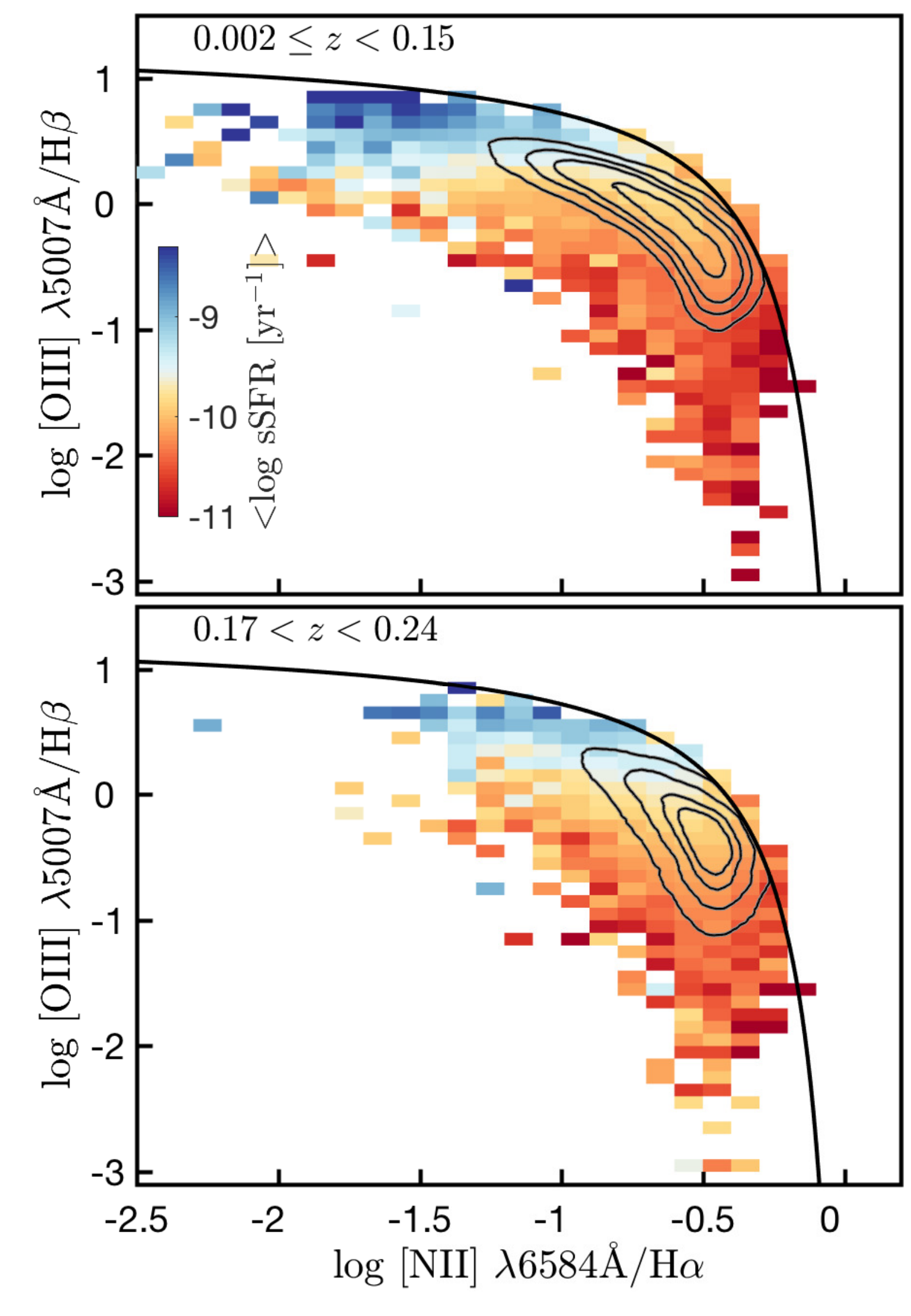}
\includegraphics[width=0.315\textwidth, height=0.5\textwidth, trim={2.1cm 0.4cm 0 0.3cm}, clip]{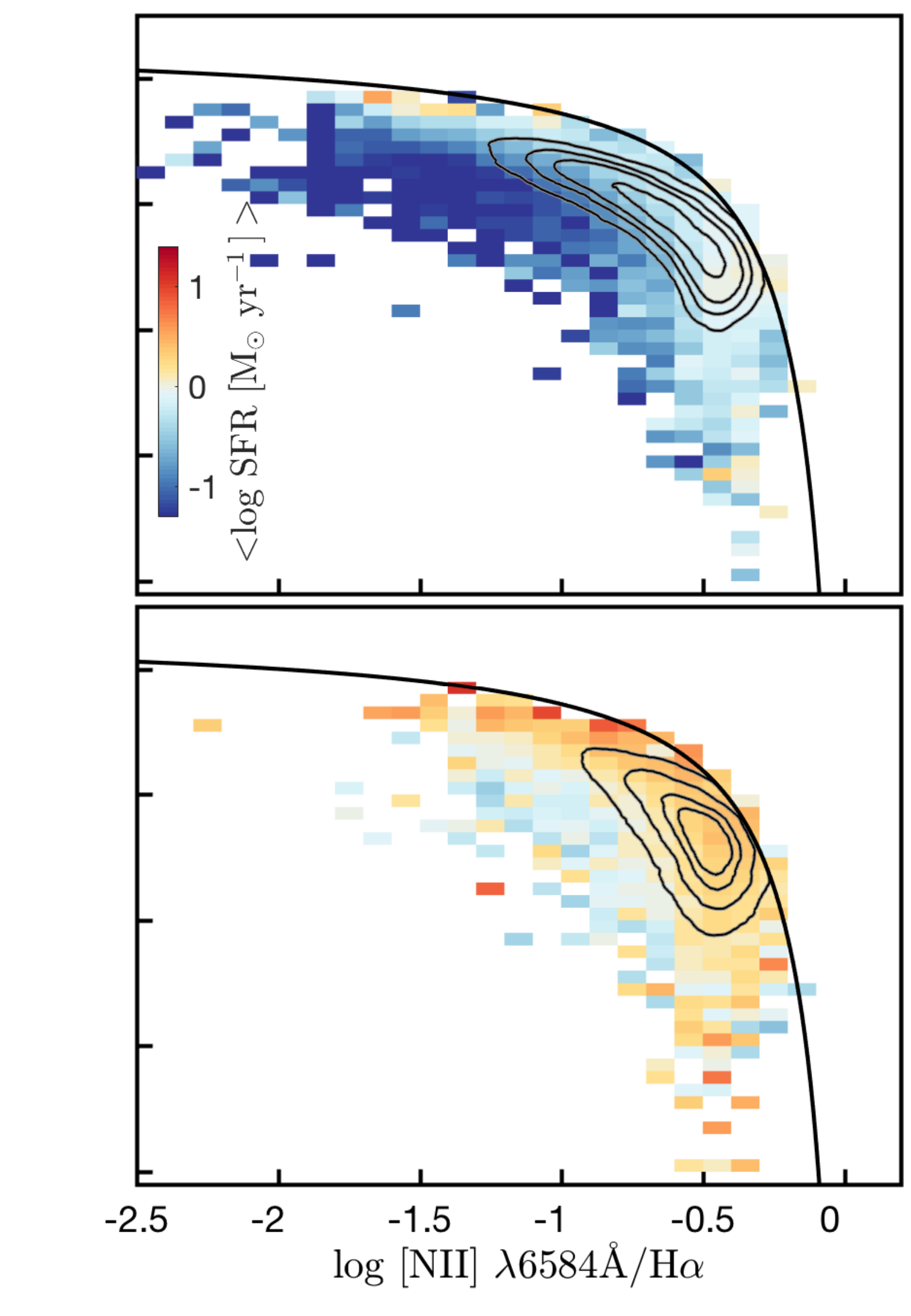}
\includegraphics[width=0.315\textwidth, height=0.5\textwidth, trim={2.1cm 0.4cm 0 0.3cm}, clip]{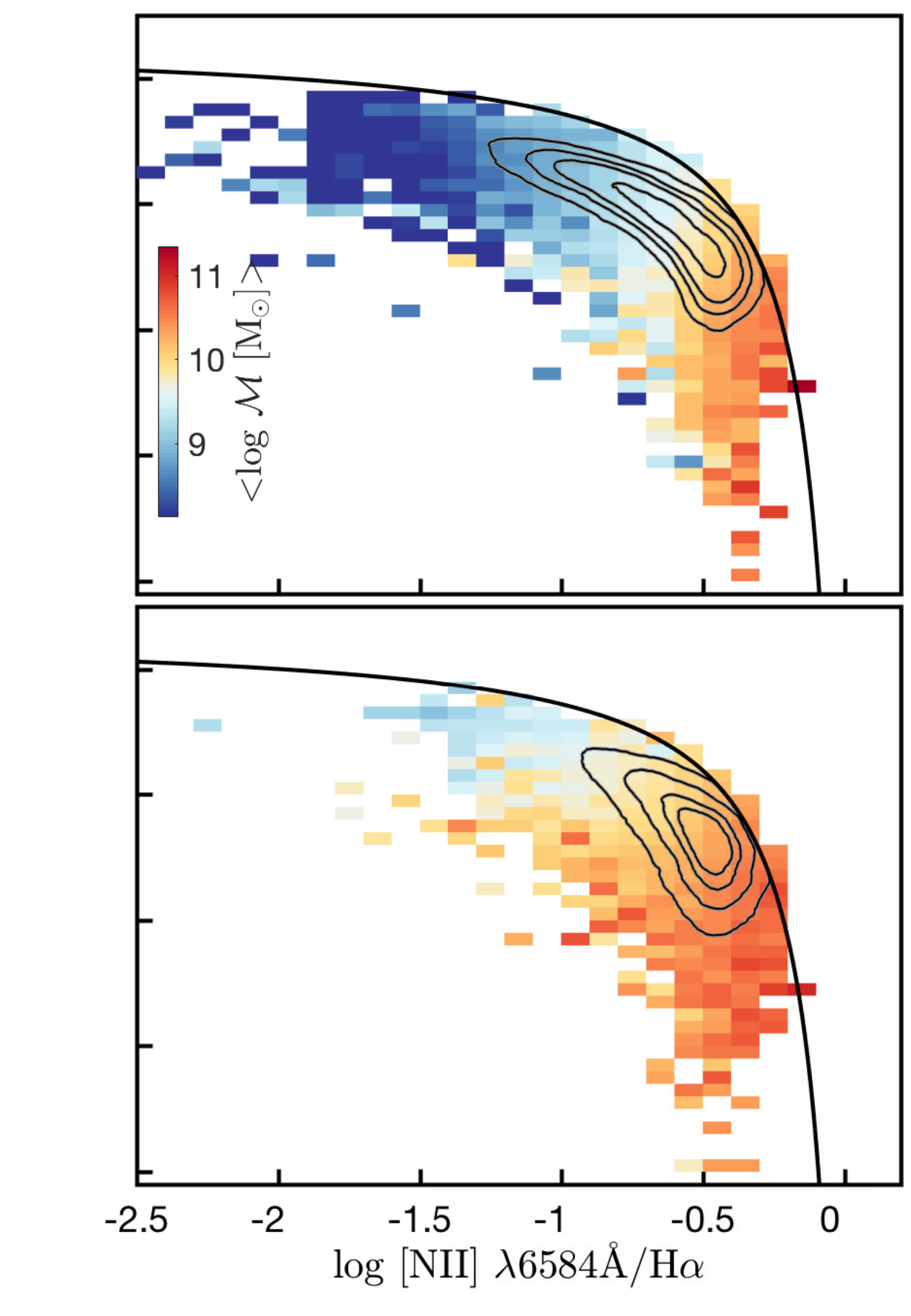}\\
\captionof{figure}{The mean variation in sSFR, SFR, stellar mass (i.e.\,$\log\,\mathcal{M}$) of star forming galaxies across the BPT plane in two redshift bins (from top-to-bottom, with the key shown in left-most panels). The mean value of each property in a given [\ion{O}{iii}]/H$\beta$ and [\ion{N}{ii}]/H$\alpha$ (i.e. the BPT diagnostics) bin is shown in colour, with the black line denoting the \citet{Kauffmann2003} AGN/SF discrimination criterion. The contours enclose $\sim$25, 50, 75, 90$\%$ of the data in each redshift range.}
\label{fig:property_dist1}
\end{figure*}
\begin{figure*}
\centering
\includegraphics[width=0.335\textwidth, height=0.5\textwidth, trim={0.6cm 0.4cm 0 0.3cm}, clip]{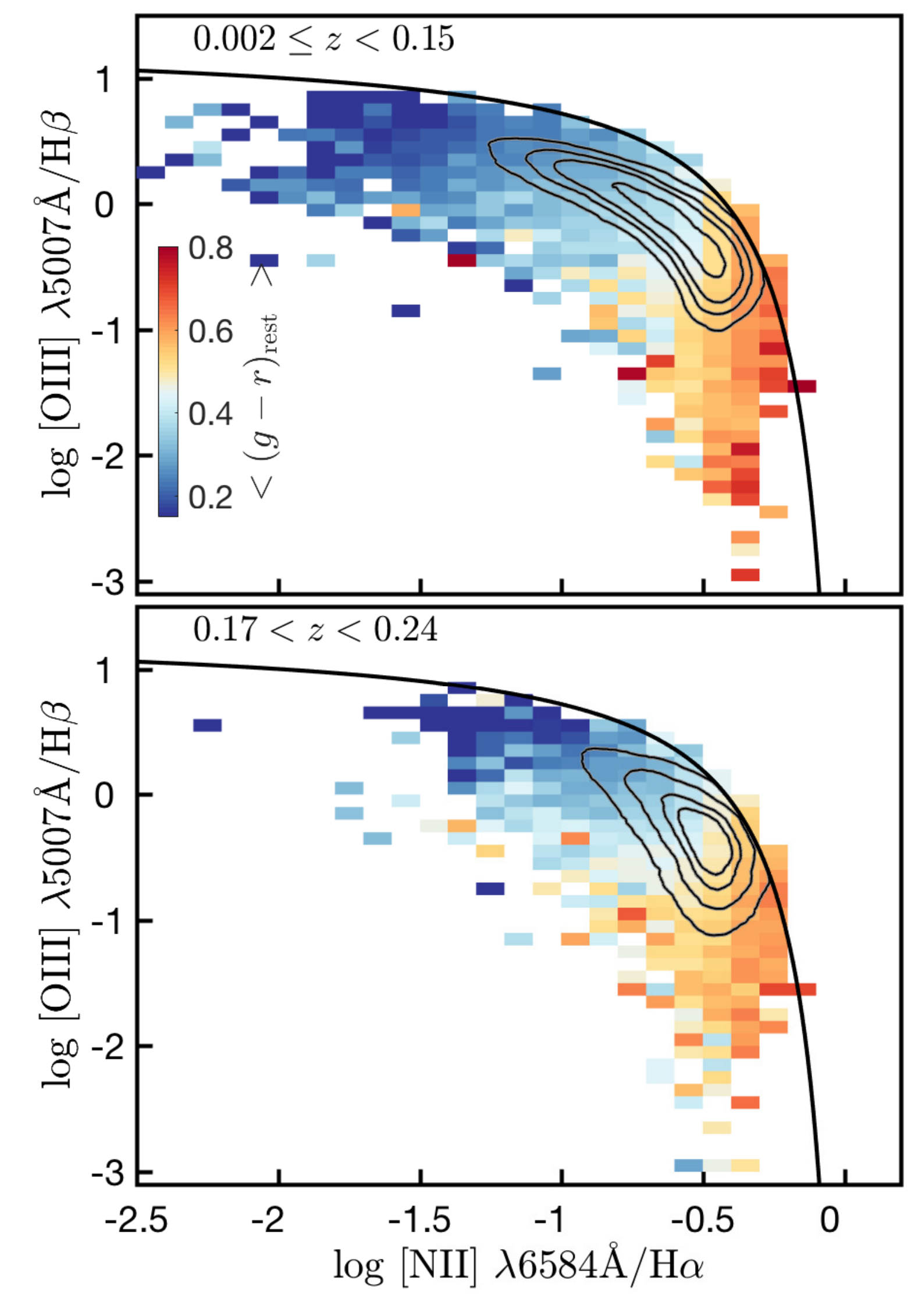}
\includegraphics[width=0.315\textwidth, height=0.5\textwidth, trim={2.1cm 0.4cm 0 0.3cm}, clip]{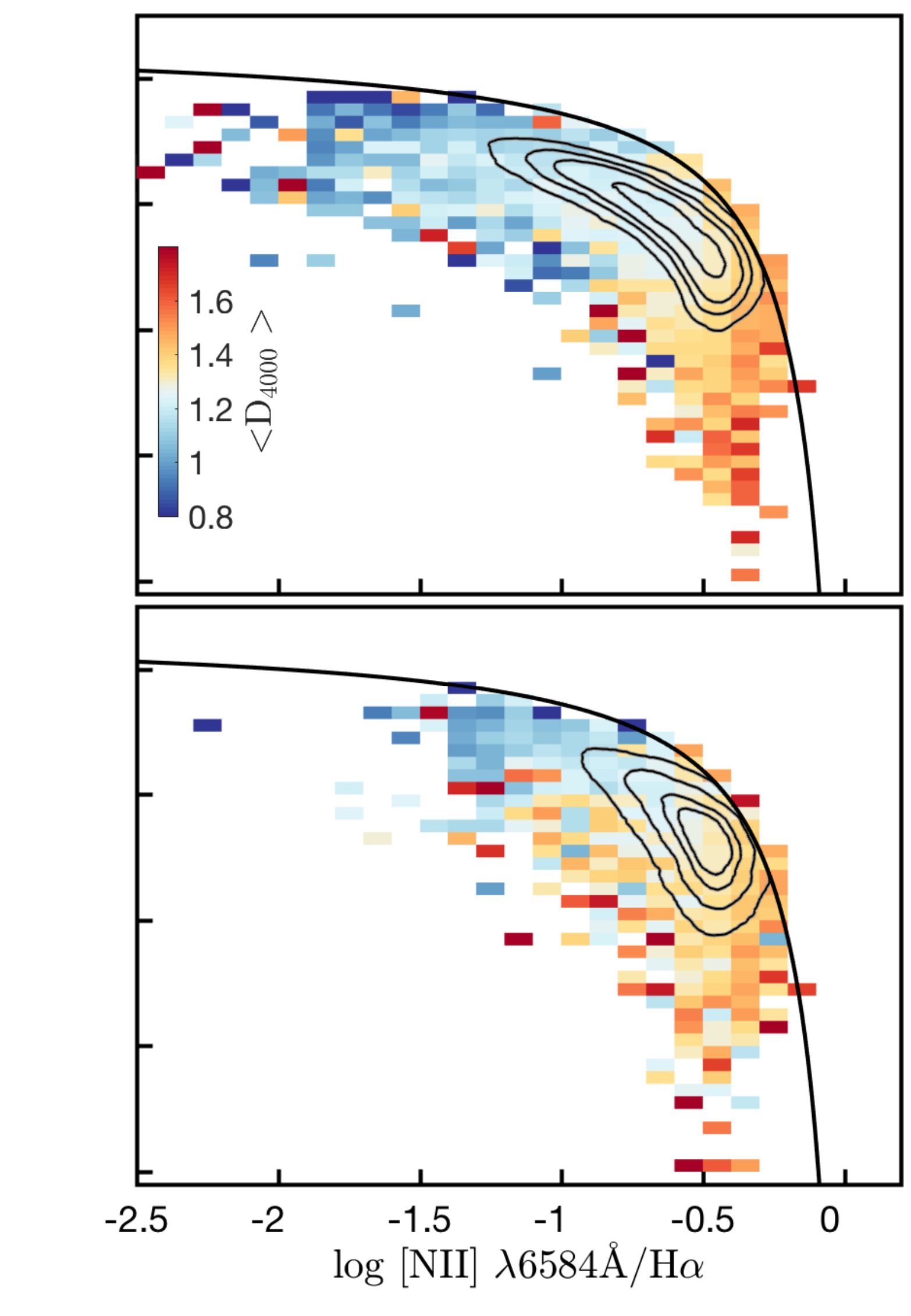}
\includegraphics[width=0.315\textwidth, height=0.5\textwidth, trim={2.1cm 0.4cm 0 0.3cm}, clip]{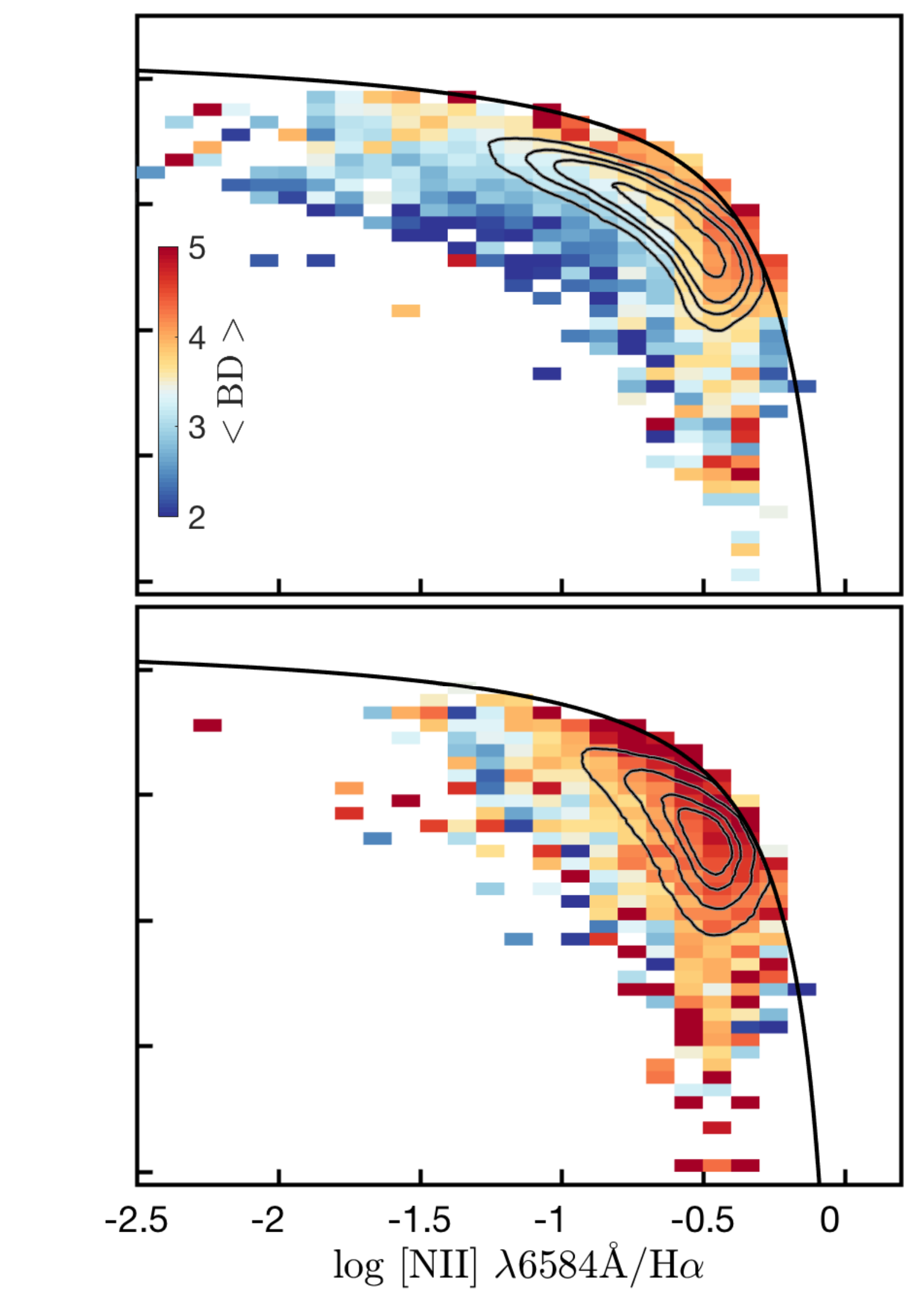}\\
\captionof{figure}{Same as above, now showing the mean variation of ($g-r$)$_{\rm{rest}}$, D$_{\rm{4000}}$ and Balmer decrement (i.e.\,BD) of star forming galaxies across the BPT plane in two redshift bins.}
\label{fig:property_dist}
\end{figure*}
\begin{figure}
\begin{center}
\includegraphics[width=0.49\textwidth]{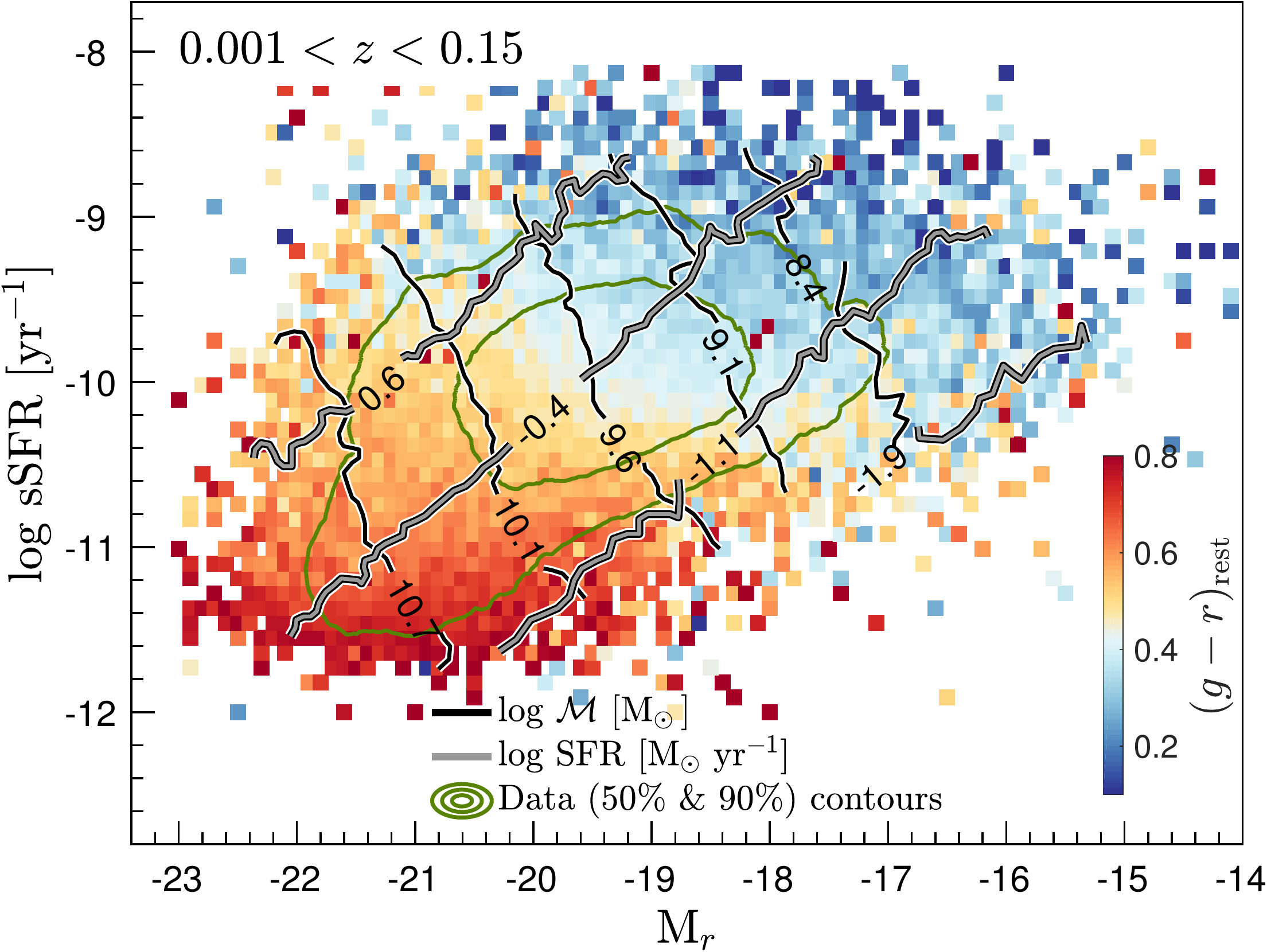}
\caption{The $\log$ sSFR [yr$^{-1}$] and M$_r$ distribution of $z<0.15$ SF complete galaxies, colour-coded by the mean ($g-r$)$_{\rm{rest}}$ of galaxies at a given $\log$ sSFR and M$_r$. The thin black and thick grey lines denote the constant stellar mass (in $\log\,\mathcal{M}$ [M$_{\odot}$]) and log SFR [M$_{\odot}$ yr$^{-1}$] contours, respectively, that span a relatively large range in both M$_r$ and log sSFR. {The green contours enclose 50 and 90$\%$ of the data.} }
\label{fig:ssfr_mass_dist}
\end{center}
\end{figure}

The enhancement of star formation, or starburst, is perhaps the most important and direct signature of a gravitational interaction \citep{Kennicutt1998, Wong2011}. There are several definitions of "starburst" galaxies. \cite{Bolton2012}, for example, define "starburst" as SF galaxies with H$\alpha$ equivalent widths (EW), a proxy for sSFR, larger than $50$\AA. \cite{Rodighiero2011, Luo2014}, and \cite{Knapen2015} use enhancement of SFR as a function of stellar mass to identify starbursts. Additionally, the evidence of certain ionised species (e.g.\,[\ion{Ne}{iii}]~$\lambda3869$\AA) indicative of the high ionisation state of gas, as well as the overall enhancement of emission features in galaxy spectra (e.g.\,[\ion{O}{ii}], [\ion{O}{iii}], H$\alpha$, H$\beta$) are other signatures of starbursts \citep{Wild2014}. {Despite the differences, most "starburst" definitions rely on spectral and/or physical properties of galaxies that are powerful tracers of SFR per unit mass.}

The sSFR and $\mathcal{M}$ distribution of star forming galaxies used in this analysis (filled contours) is presented in Figure\,\ref{fig:sfr_mass}. Over-plotted are several well known "starburst" definitions in the literature; red open contours show the distribution of starbursts \citep[H$\alpha$ EW$>50$\AA,][]{Rodighiero2011}, and the dotted and dashed dark pink lines denote the star formation main sequence \citep[solid dark pink line,][]{Elbaz2007} based starburst definitions \citep[e.g.][]{Rodighiero2011, Silverman2015}. The rest of the lines indicate the selection limits of the 30$\%$ highest sSFR galaxies of M$_b$, M$_*$ and M$_f$ samples. Note that most of the galaxies selected based on the 30$\%$ highest sSFR criterion are in fact those that qualify as starbursts according to the different starburst definitions discussed above.  

The signatures of interaction driven star formation that we consider for this analysis are sSFR, SFR, colour, D$_{\rm{4000}}$ and Balmer decrements, and we use the BPT diagnostics to show (average) variations of these properties in star forming galaxies (Figures\,\ref{fig:property_dist1} and \ref{fig:property_dist}). The BPT diagnostics themselves are indicators of gas-phase metallicities (i.e.\,oxygen abundances) in galaxies \citep{Pettini2004} that can be heavily affected by pristine gas inflows and enriched gas outflows triggered during an interaction. Overall, relatively more massive and lower sSFR galaxies in our star forming sample have higher metallicites (Figure\,\ref{fig:property_dist1}) and are characterised by redder optical colours and D$_{\rm{4000}}$ indices (Figure\,\ref{fig:property_dist})

Galaxy interactions impact dust to a lesser extent than metallicities as inflowing pristine gas cannot dilute the line-of-sight dust obscuration, though, outflows can remove dust from the interstellar medium. The dust is thought to rapidly build up during a burst of star formation \citep{daCunha2010, Hjorth2014}, giving rise to the observed relationship between dust obscuration and host galaxy SFR \citep{Garn2010, Zahid2013}. This relationship between dust obscuration and SFR is evident in Figure\,\ref{fig:property_dist} (right panels), where the increment in Balmer decrement approximately mirrors the increase in SFR.   

The observed bimodality in optical colours \citep{Baldry2004} can also be used to assess the level of star formation in galaxies. A sudden influx of new stars alters the colour of a galaxy that lasts on time scales that are considerably longer than the parent starburst itself. The trends evident in the distributions of ($g-r$)$_{\rm{rest}}$ and D$_{\rm{4000}}$ indices (left and middle panels of Figure\,\ref{fig:property_dist}) are such that high sSFR galaxies, including starbursts, are typically characterised with bluer colours. 

Overall, SFR or stellar mass alone cannot effectively discriminate a low mass galaxy undergoing a burst of star formation from a quiescently star forming high mass galaxy (see Figure\,\ref{fig:ssfr_mass_dist}). Likewise, optical colour, while indicative of the state of star formation within galaxies, taken alone is insufficient to discriminate starbursts from post starburst and/or dusty starburst systems.

\subsection{Auto correlation functions of star forming galaxies} \label{subsec:auto_corr}
\begin{figure*}
\begin{center}
\includegraphics[width=0.36\textwidth]{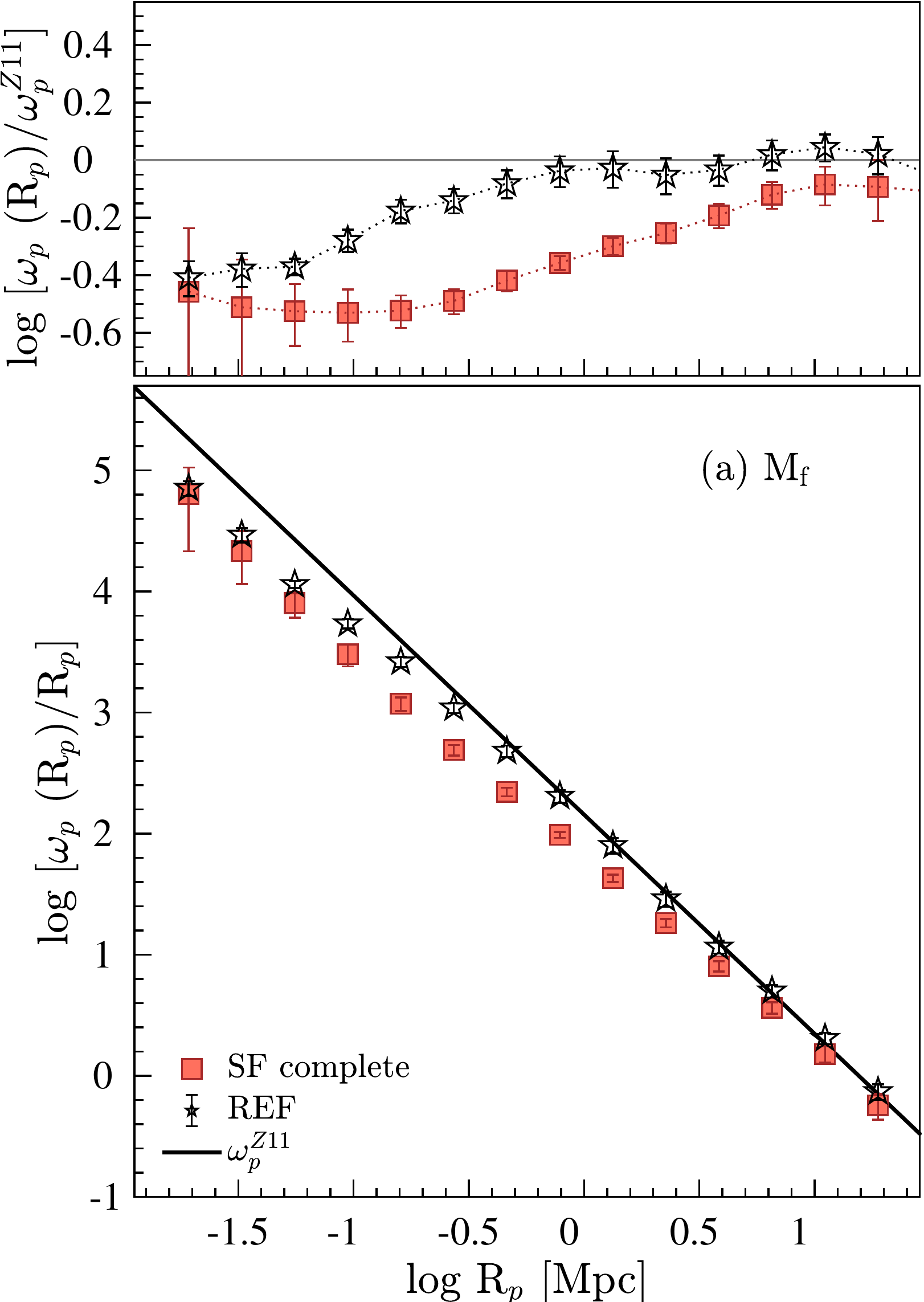}
\includegraphics[width=0.31\textwidth]{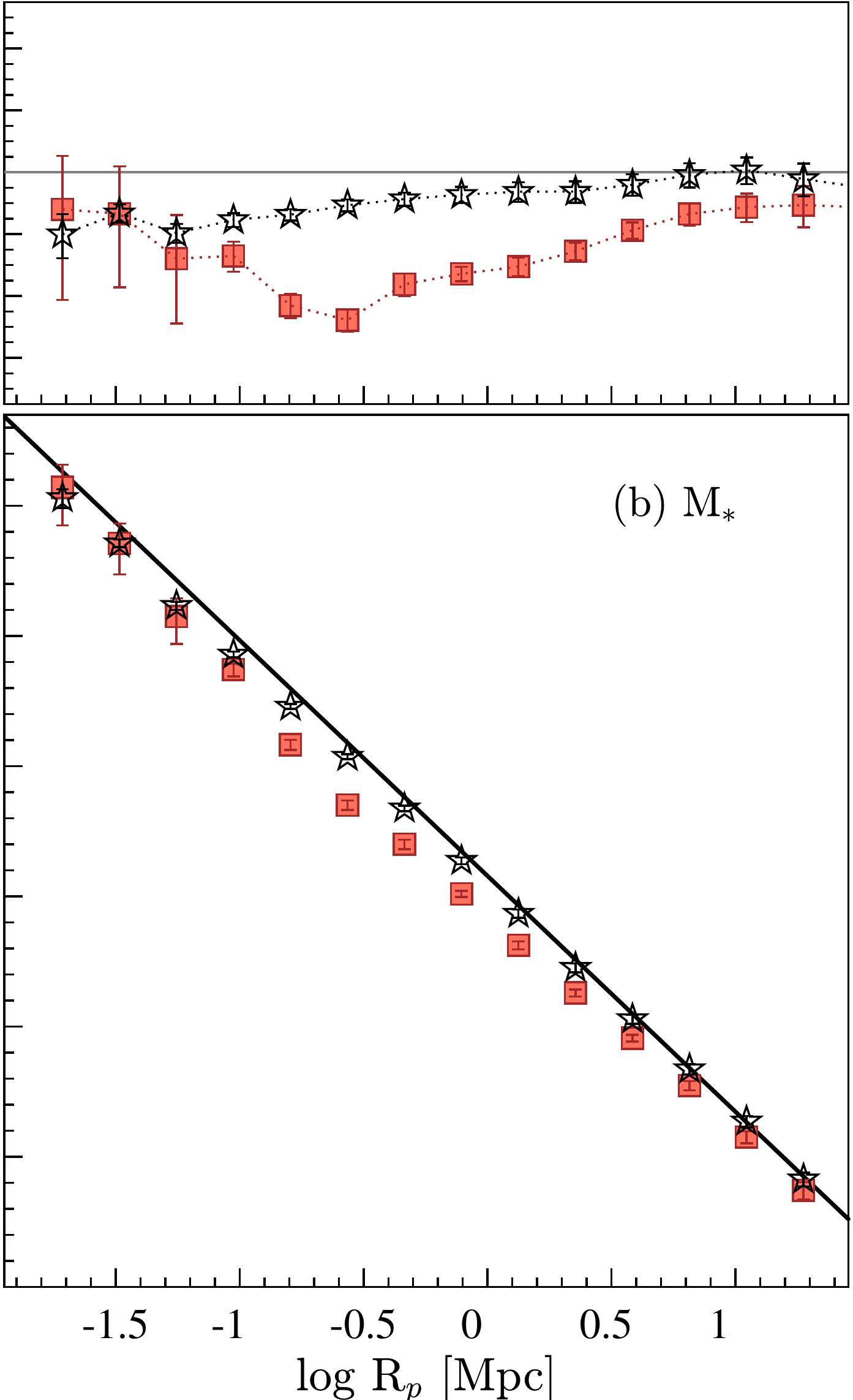}
\includegraphics[width=0.31\textwidth]{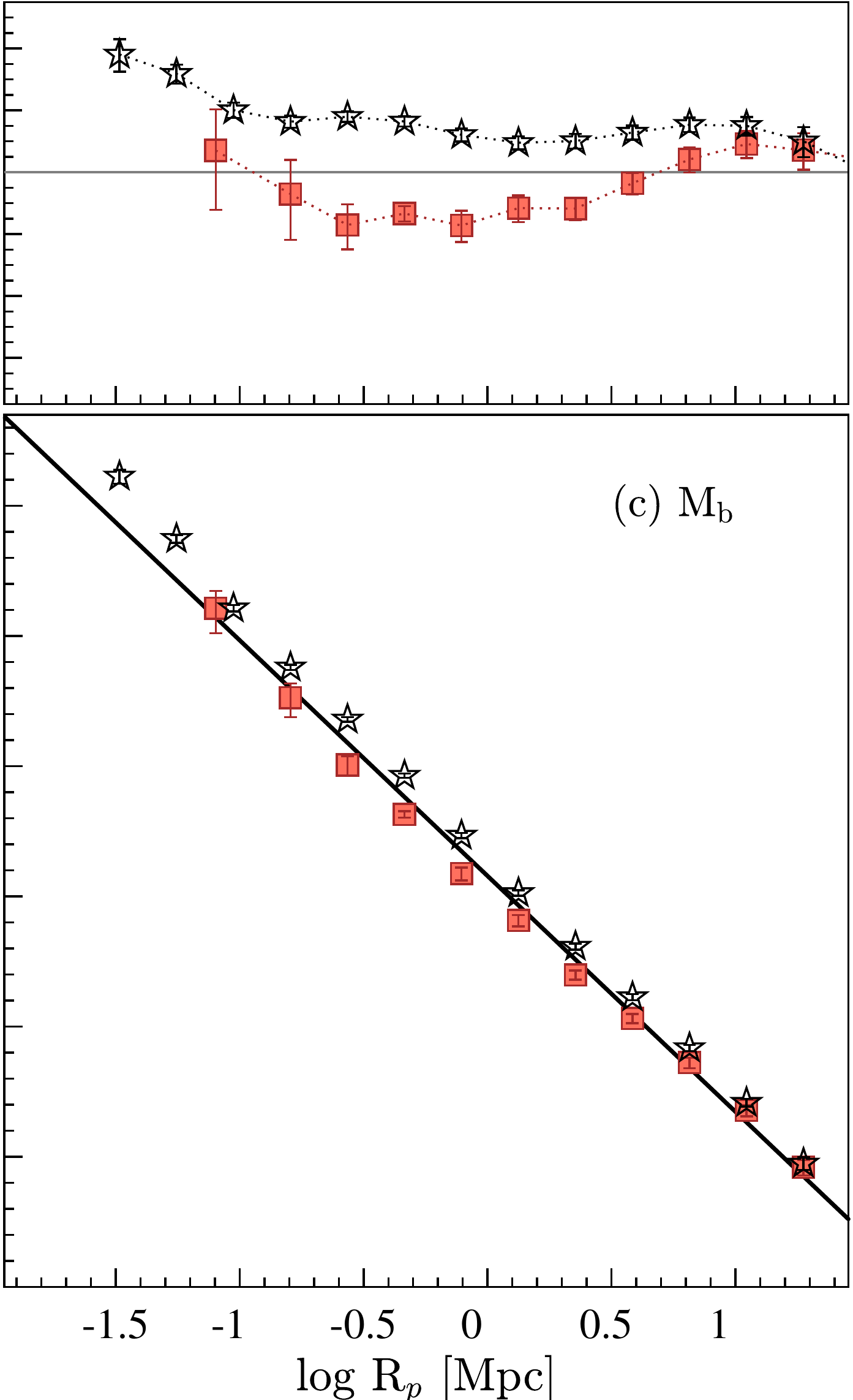}
\caption{Main panels: The GAMA projected ACFs of luminosity selected (i.e.\,M$_f$, M$_*$ and M$_b$, from left-to-right) REF (open black stars) and SF complete (orange filled squares) samples covering the $0.01\leq z\leq0.34$. The black solid line denote the empirical relation given in Eq.\,\ref{eq:zehavi} (i.e.\,$\omega_p^{Z11}$). Top panels: GAMA projected ACFs relative to $\omega_p^{Z11}$. The key is the same as that shown in the left main panel.} 
\label{fig:auto_corr_proj_main}
\end{center}
\end{figure*}
\begin{figure*}
\begin{center}
\includegraphics[width=1\textwidth]{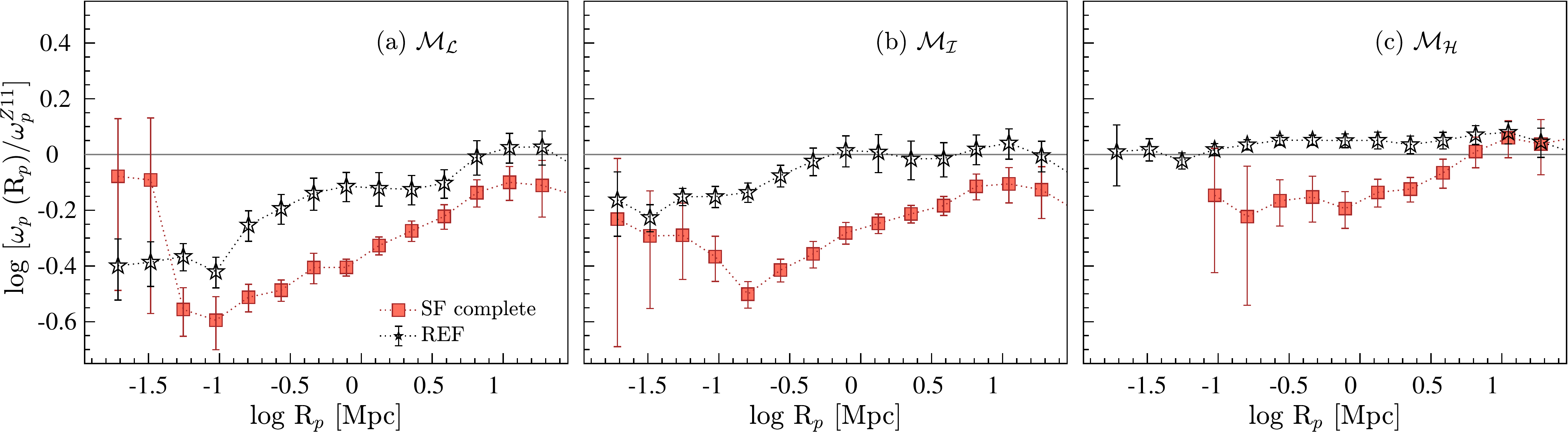}
\caption{The GAMA projected  ACFs of REF (black open symbols) and SF complete (orange filled symbols) stellar mass selected samples (i.e.\,$\mathcal{M}_{L}$, $\mathcal{M}_{I}$ and $\mathcal{M}_{H}$, from left-to-right) relative to $\omega_p^{Z11}$ (the key is shown in the left panel).}
\label{fig:auto_corr_proj_main_smass}
\end{center}
\end{figure*}

The projected auto correlation functions (ACF) of the disjoint luminosity selected samples (Table\,\ref{table:stats1}) are presented in the main panels of Figure\,\ref{fig:auto_corr_proj_main}, and the ACFs relative to \cite{Zehavi2011} power law fit ($\omega_p^{z11}$,  Eq.\,\ref{eq:zehavi}), hereafter ACF$_{\omega_p^{z11}}$, are shown in the top panels.
The ACFs of REF versus SF complete galaxies differ significantly over most scales, reflecting the differences in the clustering of the two sets of galaxy populations. These differences are in agreement with the previous clustering studies of the local Universe that find galaxies with bluer optical colours, representative of star forming systems, tend to cluster less strongly than optically redder galaxies \citep[][]{Zehavi2005b, Skibba2009, Zehavi2011, Bray2015, Farrow2015}. In our case, REF galaxies comprise of both optically bluer and redder galaxies. Likewise, the ACFs of disjoint stellar mass selected samples (Figure\,\ref{fig:auto_corr_proj_main_smass}) show a qualitative agreement with the ACFs of luminosity selected samples introduced in Figure\,\ref{fig:auto_corr_proj_main}. 

On the $-0.15\lesssim$ log R$_p$ [Mpc] $\lesssim1.3$ range, we find that the ACFs of REF and SF complete galaxies, on average, are consistent with a power-law. On smaller scales (log R$_p\lesssim-0.15$ Mpc\footnote{Corresponds to a R$_p$ of $\lesssim0.7$ [Mpc]}), however, both sets of functions show varying levels of increase in the strength of clustering with decreasing R$_p$ and optical brightness. This is most clearly evident in the ACF$_{\omega_p^{z11}}$s (i.e.\,top panels of Figure\,\ref{fig:auto_corr_proj_main}) that demonstrate that at a fixed R$_p$, the amplitude of ACF$_{\omega_p^{z11}}$s increase with increasing optical brightness. This increase in amplitude appears to be stronger in the ACF$_{\omega_p^{z11}}$s of SF complete galaxies than in REF functions on smaller scales, and vice versa on larger scales. Overall, the behaviour we see on larger scales (log R$_p\gtrsim-0.15$ Mpc) is consistent with other studies that report stronger clustering of massive and luminous galaxies than less massive, low luminosity systems \citep[e.g.][]{Norberg2001, Marulli2013, Guo2014, Zehavi2005b, Skibba2009, Zehavi2011, Bray2015}, and on smaller scales, is mostly consistent with the results of another GAMA study by \cite{Farrow2015}.

It is worth noting that even though the ACFs$_{\omega_p^{z11}}$ of SF complete galaxies show lower clustering amplitudes than their respective REF functions on most scales, the change in the strength of the ACFs$_{\omega_p^{z11}}$ of SF complete galaxies with decreasing R$_p$ is greater than that of REF functions. In other words, the ACFs$_{\omega_p^{z11}}$ of SF complete galaxies show a steeper decline (increase) in strength on log R$_p\gtrsim-0.15$ Mpc (log R$_p\lesssim-0.15$ Mpc) with decreasing R$_p$ than REF functions. This rapid increase in the clustering strength of the ACFs$_{\omega_p^{z11}}$ of SF complete galaxies on smaller scales (i.e.\,excess clustering) suggests increased galaxy-galaxy interactions. The same behaviour is also apparent in the ACFs$_{\omega_p^{z11}}$ of disjoint stellar mass samples of SF complete galaxies (Figure\,\ref{fig:auto_corr_proj_main_smass}).

Interestingly, the R$_p$ at which the ACFs$_{\omega_p^{z11}}$ of SF-complete galaxies begin to show an increase in strength seems to also be optical brightness dependent, such that higher optical luminosities correspond to larger R$_p$ and vice versa. For instance, the SF ACF$_{\omega_p^{z11}}$ of M$_{\rm{f}}$ galaxies show a turn-over in the signal at $\sim0.1$ Mpc, though the signal appears to plateau\footnote{Plateau here implies that the ACF has the same gradient as $\omega_p^{z11}$.} at a R$_p$ of $\sim0.4$ Mpc (or log R$_p$ of $-0.4$). The SF ACFs$_{\omega_p^{z11}}$ of M$_*$ and M$_{\rm{b}}$ show turn-overs at larger R$_p$ of $\sim0.31$ Mpc and $\sim0.5$ Mpc (i.e.\,log R$_p$ of $-0.51$ and $0.3$), respectively. This is in the sense that optically luminous star forming galaxies show an enhancement in clustering at relatively larger separations than their low luminosity counterparts. 
 
As mentioned earlier, R$_p$ provides an alternative metric to assess the interaction-phase of a galaxy pair through the association of large R$_p$ with time elapsed since or time to pericentric passage and small R$_p$ with galaxies currently undergoing a close encounter. One of the advantages of using ACFs to trace interaction-phase is that, aside from the initial sample selection, ACFs are not affected by the properties of galaxies. As such, it is not the net change in a property with R$_p$ that is being assessed, but the change in the clustering strength with R$_p$ within the one- and two-halo terms. Interpreting the change in the strength of the clustering of ACFs$_{\omega_p^{z11}}$ of SF complete galaxies as a signature of increased interactions between galaxies, then any correlation between optical brightness (or stellar mass) and R$_p$ in which a change in the clustering signal takes place can be taken as a signature of a halo size-interaction scale dependence. This suggests that the physical evidence of interactions between star forming galaxies within massive halos are (or ought to be) visible out to larger radii than those between star formers residing in less massive halos. This is also supported by the fact that optically bright star forming galaxies are likely hosted within massive halos. 

\subsection{Cross correlation functions of star forming galaxies} \label{subsec:cross_corr}
\begin{figure*}
\begin{center}
\includegraphics[width=0.4\textwidth]{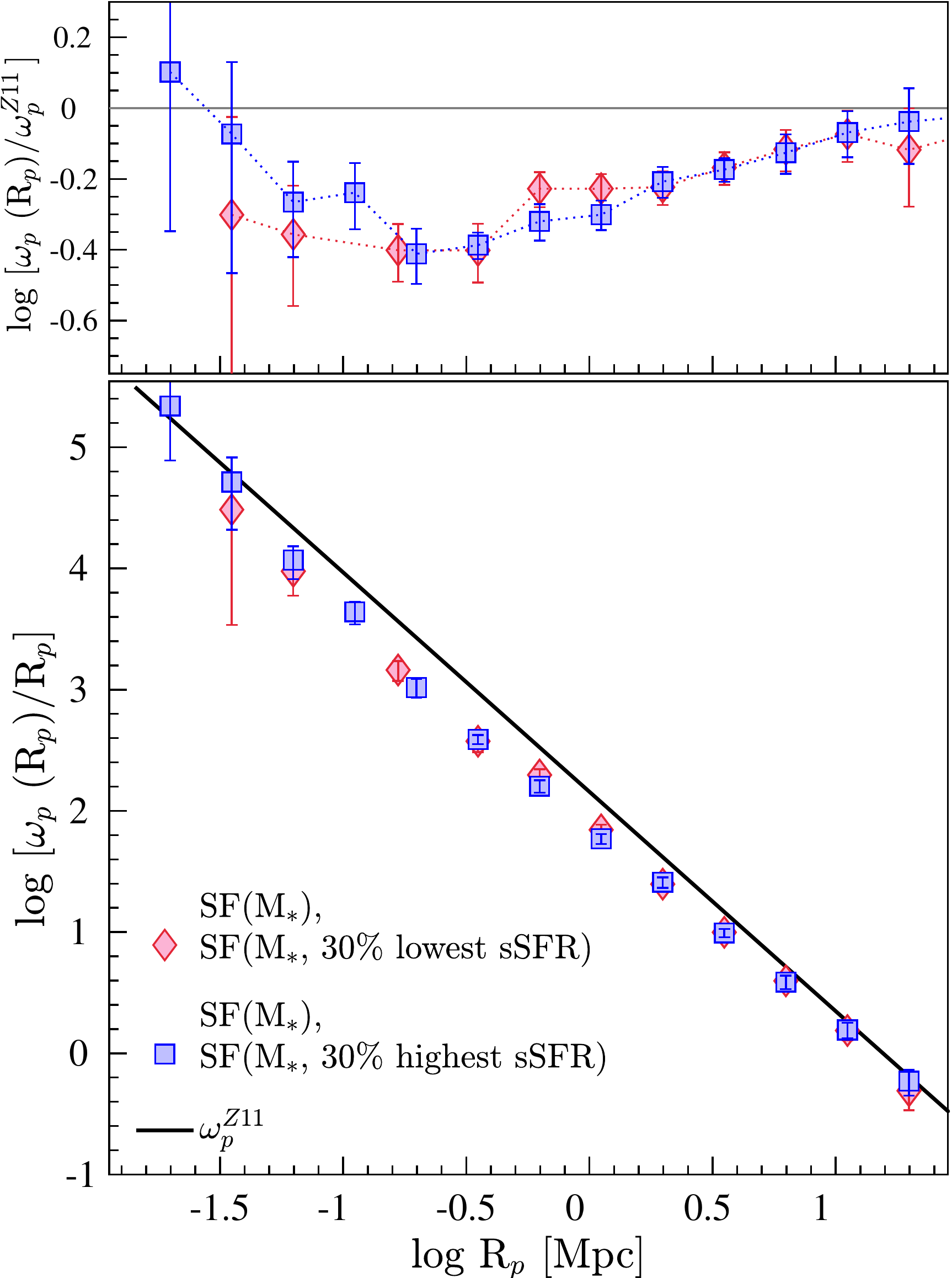}
\includegraphics[width=0.4\textwidth]{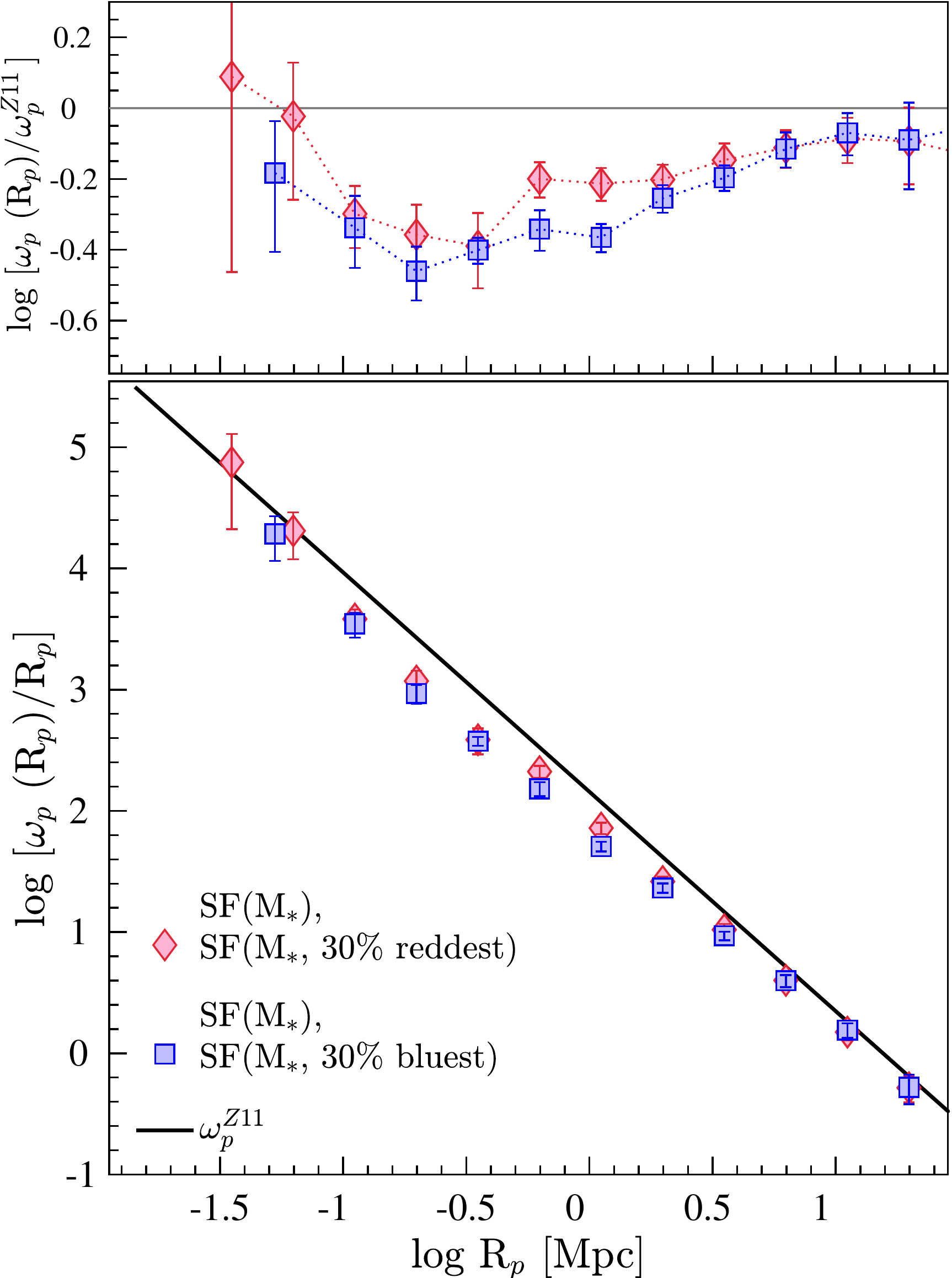}
\caption{The projected CCFs of the $30\%$ subsets of M$_*$ SF complete galaxies. Left panels: The projected CCFs of the $30\%$ highest (blue squares) and the $30\%$ lowest (red diamonds) sSFR galaxies (main panel), and the same functions relative to $\omega_p^{Z11}$ (top panel). Right panels: The projected CCFs of the $30\%$ bluest (blue squares) and the $30\%$ reddest (red diamonds) galaxies in ($g-r$)$_{\rm{rest}}$ (main panel), and the same functions relative to $\omega_p^{Z11}$ (top panel).}
\label{fig:xcrr_Mr1_color_ssfr1}
\end{center}
\end{figure*} 

In this section, we extend the above analysis to further investigate the clustering properties of star formers with respect to different galaxy properties. For this, from each disjoint luminosity (and stellar mass) selected sample, we draw subsamples containing the $30\%$ highest and the $30\%$ lowest sSFRs, ($g-r$)$_{\rm{rest}}$, D$_{\rm{4000}}$ and Balmer decrements. This selection is detailed in \S\,\ref{subsec:clustering_samples}. The smaller $30\%$ samples increase the susceptibility of auto correlation results to the effects of small number statistics, hence we utilise cross correlation techniques for the analyses presented in the subsequent sections. Note that all the CCF results shown in the main paper correspond to cross correlations between a given $30\%$ sample and its parent SF complete sample. As part of this analysis, we also investigated the cross correlations between a given $30\%$ sample and its parent REF sample, and we refer readers to Appendix\,\ref{App:XC_ref} for a discussion of that investigation. 

The CCFs of the $30\%$ highest and the lowest sSFR M$_*$ galaxies, and the $30\%$ bluest and the reddest ($g-r$)$_{\rm{rest}}$ M$_*$ galaxies are presented in the left and right panels of Figure\,\ref{fig:xcrr_Mr1_color_ssfr1}, respectively, where each $30\%$ sample is cross correlated with its parent SF complete sample. Also shown in the top panels of Figure\,\ref{fig:xcrr_Mr1_color_ssfr1} are the CCFs relative to $\omega_p^{Z11}$, hereafter CCF$_{\omega_p^{Z11}}$.

Most notable in Figure\,\ref{fig:xcrr_Mr1_color_ssfr1} is, perhaps, the similar clustering excesses on small scales observed for the $30\%$ M$_*$ samples of high sSFR, optically blue, and optically red galaxies. The overlap in clustering amplitudes between high sSFR and optically blue galaxy populations is expected given the correlation between sSFR and optical colour (Figure\,\ref{fig:ssfr_mass_dist}). The overlap between the optically reddest and the highest star forming populations suggests again that a significant fraction of the reddest M$_*$ galaxies in SF complete sample are in fact likely highly dust obscured high sSFR galaxies or starbursts.     
     
\begin{figure*}
\begin{center}
\includegraphics[width=1\textwidth]{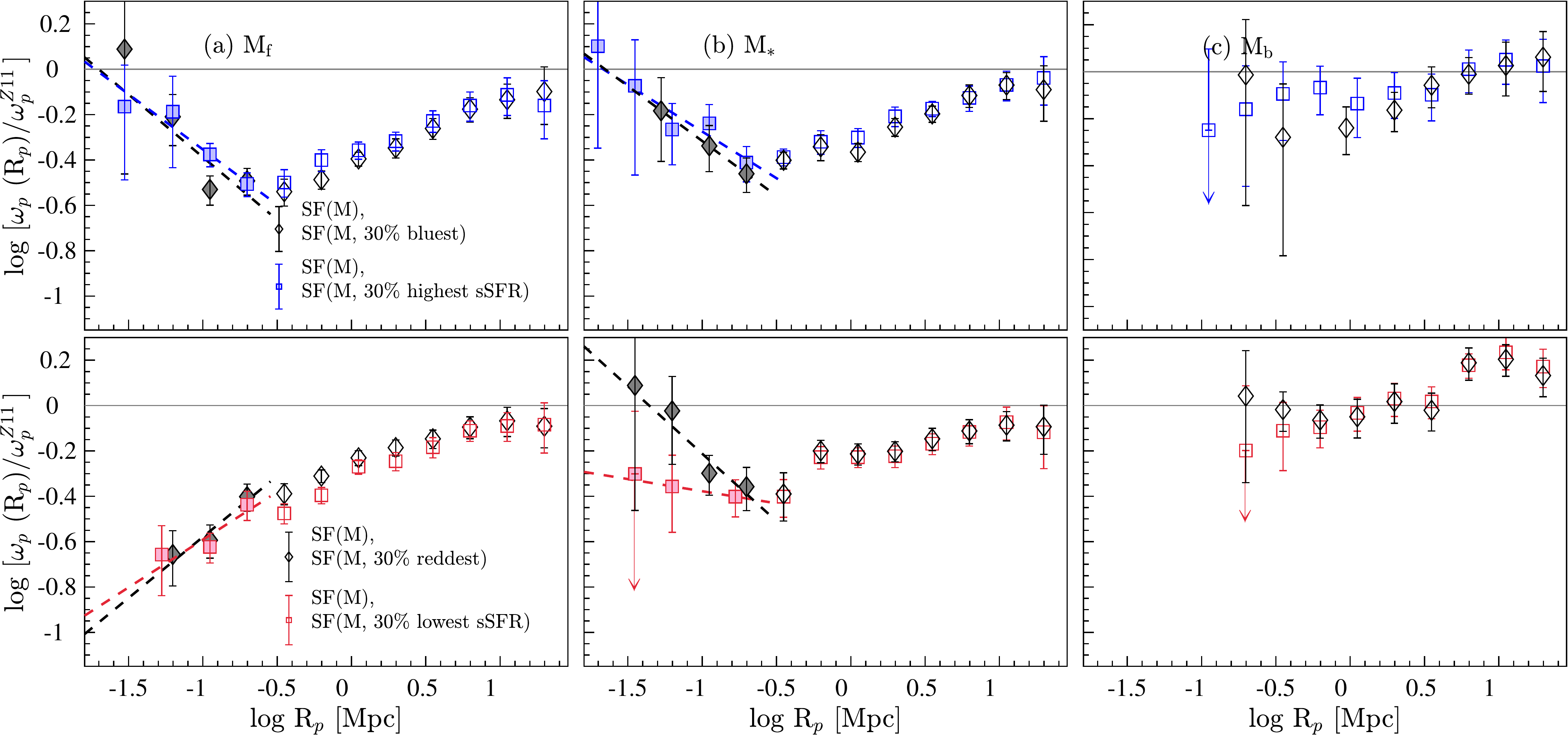}
\caption{The projected CCFs of high (low) sSFR and optically blue (red) galaxies of luminosity selected SF complete samples relative to $\omega_p^{Z11}$ (optical luminosity increases left-to-right). Top panels: the CCFs of optically blue (black diamonds) and high sSFR (blue squares) galaxies. Bottom panels: the CCFs of optically red (black diamonds) and low sSFR (red squares) galaxies. The dashed lines of the same colours denote the best-fitting linear relations to the R$_p<0.23$ (log R$_p<-0.64$) Mpc data of the same colour. The data points used for the fitting are shown as filled squares, and the arrows denote the data with significant uncertainties.}
\label{fig:xcrr_Mr1_color_ssfr_func}
\end{center}
\end{figure*}

\begin{figure*}
\begin{center}
\includegraphics[width=1\textwidth]{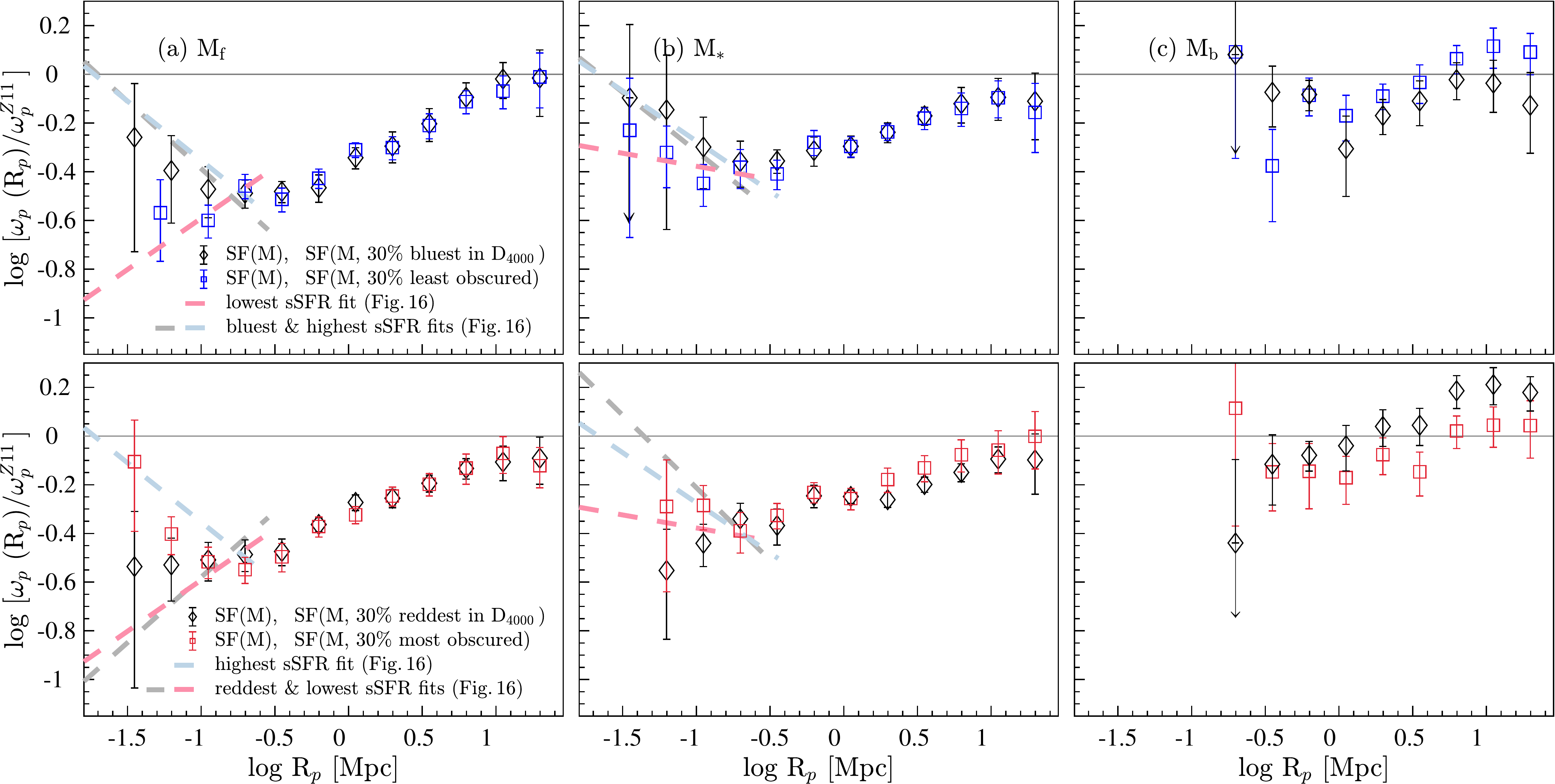}
\caption{The projected CCFs of least (most) dust obscured and spectroscopically blue (red) galaxies of luminosity selected SF samples relative to $\omega_p^{Z11}$ (optical luminosity increases from left to right). Top panels: the CCFs of least dust obscured (the $30\%$ of galaxies with the lowest Balmer decrement measures; blue squares) and spectroscopically blue (the $30\%$ with the lowest D$_{\rm{4000}}$ indices; black diamonds) galaxies. Bottom panels: the CCFs of most dust obscured (high Balmer decrement measures; red squares) and spectroscopically red (high D$_{\rm{4000}}$ indices; black diamonds) galaxies. For reference, we show the best-fitting linear relations shown in Figure\,\ref{fig:xcrr_Mr1_color_ssfr_func} as dashed lines.}
\label{fig:xcrr_Mr1_dust_d4000_func}
\end{center}
\end{figure*}

In Figures\,\ref{fig:xcrr_Mr1_color_ssfr_func} and \ref{fig:xcrr_Mr1_dust_d4000_func}, we compare the CCFs$_{\omega_p^{Z11}}$ of all $30\%$ subsamples drawn from the three disjoint luminosity selected SF complete samples. The top panels of Figure\,\ref{fig:xcrr_Mr1_color_ssfr_func} present the CCFs$_{\omega_p^{Z11}}$ of high sSFR galaxies and optically blue galaxies (blue and black symbols), and those of low sSFR and optically red galaxies (red and black symbols) are presented in the bottom panels. To emphasise the degree of the enhancement of the clustering signal on small scales, we fit a linear relation to the log R$_p<-0.64$ Mpc data, where the dashed lines of the same colour denote the best-fitting linear relations to the data of the same colour. Likewise, the CCFs$_{\omega_p^{Z11}}$ of galaxies with low (high) D$_{\rm{4000}}$ indices and low (high) dust obscurations are presented in the top (bottom) panels of Figure\,\ref{fig:xcrr_Mr1_dust_d4000_func}. In this figure, for reference, we over plot the best fitting linear relations to the log R$_p<-0.64$ Mpc data shown in Figure\,\ref{fig:xcrr_Mr1_color_ssfr_func} as dashed lines. 

For $-0.52\lesssim$ log R$_p$ [Mpc] $\lesssim1.3$, all CCFs$_{\omega_p^{Z11}}$ show a progressive decline in the strength of clustering with decreasing R$_p$. The clustering amplitudes of low sSFR and optically red galaxies over this range are, on average, higher than that of their respective high sSFR and optically blue counterparts, in agreement with the studies that find high sSFR galaxies are less clustered than their low sSFR counterparts \citep[e.g.][]{Mostek2013, Coil2016}, as well as with the studies that find higher clustering strengths for optically redder galaxies versus optically bluer systems \citep[e.g.][]{Zehavi2011, Ross2014, Favole2016}.

Interestingly, on small scales (i.e.\,log R$_p\lesssim-0.52$ Mpc), we see a discrepant behaviour between the CCFs$_{\omega_p^{Z11}}$ of optically red galaxies of different luminosity selected samples. The most notable is the CCF$_{\omega_p^{Z11}}$ of M$_f$ galaxies that show a continuous decline in clustering strength with decreasing R$_p$, whereas the respective CCFs$_{\omega_p^{Z11}}$ of M$_*$ and M$_b$ galaxies suggest otherwise. These differences can shed light into dust build up and destruction mechanisms in optically faint (low mass) versus bright (massive) star forming galaxies. At a fixed SFR, an optically faint galaxy would be classified as a starburst, while a luminous system would appear as a normal (or a low) star former (see the distribution of the constant log SFR contours in Figure\,\ref{fig:ssfr_mass_dist}). Therefore to gain further insights into these differences, we add an analysis based on D$_{\rm{4000}}$ and Balmer decrements (Figure\,\ref{fig:xcrr_Mr1_dust_d4000_func}), which are complementary to sSFR and ($g-r$)$_{\rm{rest}}$, to this study.

\begin{figure*}
\begin{center}
\includegraphics[width=1\textwidth]{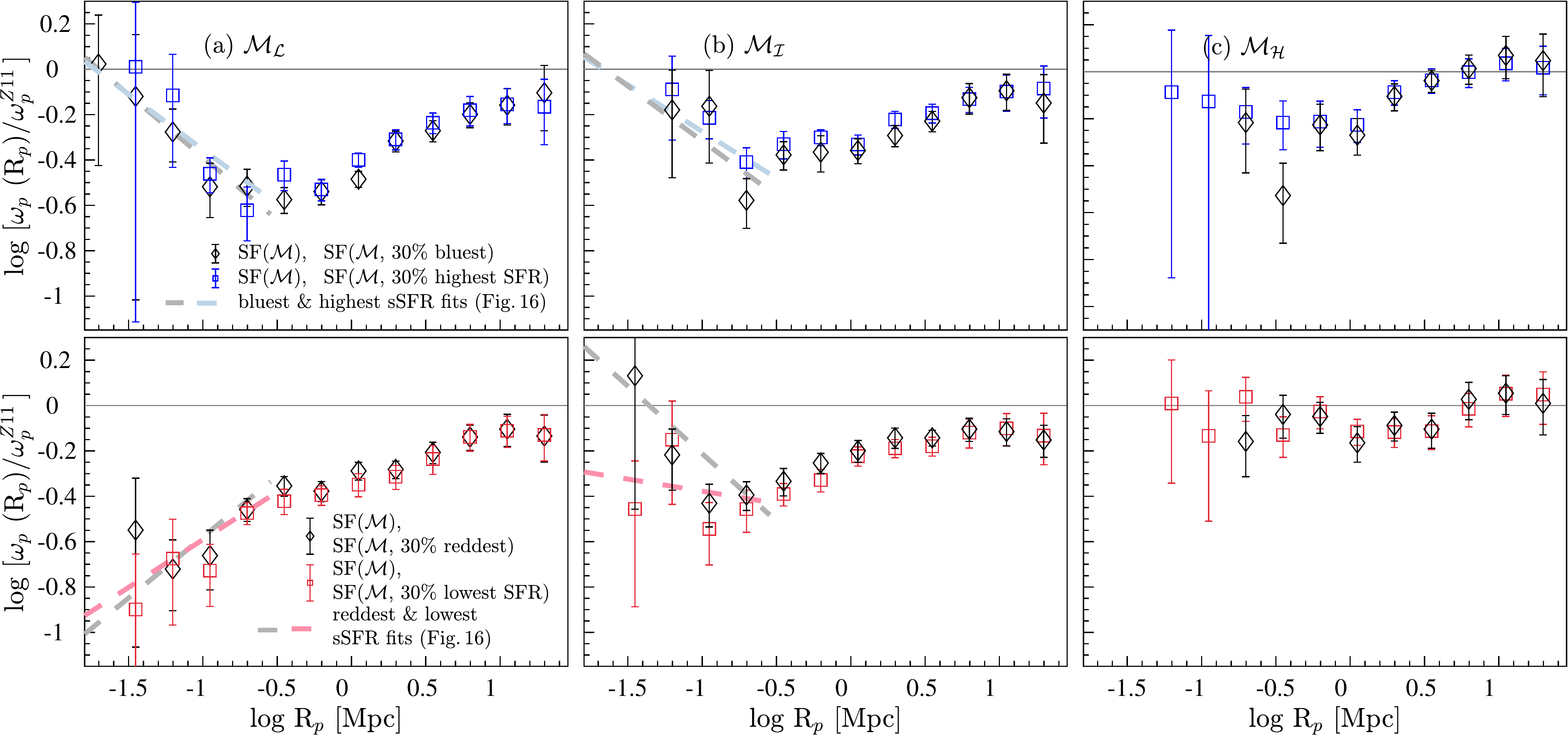}
\caption{The projected CCFs of high (low) SFR and optically blue (red) galaxies of stellar mass selected star forming galaxies  relative to $\omega_p^{Z11}$ (Table\,\ref{table:stats3}; stellar mass increases left-to-right). Top panels: the CCFs of high SFR (blue squares) and optically blue (black diamonds) galaxies. Bottom panels: the CCFs of low (red squares) and optically red (black diamonds) galaxies. For reference, we show the best-fitting linear relations shown in Figure\,\ref{fig:xcrr_Mr1_color_ssfr_func} as dashed lines.}
\label{fig:xcrr_Mass1_color_ssfr_func}
\end{center}
\end{figure*}

The D$_{\rm{4000}}$ spectral index is a diagnostic of cumulative star formation history of a galaxy, where lower D$_{\rm{4000}}$ indices are indicative of younger stellar populations and vice versa, and is therefore considered a proxy for ($g-r$)$_{\rm{rest}}$ (Figure\,\ref{fig:property_dist}). For ease of comparison with the optical colour based analysis discussed above, we hereafter refer to galaxies with lower (higher) D$_{\rm{4000}}$ indices as spectroscopically blue (red).   

The principal advantage of using D$_{\rm{4000}}$ is that it is less sensitive to dust reddening than ($g-r$)$_{\rm{rest}}$\footnote{The D$_{\rm{4000}}$ measures used for this study are  based on the \cite{Balogh1999} definition, which samples a very narrow range in wavelength.}. Secondly, it is a spectroscopy-based quantity. In the case of single fibre spectroscopy, the spectrum of a galaxy represents the central region where interaction triggered starbursts are likely to occur \citep{Mihos1996, DiMatteo2007, Montuori2010}, whereas photometry based colours represent the light from the whole galaxy. Therefore in galaxies undergoing interactions, where the SFR of centrally triggered starburst, that may also be highly dust obscured, likely mostly contribute to the total SFR, the correlation between SFR and D$_{\rm{4000}}$ can be stronger than that between SFR and ($g-r$)$_{\rm{rest}}$. Indeed this is evident in Figure\,\ref{fig:xcrr_Mr1_dust_d4000_func}. The CCF$_{\omega_p^{Z11}}$ of spectroscopically red M$_*$ galaxies shows a continuous decline in strength on log R$_p\lesssim-0.64$ Mpc, whereas the opposite is observed for optically red M$_*$ galaxies. In comparison to the CCF$_{\omega_p^{Z11}}$ of optically red M$_f$ galaxies, the CCF$_{\omega_p^{Z11}}$ of spectroscopically red M$_f$ galaxies too show some differences, though within uncertainties the two CCFs$_{\omega_p^{Z11}}$ are in agreement. According to these results, the D$_{\rm{4000}}$ index appears to be more useful in discriminating starbursts than optical colours. Even though fibre colours are still more susceptible to dust effects than D$_{\rm{4000}}$, the correlation between fibre colour and SFR can be stronger than that between global colour and SFR. 
  
The dust obscuration in star forming galaxies has been observed to depend on both galaxy SFR and stellar mass \citep[e.g.][]{Brinchmann2004, Garn2010, Zahid2013}. Dust is theorised to build up rapidly during a starburst \citep{Hjorth2014}, while a quiescently star forming galaxy experiences a simultaneous decline in dust and SFR as a result of dust destruction and diminishing gas supply \citep{daCunha2010}. The CCFs$_{\omega_p^{Z11}}$ of both most and least dust obscured galaxies (Figure\,\ref{fig:xcrr_Mr1_dust_d4000_func}) show enhancements in clustering amplitudes with decreasing log R$_p$ on $\lesssim-0.64$ Mpc. As shown in Figure\,\ref{fig:BD_vs_lum}, the star forming galaxies can have a range of dust obscurations, which can explain the similar enhancements in clustering observed for most and least dust obscured star forming populations of the luminosity selected samples. The clustering excess observed for most dust obscured M$_*$ galaxies further supports our earlier assertion that the increase in clustering amplitude of optically red M$_*$ galaxies (Figure\,\ref{fig:xcrr_Mr1_color_ssfr1}) is, at least in part, caused by the presence of dusty starbursts.  
 
For $-0.52\lesssim$ R$_p$ [Mpc] $\lesssim1.3$, the CCFs$_{\omega_p^{Z11}}$ of the most (least) dust obscured M$_{\rm{f}}$ and M$_*$ galaxies agree qualitatively with the high (low) sSFR counterparts, as well as with the CCFs$_{\omega_p^{Z11}}$ of spectroscopically blue (red) galaxies. The M$_{b}$ CCFs$_{\omega_p^{Z11}}$ of the most and least dust obscured star formers, on the other hand, show an agreement with that of high sSFR M$_{b}$ galaxies. 

\begin{figure*}
\begin{center}
\includegraphics[width=1\textwidth,trim={0.2cm 1.75cm 0.4cm 0.0cm},clip]{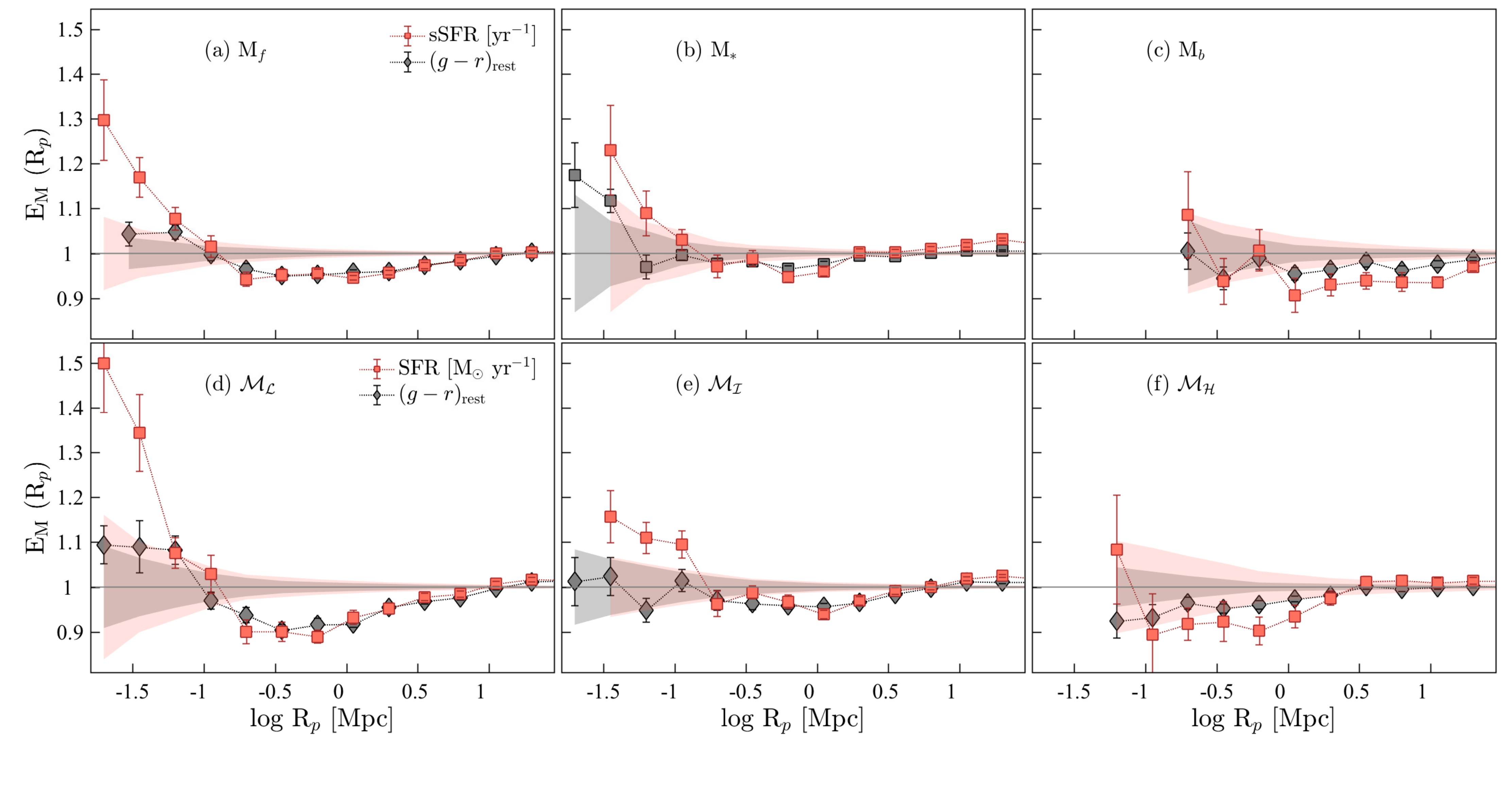}
\caption{The rank-ordered MCFs of luminosity (top row; optical luminosity increases left-to-right) and stellar mass (bottom row; stellar mass increases left-to-right) selected SF complete samples. The orange and grey filled symbols on the top panels denote the rank-ordered sSFR and ($g-r$)$_{\rm{rest}}$  MCFs, respectively, and on bottom panels SFR and ($g-r$)$_{\rm{rest}}$ MCFs, respectively. The shaded regions indicate the scatter from randomising the marks.}
\label{fig:mcrr_ranked_Mr1_color_ssfr_func}
\end{center}
\end{figure*}

For completeness, we present the CCFs$_{\omega_p^{Z11}}$ of high (low) SFR and optically blue (red) galaxies of the three disjoint stellar mass selected samples (Table\,\ref{table:stats3}) in Figure\,\ref{fig:xcrr_Mass1_color_ssfr_func}. The dashed lines are the same as in previous figures. These results are, as expected, largely comparable to that observed for high (low) sSFR and optically blue (red) galaxies of luminosity selected samples, and as such, we do not discuss them separately here.  

Finally, we also perform a volume limited cross correlation analysis, the results of which are presented and discussed in Appendix\,\ref{app:vanalysis}. Briefly, the CCFs$_{\omega_p^{Z11}}$ of volume limited samples show a qualitative agreement with their respective non-volume limited counterparts on most scales. There are some quantitive differences between the two sets of CCFs$_{\omega_p^{Z11}}$ on smaller scales that rise as a result of small number statistics.  

\subsection{The rank-ordered mark correlation functions of star forming galaxies}\label{subsec:mark_cfs}

The mark clustering statistics are different than the auto- and cross-correlation techniques discussed in the previous sections. The mark statistics can shed light on the dependence of a given physical property on the separation of a galaxy pair by weighting each galaxy in that pair by that physical property. Given this sensitivity of mark correlation functions (MCF) to environmental effects, they form a useful tool in identifying and quantifying underlying correlations of various galaxy properties with environment. 

In conventional mark two-point clustering statistics, the correlation function is directly weighted by a given mark, i.e.\,a physical property (e.g.\,SFR, sSFR). Consequently, the amplitude of a MCF depends not only on the distribution of marks \citep{Skibba2006, Skibba2009}, but also on the differences in the formulation of a mark \citep[e.g.\,log or linear;][]{Skibba2006, Skibba2009}. Therefore, unless the distributions of different marks are similar, different MCFs cannot be compared with each other to understand the dependence of different galaxy properties with galaxy separation. In our case, the SFR, sSFR, and ($g-r$)$_{\rm{rest}}$ distributions of SF complete samples used differ in shape, in magnitude, and in range. As such, in order to compare the SFR, sSFR and  ($g-r$)$_{\rm{rest}}$ MCFs, we rank-order the marks and use the rank as the mark. This method, introduced in \cite{Skibba2013}, allows the effects of the shape of the distribution on the strength of the MCF to be removed, such that mark correlation signal can be compared between different marks. The caveat of this method is that any information contained in the shape of a distribution will be lost.   

We present the rank-ordered sSFR and ($g-r$)$_{\rm{rest}}$ MCFs of luminosity selected (top panels), and rank-ordered SFR and ($g-r$)$_{\rm{rest}}$ MCFs of stellar mass selected (bottom panels) SF complete samples in Figure\,\ref{fig:mcrr_ranked_Mr1_color_ssfr_func}. On small scales, the rank-ordered sSFR and SFR MCFs indicate a clear enhancement in amplitude compared to that of ($g-r$)$_{\rm{rest}}$ MCFs. This suggests that sSFR and SFR correlation signals indeed correlate more strongly with the environment than optical colour. The decrement in sSFR, SFR and  ($g-r$)$_{\rm{rest}}$ mark correlation signals between the $-0.82\lesssim$ log R$_p$ [Mpc] $\lesssim0.6$, which is more strongly evident in rank-ordered MCFs of M$_f$ and $\mathcal{M}_{\mathcal{L}}$ galaxies, likely demonstrates the effects of post-starbursts, where certain physical properties of a galaxy, e.g.\,SFR and colour, are affected by the increased presence of now ageing stellar population produced during a starburst.
       
For completeness, we also present and discuss the conventional MCFs in Appendix\,\ref{app:mark_CFs} (Figures\,\ref{fig:mcrr_Mr1_color_ssfr_func} and \ref{fig:mcrr_Mass_color_ssfr_func}). The most notable in the conventional case is the strengthening in clustering amplitude with increasing optical brightness observed for sSFR and SFR populations, which mirrors that observed in auto- and cross-correlation functions presented in previous sections.   

\vspace{-0.6cm}
\section{Discussion}\label{sec:discuss}

In this study, we considered several different star forming properties of galaxies (i.e.\,SFR, sSFR, ($g-r$)$_{\rm{rest}}$, D$_{\rm{4000}}$ and Balmer decrement), which are most likely to be affected by galaxy-galaxy interactions.  We utilised the [\ion{O}{iii}]~$\lambda5007$\AA/H$\beta$ and [\ion{N}{ii}]~$\lambda6584$\AA/H$\alpha$ diagnostics \citep[i.e.\,BPT,][]{Baldwin1981}, which can be used as an indicator of gas phase metallicity, to demonstrate the variation of the physical properties considered with metallicity in star forming galaxies (Figures\,\ref{fig:property_dist1}-\ref{fig:ssfr_mass_dist}). In general, the variation in sSFR largely mirrors that of ($g-r$)$_{\rm{rest}}$ and D$_{\rm{4000}}$, where low sSFR are typically characterised by lower metallicities. Dust obscuration, on the other hand, indicates a variation similar to that seen with SFR, where high SFR galaxies show a higher dust obscuration than low SFR systems.    

Below we discuss the main findings of this study and is structured as follows. A discussion of the results of auto, cross and mark correlation analysis of star forming galaxies is presented in \S\ref{subsec:intscales_auto} to \S\ref{subsec:discuss_mark}, and in \S\ref{sebsec:sdss_gama_comp}, we compare the GAMA results of this study with that of SDSS. 

\subsection{On the potential interaction-scale halo-size dependence of interaction-driven disturbances}\label{subsec:intscales_auto}

The role that large-scale environment plays in driving and sustaining changes induced during a galaxy-galaxy interaction is understood to a lesser extent than the role of the interaction itself. Generally, the net changes in physical properties of galaxies are used as \textit{direct} indicators of interactions and environmental effects. As mentioned before, the focus of our study is to explore the suitability of utilising \textit{two-point correlation statistics} to shed light on any dependence of galaxy-galaxy interactions on their large scale (i.e.\,halo-scale) environment.  For this, we have computed two-point auto-, cross- and mark-correlation functions of star forming galaxies as a function of both optical luminosity and stellar mass, which approximately correlate with halo mass. 

In order to quantify the R$_p$ out to which signatures of interactions ought to persist, in \S\,\ref{subsec:auto_corr}, we make the assumption that any change in the relative strength of clustering of a given population reflects its interaction scale\footnote{We use the term `interaction scale' to denote the R$_p$ out to which changes in physical properties ought to be evident instead of `observable timescale' to avoid confusion, as this term is generally used by studies that rely on net changes in physical properties to trace interactions.}. For example, the ACFs of both luminosity and stellar mass selected star forming galaxies are consistent with a power-law on $-0.15\lesssim$ log R$_p$ [Mpc]$\lesssim1.3$. On log R$_p\lesssim-0.15$ Mpc, they show a significant clustering excess (Figure\,\ref{fig:auto_corr_proj_main}). This is best seen in ACFs$_{\omega_p^{Z11}}$, where this change appears as a turn-over in the signal. It is this `turn-over' that we consider to approximately correspond to the interaction scale of that galaxy population. The interaction scales estimated this way appear to depend on galaxy luminosity. This is in the sense that the interaction scale of optically brighter star-forming galaxies is greater than that of optically faint galaxies. \textit{This could be interpreted as a signature of a halo size-interaction scale dependence, where the evidence of interactions between star formers residing in massive halos are visible out to larger radii than those between star formers residing in low mass halos.} This can be, in part, due to massive halos playing a greater role in enhancing and sustaining the effects of galaxy interactions than their less massive counterparts. Equally, this could also be an artefact of high-mass inhabitants of massive halos being able to form stars more efficiently than low-mass galaxies in interactions (Ferreras et al.\, accepted).

The ACFs of both luminosity and stellar mass selected REF galaxies also show similar changes in the small scale clustering. These changes are, however, not as significant as those observed in star-forming galaxies. In comparison to the ACFs of REF galaxies, the star formers show lower clustering amplitudes over most scales, except on log R$_p\lesssim-0.15$ Mpc. On log R$_p\lesssim-0.15$ Mpc, the ACFs of star forming galaxies show a rapid increase in the amplitude of clustering with decreasing R$_p$. Consequently, over these scales, the clustering of star forming galaxies appears to be similar to that of REF. Both of these results are consistent with the findings of previous studies; the former with the studies that find optically redder galaxies are more strongly clustered than their bluer counterparts \citep[e.g.][]{Zehavi2005b, Skibba2009, Zehavi2011, Bray2015}, and the latter with the \cite{Farrow2015} clustering study of optically selected red and blue galaxies, finding an up-turn in the clustering of the blue systems on small scales, as well as with the results of \cite{Heinis2009} and \cite{Mostek2013}.  

\subsection{On the direct indicators versus two-point correlation statistics tracing interaction scales}

Here we discuss the potential reasons for the differences in R$_p$ reported by the studies that utilise direct probes of interactions (see \S\,\ref{sec:intro} for a discussion), as well as between those and the predictions of our auto correlation analysis. 

A vast number of competing factors can influence both the strength of an interaction-induced physical change and the R$_p$ out to which the net effect is observable. The orbital parameters, for instance, can play a significant role in moderating the SFR response. Both observational and theoretical studies suggest that retrograde encounters lead to higher star formation efficiencies, and thus higher SFR enhancements, than prograde encounters \citep{DiMatteo2007, Mesa2014}. The ratio of the stellar masses of the progenitors and their gas fractions are two other factors that can significantly influence the strengths of direct indicators. Galaxy pairs with mass ratios between $1-3$ are observed to have the strongest SFR enhancements \citep[e.g.][]{Ellison2008, Cox2006}. Likewise, lower gas fractions are theorised to lead to lower SFR enhancements \citep{DiMatteo2007}. While starbursts with the shortest durations tend to typically show the strongest enhancements \citep{DiMatteo2007} and tend to occur over the smallest separations \citep[typically $<30\,h^{-1}_{70}$ kpc; e.g.][]{Ellison2008, Li2008, Wong2011, Scudder2012, Patton2013}, the smallest separations can, also, inhibit SFR if the tidal forces are strong enough to eject molecular gas to tidal tails without allowing the gas to funnel to the centres of galaxies \citep{DiMatteo2007}. Overall, these competing effects can ``wash out'' the net signal of direct indicators, thus affecting the observability of a physical change. 

The differences in dynamical timescales associated with different star formation probes is another factor that must be considered when using physical properties as tracers of interactions. \cite{Davies2015}, based on the GAMA survey data, report that short-duration star formation indicators show stronger signs of enhancement/suppression than long-duration tracers. The H$\alpha$ SFR, for example, is a direct tracer of on-going star formation in galaxies, probing on average the star formation over a shorter timescale (i.e.\,$\sim10$ Myr) than broadband photometry, e.g.\,($g-r$) probes star formation over much longer timescales of $\sim1$ Gyr. The short-duration indicators are, therefore, expected to be most vulnerable to recent dynamical events. This suggests that the dynamical timescales of processes that likely trigger short-duration star formation events are also shorter than those of processes that likely trigger long-duration events. The implication being that analyses which rely on observations of net changes are susceptible to the differences in the dynamical timescales of physical processes that trigger and sustain different changes. This can, perhaps, further explain the differences in the reported interaction scales. In this sense, auto correlation techniques offer an alternative to trace interaction scales which is almost\footnote{As we have used H$\alpha$ fluxes to select the star forming sample used for this analysis, our results are not completely independent.} independent of the influences of interaction induced direct observables.  

A dependence on the interaction scale and the size of a halo suggests that star formation activity evolves differently in different environments. \cite{Elbaz2007} and \cite{Ziparo2014} report a reversal of the SFR-density relationship at $z\sim1$, from high-density environments hosting high-SFR galaxies at earlier times to low-density environments hosting high-SFR systems at later times. \cite{Popesso2015a, Popesso2015b} interpret SFR-density relation and "galaxy downsizing" \citep{Cowie1996} in terms of "halo downsizing", where the SFR contribution of massive halos to the cosmic SFR density becomes progressively less significant with increasing cosmic time. In the local Universe, the bulk of the stellar mass is locked in galaxy-groups \citep{Eke2005} so that group-sized halos are the most common type of halos for a star forming galaxy to inhabit. Therefore it is likely that most of the aforementioned studies preferentially selected galaxies residing in one type of a halo (i.e.\,group-sized halos) over the others. In our study, by using disjoint luminosity and stellar mass samples, we attempt to minimise this preferential selection, as well as the overlap between halos of different sizes, thereby giving insight into interactions between star formers in relatively low- versus high-mass halos. 

\subsection{On the use of cross-correlation techniques in the determination of interaction scales with respect to galaxy properties}

As mentioned earlier, galaxy-galaxy interactions have been observed to drive many physical changes in galaxies. The best physical tracer of an interaction can, however, differ depending on the progenitors, the environment and the interaction itself. It has been shown, both theoretical and observationally, that SF--SF galaxy pairs largely favour low-to-moderate density environments, which are typical hosts to low mass galaxies with higher gas fractions, whereas non-SF -- non-SF  and SF -- non-SF galaxy pairs are preferentially found in high-density environments \citep[e.g.][]{Lin2010, Ellison2010}.
As such, while interactions still occur in high-density environments, they may not always lead to an enhancement in star formation \citep[][]{Ellison2008, Ellison2010}, though can, perhaps, lead to a change in another property such as optical colours. Below we discuss the clustering properties of star forming galaxies in different environments with respect to different star forming properties of galaxies obtained from the cross correlation analysis presented in \S\,\ref{subsec:cross_corr}.

On small scales, the CCFs$_{\omega_p^{Z11}}$ of M$_{f}$ star formers of high (low) sSFRs, high (low) dust obscurations, and bluer (redder) optical and spectroscopic colours show enhancements (decrements) of varying degree in clustering amplitudes with decreasing R$_p$ (Figures\,\ref{fig:xcrr_Mr1_color_ssfr_func} and \ref{fig:xcrr_Mr1_dust_d4000_func}). In contrast, all CCFs$_{\omega_p^{Z11}}$ of star forming M$_*$ galaxies, except spectroscopically red objects, show enhancements. The most notable are the opposing clustering trends observed between optically red M$_f$ versus M$_*$ populations, and likewise between spectroscopically red (i.e.\,higher D$_{\rm{4000}}$ indices) and optically red M$_*$ populations. As D$_{\rm{4000}}$ is less sensitive to dust effects than optical colours, one of the potential drivers of these discrepancies is dust obscuration. 

The clustering excess observed for highly dust obscured M$_*$ galaxies and the dearth in clustering observed for spectroscopically red galaxies support the assertion that a large fraction of optically red star forming M$_*$ galaxies are likely dusty starbursts. To illustrate this further, in Figure\,\ref{fig:dust_effects}, we show the distribution of sSFRs as a function of $250\mu$m luminosity\footnote{The {\textsc{Herschel}} 250$\mu$m photometry is drawn from {\textsc{HATLASCatv03}} \citep{Smith2011}.} (L$_{\rm{250}}$ [W/Hz]) for all optically red M$_{f}$ and M$_*$ star formers detected in {\textsc{Herschel}} 250$\mu$m. The colour code denotes mean dust obscuration as measured by Balmer decrement, and black and red contours indicate the distribution of the $30\%$ reddest M$_{f}$ and M$_*$ galaxies, respectively. The sSFR cuts used to select the $30\%$ highest sSFR galaxies from M$_{f}$ and M$_*$ samples are shown by the dashed lines. The significant overlap between the high sSFR and optically red M$_*$ populations demonstrates that the redder optical colours of these star forming systems have been enhanced by dust. 
\begin{figure}
\begin{center}
\includegraphics[scale=0.38,trim={0.2cm 0.75cm 2cm 1.5cm},clip]{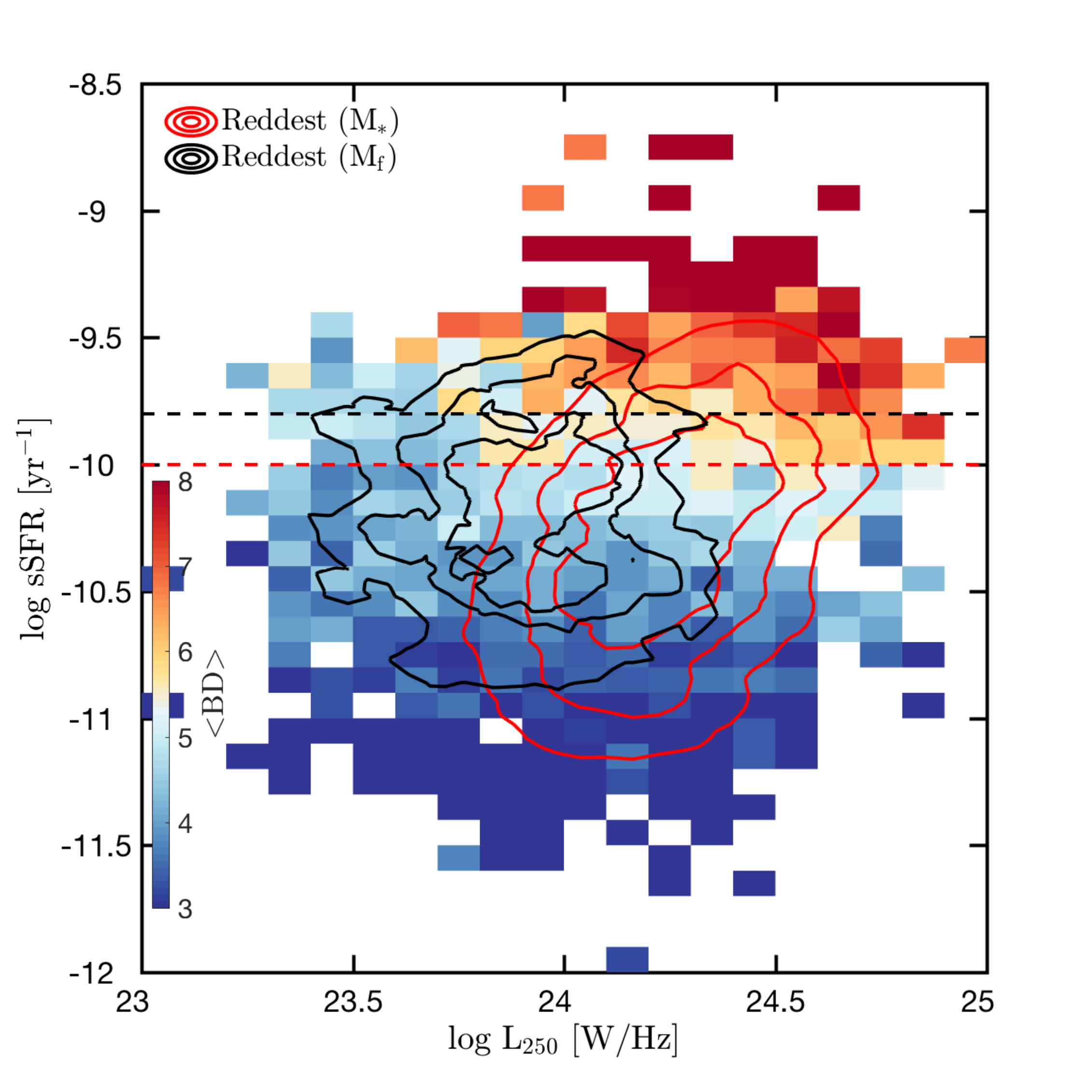}
\caption{The log L$_{\rm{250}}$ [W/Hz] and log sSFR [yr$^{-1}$] distribution of the $30\%$ reddest M$_{\rm{f}}$ and M$_*$ galaxies of SF complete sample, colour coded by mean dust obscuration (as measured by Balmer decrement, i.e.\,BD). The contours enclose 25, 50 and 75$\%$ of the data, and the dashed lines denote the approximate cuts in log sSFR used to select the $30\%$ highest sSFR M$_{\rm{f}}$ (black) and M$_*$ (red) galaxies.}
\label{fig:dust_effects}
\end{center}
\end{figure}

Finally, the differences in the environments typically inhabited by optically faint versus bright galaxies provide another explanation for the differences between CCFs. The galaxy-galaxy interactions in higher density environments have been observed to lead to quenching of on-going star formation, thus amplifying their redder colours \citep{Patton2011, Ellison2010}. This likely also plays a role in enhancing redder colours of M$_*$ galaxies that generally reside in denser environments than fainter systems.  

\subsection{On the use of mark-correlation techniques in the determination of interaction scales} \label{subsec:discuss_mark}

The mark correlation statistics allow for the dependence of interaction scale on galaxy properties to be investigated. In mark statistics, unlike in cross correlation analysis, the galaxies are either weighted directly by a given physical property or by a rank-order assigned to them based on the distribution of a given physical property. We compute MCFs using both of these methods. The results based on the former (i.e.\,the conventional) method are presented in Appendix\,\ref{app:mark_CFs}, and they allow the comparison of MCFs of the same mark between different galaxy samples. Those based on the latter are shown in Figure\,\ref{fig:mcrr_ranked_Mr1_color_ssfr_func}, and permits the comparison of MCFs of different marks between different galaxy samples.  

The sSFR and ($g-r$)$_{\rm{rest}}$ MCFs based on the conventional method show a strengthening in the mark correlation signal with decreasing R$_p$, and at a fixed R$_p$, the strength increases with increasing optical brightness. The same trend is also evident with increasing stellar mass in the SFR and ($g-r$)$_{\rm{rest}}$ MCFs of stellar mass selected SF complete samples. The greater enhancement in sSFR observed in the MCF of M$_*$ galaxies than that of M$_{\rm{f}}$ galaxies (Figure\,\ref{fig:mcrr_Mr1_color_ssfr_func}) is in agreement with that expected if the fraction of dusty starbursts with M$_*$ luminosities is higher than those with optically fainter luminosities. 

Finally, the comparison of rank-ordered MCFs of sSFR, SFR and ($g-r$)$_{\rm{rest}}$ galaxies show that the relative mark correlation strengths of sSFR MCFs are higher than that of the respective ($g-r$)$_{\rm{rest}}$ functions across all luminosity selected SF complete samples. Likewise, the mark correlation strengths of SFR MCFs are higher that of the respective ($g-r$)$_{\rm{rest}}$ across all stellar mass selected SF complete samples. This suggests that sSFR and SFR are more sensitive probes of the effects of interactions than optical colours, in agreement with the findings of \cite{Heinis2009}.  

\subsection{A comparison between SDSS and GAMA} \label{sebsec:sdss_gama_comp}
\begin{figure*}
\begin{center}
\includegraphics[width=.9\textwidth]{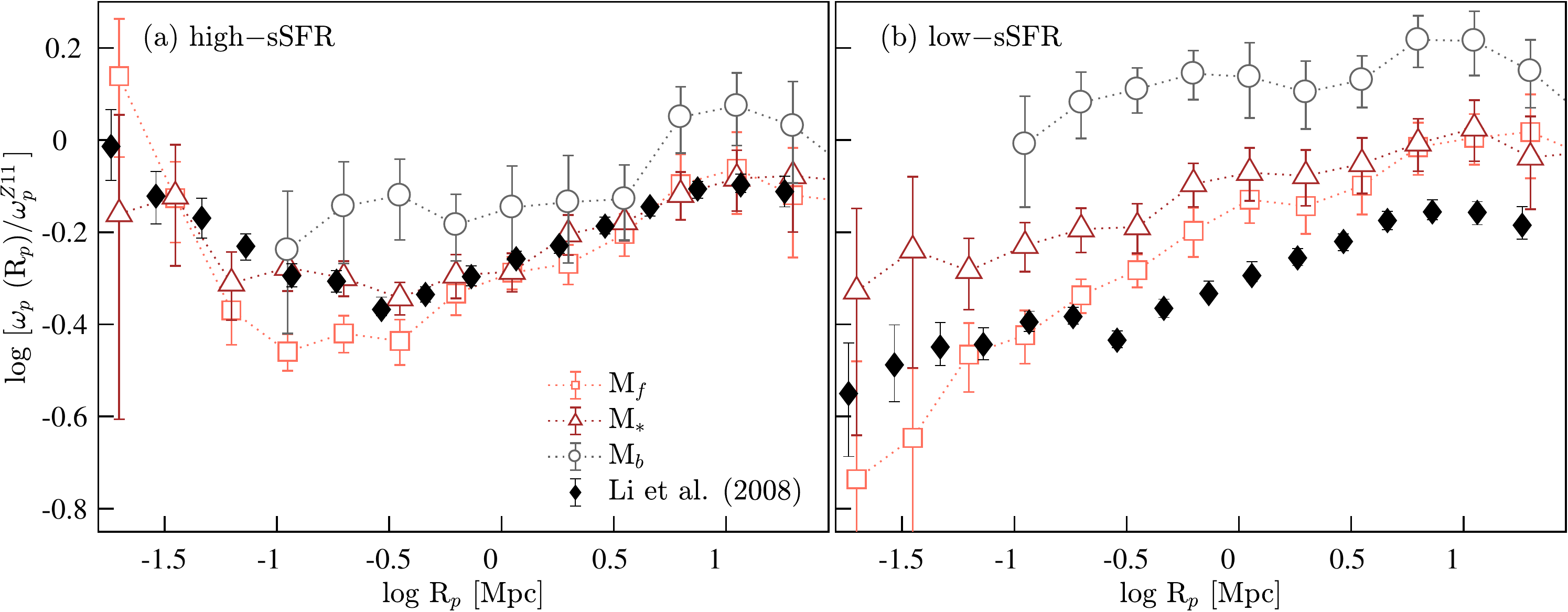}
\caption{The projected REF CCFs high sSFR (open squares; left panel) and low sSFR (open squares; right panel) M$_f$, M$_*$ and M$_b$ galaxies relative to $\omega^{Z11}_p$. The filled symbols denote the \citet{Li2008} CCFs of high sSFR (left panel) and low sSFR (right panel) SDSS galaxies. \citet{Li2008} define high and low sSFR galaxies as galaxies contained in the upper and lower $25^{\rm{th}}$ percentiles of the distribution of sSFRs of galaxies over the $0.01\leq z\leq 0.3$ and $-23\leq$ M$_r\leq-17$ ranges. }
\label{fig:cross_lit3}
\end{center}
\end{figure*}

The CCFs$_{\omega_p^{Z11}}$ of GAMA and SDSS \citep{Li2008} high and low sSFR galaxies are presented in Figure\,\ref{fig:cross_lit3}. \cite{Li2008} define high and low sSFR galaxies as those within the upper and lower 25$^{\rm{th}}$ percentiles of the sSFR distribution, and they cross correlated with a reference sample containing galaxies in the $0.01\leq z \leq0.3$ and $-23\leq$ M$_r\leq-17$. Therefore in order to make this comparison as fair as possible, the GAMA CCFs$_{\omega_p^{Z11}}$ shown in Figure\,\ref{fig:cross_lit3} are the CCFs$_{\omega_p^{Z11}}$ obtained from cross correlating SF complete samples with their respective REF samples (see Appendix\,\ref{App:XC_ref} for the cross correlation analysis involving REF samples). In general, the SDSS CCFs$_{\omega_p^{Z11}}$ of high sSFR show a good agreement with that of GAMA high sSFR M$_*$ galaxies. The low sSFR SDSS function, on the other hand, exhibits a lower clustering strength than the GAMA functions, which is most likely a result of the differences between galaxy samples used for the two studies. For example, even though we show the results of the cross correlation between star forming and reference samples in Figure\,\ref{fig:cross_lit3}, the redshifts and optical luminosities spanned by the galaxy samples used by \cite{Li2008} are still larger than the ranges that we considered for our analysis. 

In Figure\,\ref{fig:mark_lit_comp1}, we compare the GAMA sSFR MCFs with the SDSS measures provided in \cite{Li2008}. Relative to GAMA, the enhancement in sSFR of SDSS galaxies occurs at a smaller R$_p$, and the amplitude at a fixed R$_p$ is lower than that of GAMA M$_f$ and M$_*$  sSFR MCFs. Moreover, the GAMA sSFR MCFs of M$_f$ and M$_*$ galaxies show a strengthening in the mark correlation signal with increasing optical luminosity and/or redshift. The enhancement in sSFRs of SDSS galaxies also appears to support this trend, suggesting that interactions between luminous galaxies trigger more intense starbursts than those between faint systems. 

\begin{figure}
\begin{center}
\includegraphics[width=0.45\textwidth]{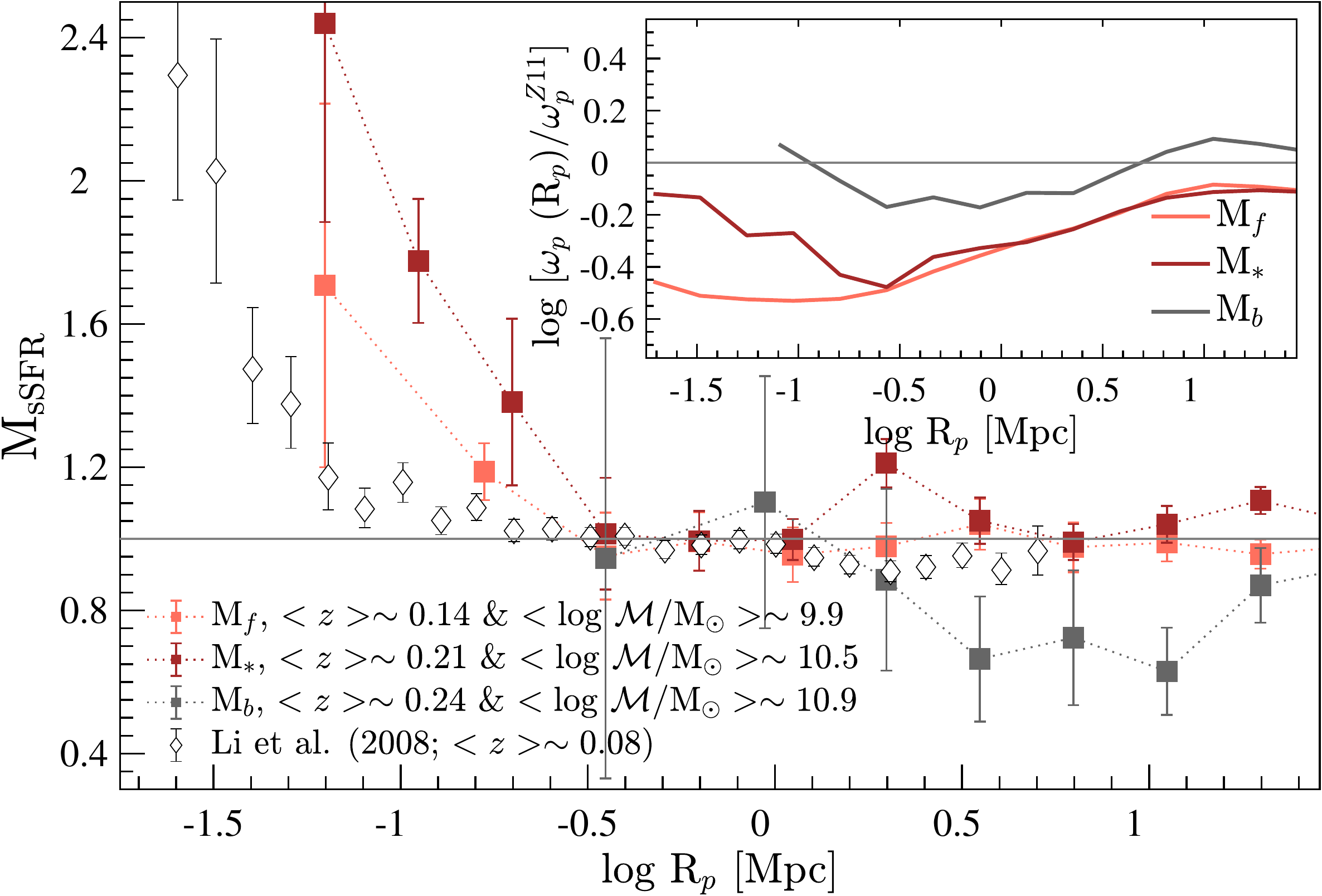}
\caption{GAMA versus SDSS sSFR MCFs. The GAMA  MCFs of M$_f$, M$_*$ and M$_b$ galaxies (filled symbols) in comparison to the sSFR enhancement functions of SDSS $-23\leq M_{r_{0.1}} \leq -17$ galaxies extended over the $0.01\leq z\leq0.3$ range, where M$_{r_{0.1}}$ is $r$-band absolute magnitudes $k$-corrected to $z=0.1$ \citep[][]{Li2008}. The mean redshift and stellar mass coverages of each galaxy sample are given in the legend. }
\label{fig:mark_lit_comp1}
\end{center}
\end{figure}

We show the GAMA  ACFs$_{\omega_p^{Z11}}$ (\S\,\ref{subsec:auto_corr}) in the inset of Figure\,\ref{fig:mark_lit_comp1} for comparison. On average, the R$_p$ at which the sSFR MCFs of the three luminosity selected SF complete samples show an enhancement in sSFR appear to coincide with the R$_p$ at which the respective ACFs$_{\omega_p^{Z11}}$ begin to show a change. This result is not unexpected as MCFs are in a sense ACFs with weights based on the galaxy properties applied. 
}}

\section{Summary}\label{sec:conc}

We have used a sample of galaxies with detected H$\alpha$ emission drawn from the GAMA survey to study the small scale clustering properties of star forming galaxies as a function of both optical luminosity and stellar mass. In the process, we provide a method with which the random clones of galaxies computed by \cite{Farrow2015} for the GAMA survey (i.e.\,computed for a galaxy survey with a univariate primary selection)  to be applied to a bivariately selected sample of galaxies (e.g.\,a star forming sample of galaxies drawn from a broadband survey). 
The auto, cross and mark two-point correlation techniques have been used in the computation of galaxy correlation functions for each luminosity and stellar mass selected sample, and below we summarise the main conclusions of this study. 

\begin{itemize}
\item {The strengthening of clustering on small scales observed in GAMA ACFs of star formers (\S\,\ref{subsec:auto_corr}) is a signature of galaxy-galaxy interactions. 
}

\item {With increasing optical brightness, both the increase in clustering amplitude of star forming population with decreasing R$_p$ at a given R$_p$ (log R$_p\lesssim-0.15$ [Mpc]) becomes progressively more significant, and the R$_p$ at which the clustering signal of the ACFs of star forming galaxies relative to the fiducial power law show a turn-over becomes progressively larger.  This behaviour of star forming galaxies can be interpreted as evidence of an existence of an interaction scale, where physical changes induced in an interaction are, or rather ought to be, evident out to the R$_p$ at which the clustering signal of a given star forming population relative to the fiducial power law starts to alter. This is in the sense that the interactions between optically bright galaxy pairs induce changes that are evident out to larger separations than those between optically faint galaxies. 
}

\item {The main advantage of utilising auto correlation techniques to map interaction scales is that they are much less susceptible to (1) fluctuations (i.e.\,enhancements, and decrements) in measured properties, (2) the observability of a change (i.e.\,the change in a physical property can be too subtle to be observable over some scales), and (3) to the differences arising from the type of star formation indicator used (e.g.\,short versus long-duration star formation indicators) than methods that employ net changes in properties to trace interactions.        
}

\item {Out of the different potential signatures of interactions (e.g.\,sSFR, SFR, optical colour, D$_{\rm{4000}}$ and Balmer decrement) considered in this study, the clustering with respect to ($30\%$) sSFR and SFR, both based on H$\alpha$ emission, on average show the strongest small scale enhancements across all magnitude and stellar mass ranges considered. Likewise, the $30\%$ lowest sSFR and SFR galaxies show a decrement in clustering across all magnitude and stellar mass ranges. 
}

\item{The optical colours, i.e.\,($g-r$)$_{\rm{rest}}$, can be affected by the dust obscuration in galaxies. The spectroscopically based D$_{\rm{4000}}$ indices, a proxy for colour that less affected by dust and in single-fibre spectroscopy represents the changes in central regions of galaxies, can provide a clearer picture of the effects of interactions than optical colours.        
}

\item {The comparison between rank ordered sSFR and ($g-r$)$_{\rm{rest}}$ MCFs show that the small scale enhancement in sSFR is stronger than that of ($g-r$)$_{\rm{rest}}$, supporting the aforementioned conclusion that sSFR is a better tracer of interactions between star forming galaxies than other tracers considered.  
}

\item {The sSFR MCFs show an increase in small scale clustering, and the amplitude at a given R$_p$ of the MCF of optically bright (e.g.\,M$_*$) sSFR galaxies greater than that of optically faint (e.g.\, M$_{\rm{f}}$) sSFR systems. This suggests that optically brighter star forming systems are characterised by higher SFR than fainter objects. Based on the comparison of dust properties of different star forming populations, it is clear that optically bright high SFR systems contain higher dust contents than their fainter counterparts.    
}

\end{itemize}

{Highly complete datasets with large redshift coverage that will be provided by the future/planned galaxy surveys will allow further insights into the relationship between interaction scale and optical brightness, and into underlying physical processes (galaxy and/or cluster-scale) that are responsible for it. Moreover, these datasets will allow any evolution in interaction scales to be tightly constrained, thereby shedding light on the evolution of physical properties and processes of galaxies across time.}

\section*{Acknowledgments}

We thank Ignacio Ferreras, Jarle Brinchmann and Nelson Padilla for valuable discussions. We also thank the anonymous referee for their careful reading of the manuscript and helpful comments which improved the presentation of the paper. 

M.L.P.G.\,acknowledges the support from a CONICYT-Chile grant FONDECYT 3160492, and has also received funding from the European Union's Horizon 2020 research and innovation programme under the Marie Sklodowska-Curie grant agreement No 707693. M.L.P.G.,\, P.N.\,and I.Z.\,acknowledge support from a European Research Council Starting Grant (DEGAS-259586). P.N.\,acknowledges the support of the Royal Society through the award of a University Research Fellowship and the support of the Science and Technology Facilities Council (ST/L00075X/1). I.Z.\,is supported by NSF grant AST-1612085. 

GAMA is a joint European-Australasian project based around a spectroscopic campaign using the Anglo-Australian Telescope. The GAMA input catalogue is based on data taken from the Sloan Digital Sky Survey and the UKIRT Infrared Deep Sky Survey. Complementary imaging of the GAMA regions is being obtained by a number of independent survey programs including GALEX MIS, VST KIDS, VISTA VIKING, WISE, Herschel-ATLAS, GMRT and ASKAP providing UV to radio coverage. GAMA is funded by the STFC (UK), the ARC (Australia), the AAO, and the participating institutions. The GAMA website is http://www.gama-survey.org/. 

Data used in this paper will be available through the GAMA website (\url{http://www.gama-survey.org/}) once the associated redshifts are publicly released. 

This work used the DiRAC Data Centric system at Durham University, operated by the Institute for Computational Cosmology on behalf of the STFC DiRAC HPC Facility (www.dirac.ac.uk. This equipment was funded by a BIS National E-infrastructure capital grant ST/K00042X/1, STFC capital grant ST/K00087X/1, DiRAC Operations grant ST/K003267/1 and Durham University. DiRAC is part of the National E-Infrastructure.

\footnotesize
{
\bibliographystyle{apsrmp}
\bibliography{references}
}
\bsp

\appendix
\section{On the modelling of the survey selection function and on the impact of sample systematics} \label{app:modelling}

\subsection{Modelling of the survey selection function in redshift bins} \label{app:randoms_weights}

\begin{figure}
\begin{center}
\includegraphics[width=0.5\textwidth]{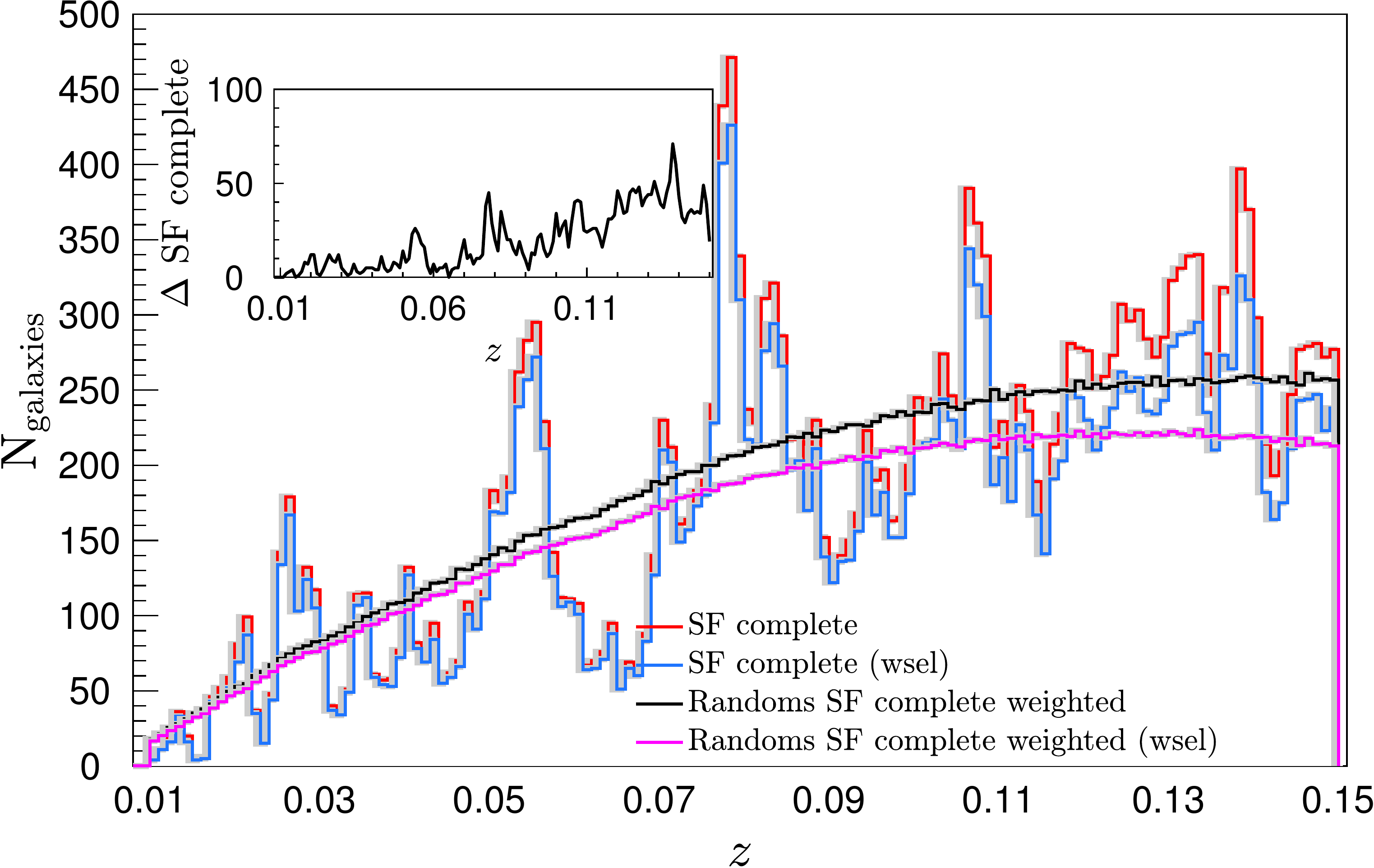}
\includegraphics[width=0.5\textwidth]{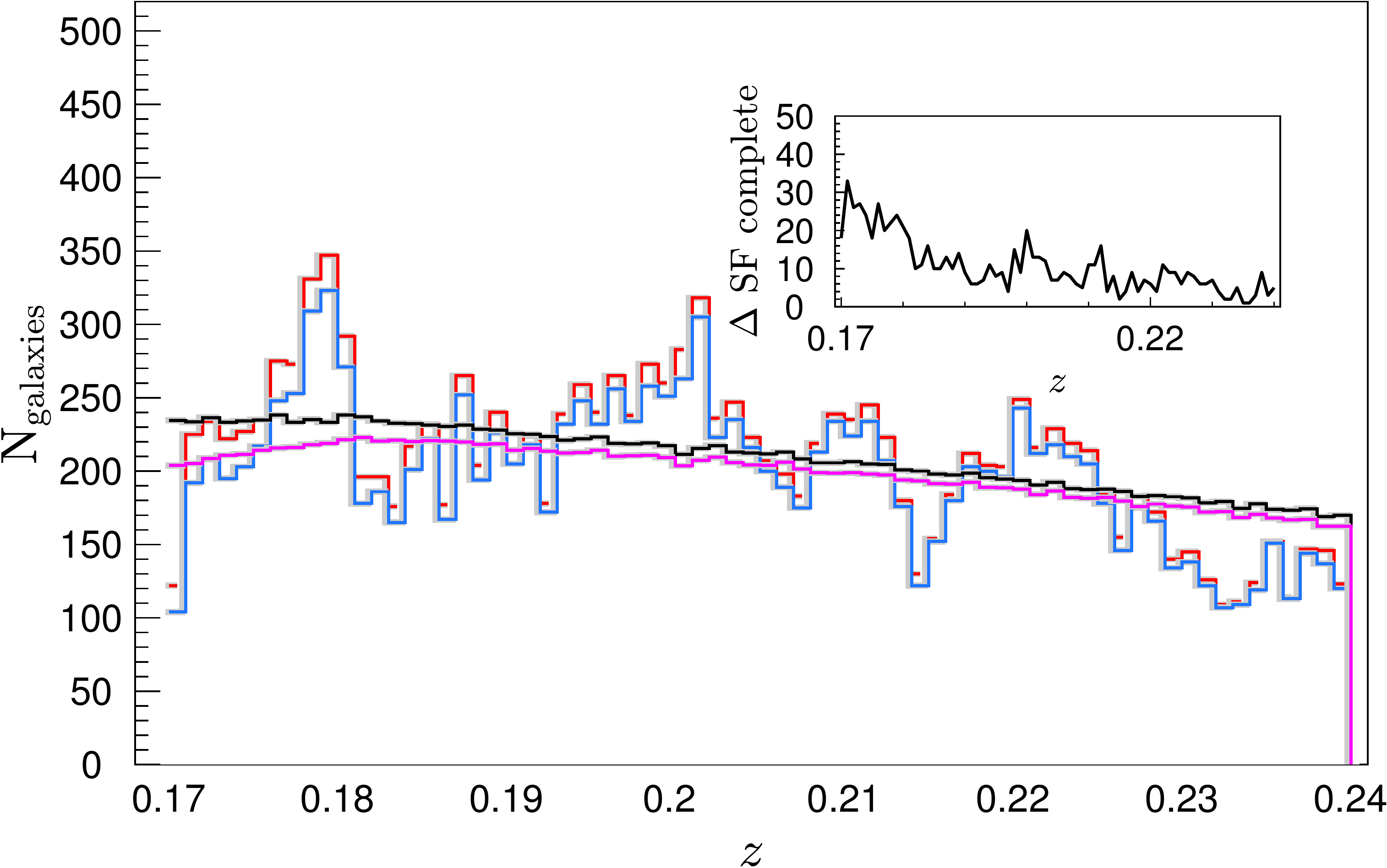}
\includegraphics[width=0.5\textwidth]{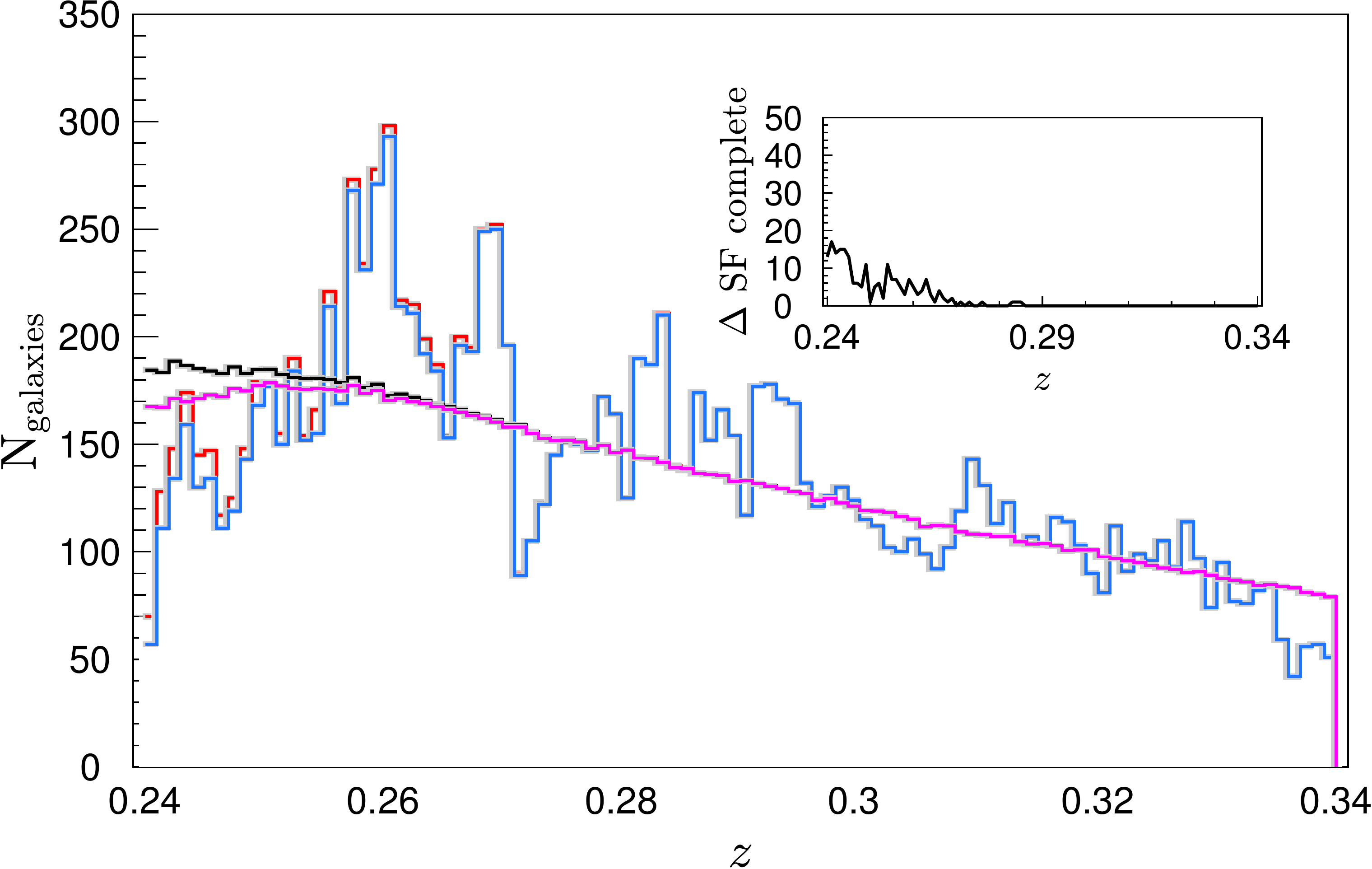}
\caption{The redshift distributions of SF complete galaxies in comparison to the weighted and weight-selected, where galaxies with $N_{\rm{weights}}>10$ are removed from the sample, distributions of the equivalent random galaxies. The randoms are weighted by $N_{\rm{weights}}$ computed in redshift slices, and each inset show the difference between SF complete and SF complete (wsel).}
\label{fig:redshift_distributions_all_appendix}
\end{center}
\end{figure}

We describe the modelling of the selection function used for the analysis presented in the main paper in \S\,\ref{subsec:modelling_randoms}, which is based on Eq.\,\ref{eq:weights_def} with V$_{\rm{zlim}}$ set either to the redshift detection limit of $0.34$ of the H$\alpha$ spectral feature in GAMA spectra or to $z\sim0.24$. Alternatively, the $N_{\rm{weights}}$ can also be computed in redshift slices such that V$_{\rm{zlim}}$ defines the volume of a given redshift slice. Figure\,\ref{fig:redshift_distributions_all_appendix} shows the distributions $N_{\rm{galaxies}}$ in redshift bins, where $N_{\rm{weights}}$ computed in redshift slices are used to weight the random galaxies. Note that the redshift ranges are defined such that none include the redshift band centred around $z\sim0.16$.   
\begin{figure*}
\begin{center}
\includegraphics[width=0.35\textwidth]{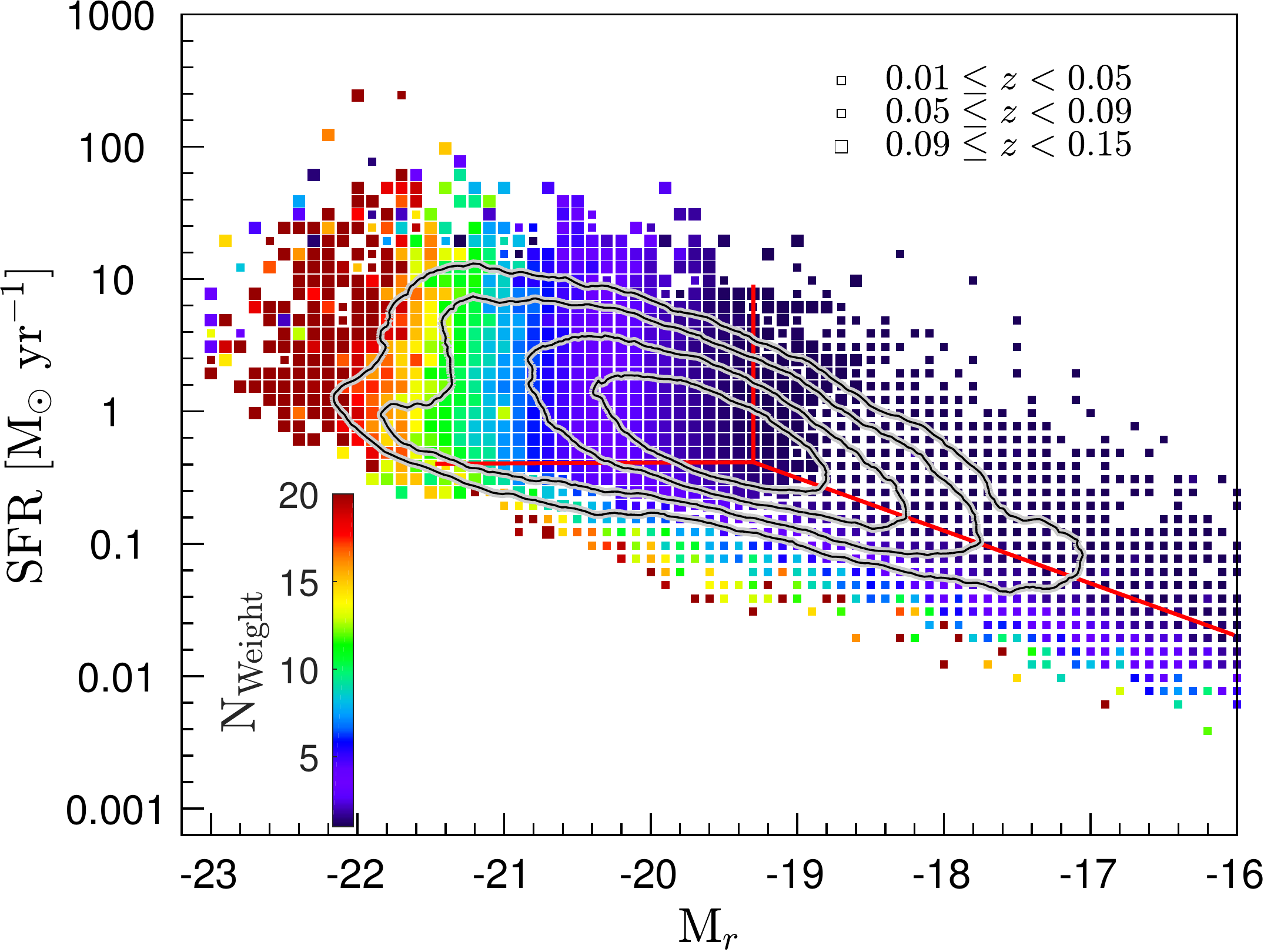}
\includegraphics[width=0.32\textwidth]{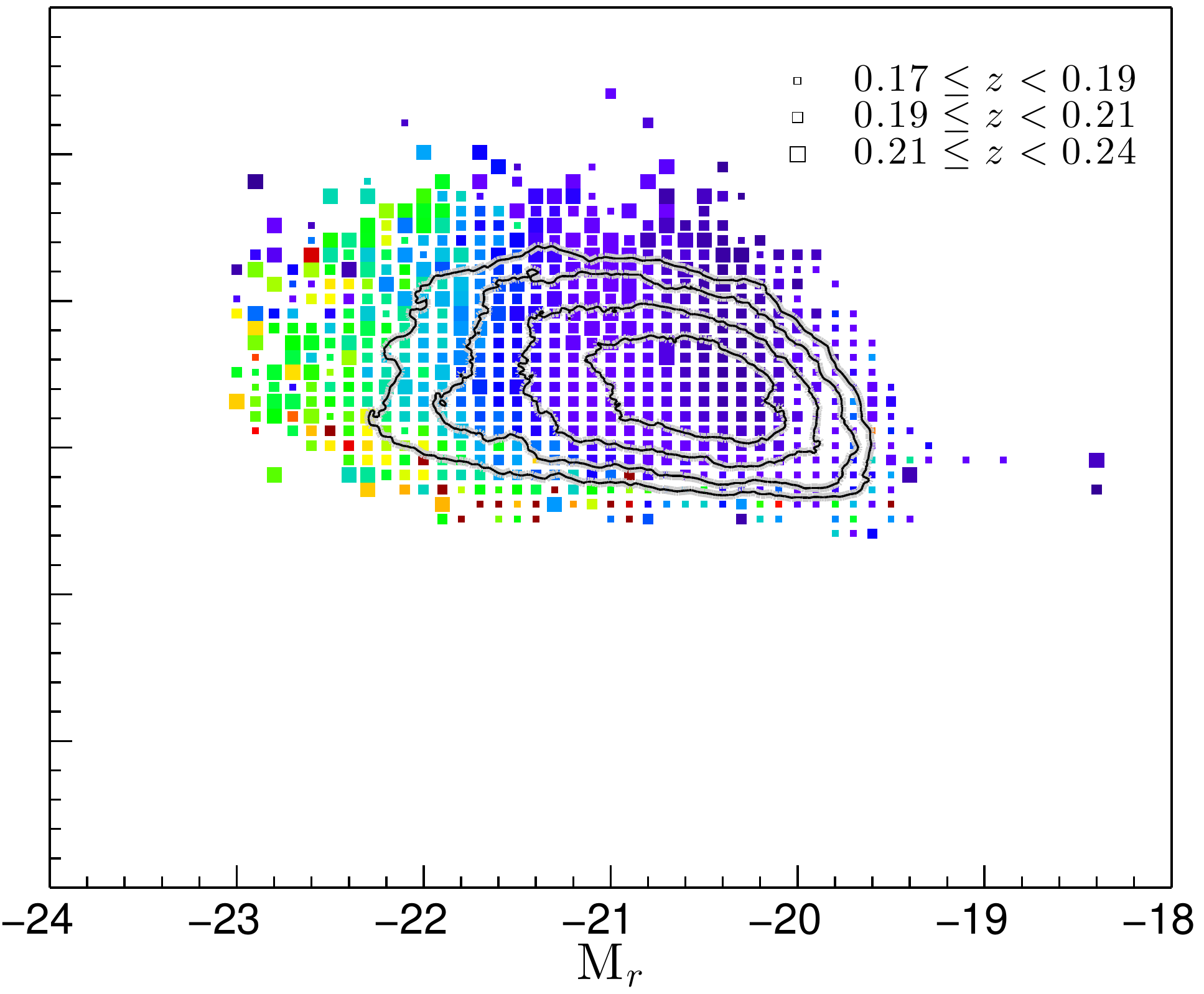}
\includegraphics[width=0.32\textwidth]{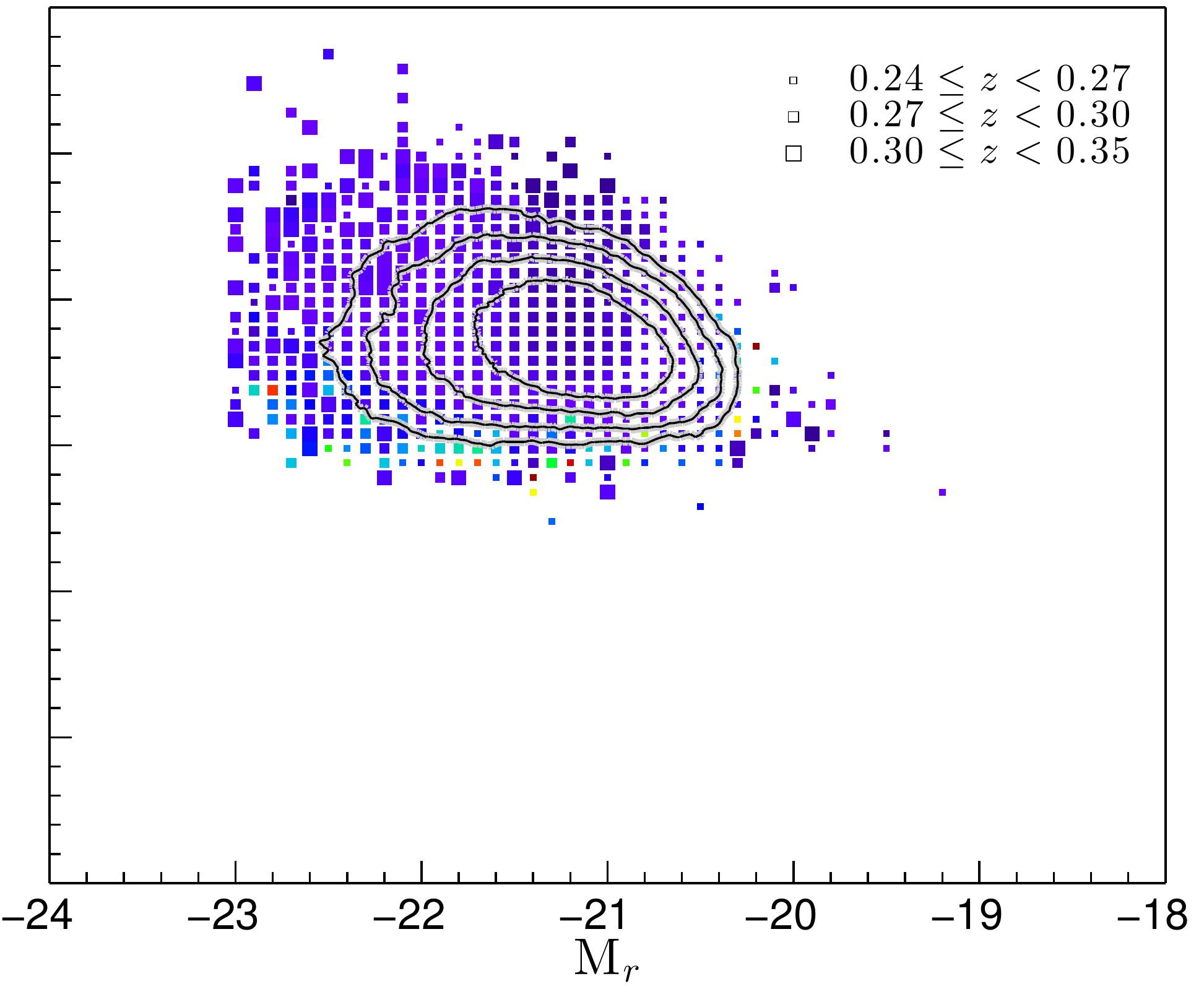}
\caption{The weights applied to the galaxies in random SF complete as a function of SFR and M$_r$ in $0.01\lesssim z<0.15$, $0.17<z<0.24$ and $0.24\lesssim z<0.35$ ranges. The weights shown are estimated using Eq.\,\ref{eq:weights_def}, where V$_{\rm{zlim}}$ is the volume of a given redshift slice. From left-to-right, the redshift slices are $0.01\leq z\leq0.15$, $0.17<z\leq0.24$ and $0.24<z\leq0.34$. The marker size is indicative of $<z>$ of SF complete galaxies with a given SFR and M$_r$ measures. The closed contours (from inwards-to-outwards) enclose 25, 50, 75 and 90$\%$ of the data, respectively. The red solid lines shown in the left panel approximately indicate the different regions in the SFR and M$_r$ plane where V$_{\rm{max}}$ of a given galaxy is mostly limited by its measured H$\alpha$ flux (lower region), or by its $r-$band magnitude (upper region) or by zlim (left most region). }
\label{fig:randoms_weights_appendix}
\end{center}
\end{figure*}
\begin{figure*}
\begin{center}
\includegraphics[width=0.48\textwidth]{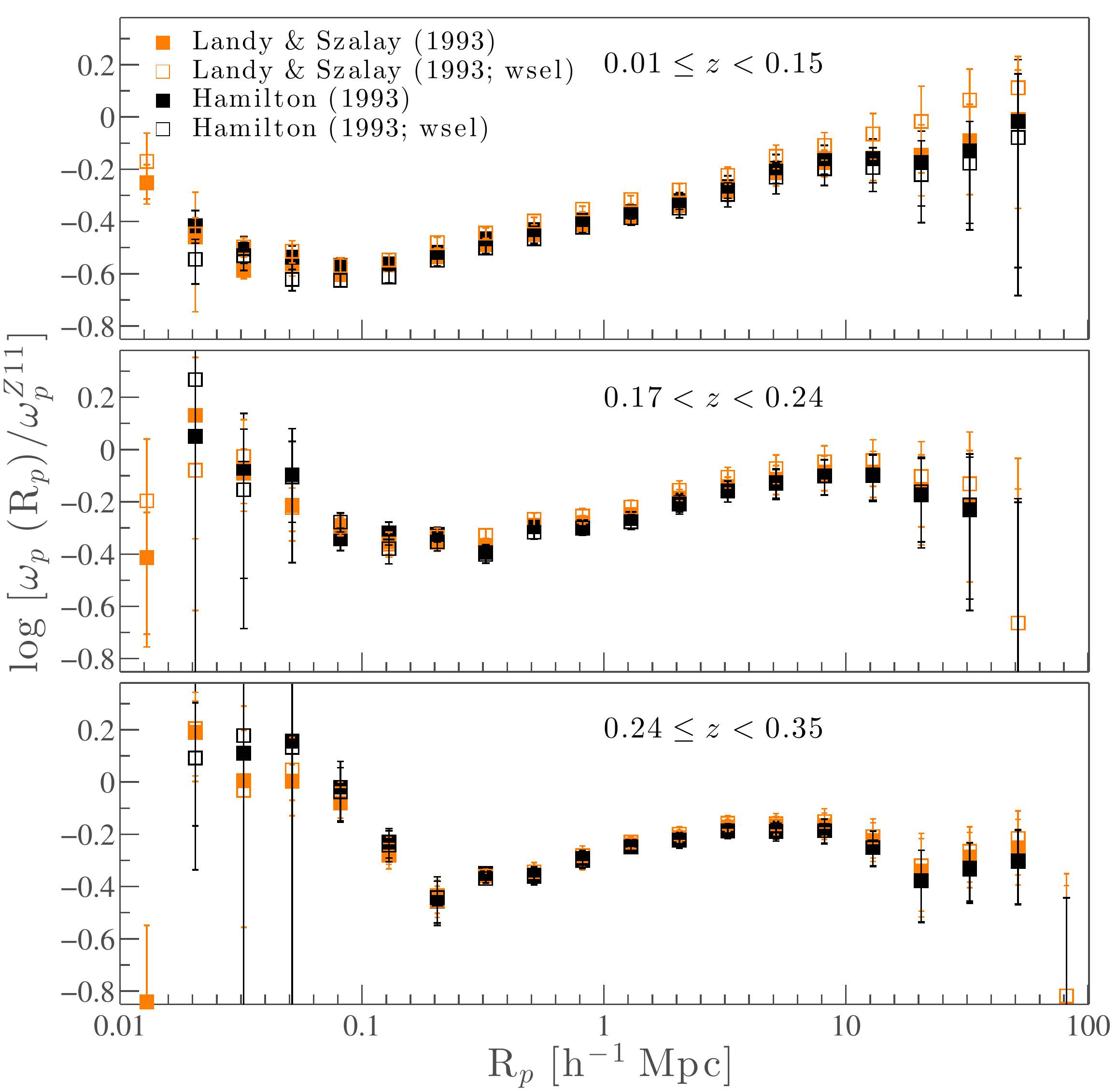}
\includegraphics[width=0.48\textwidth]{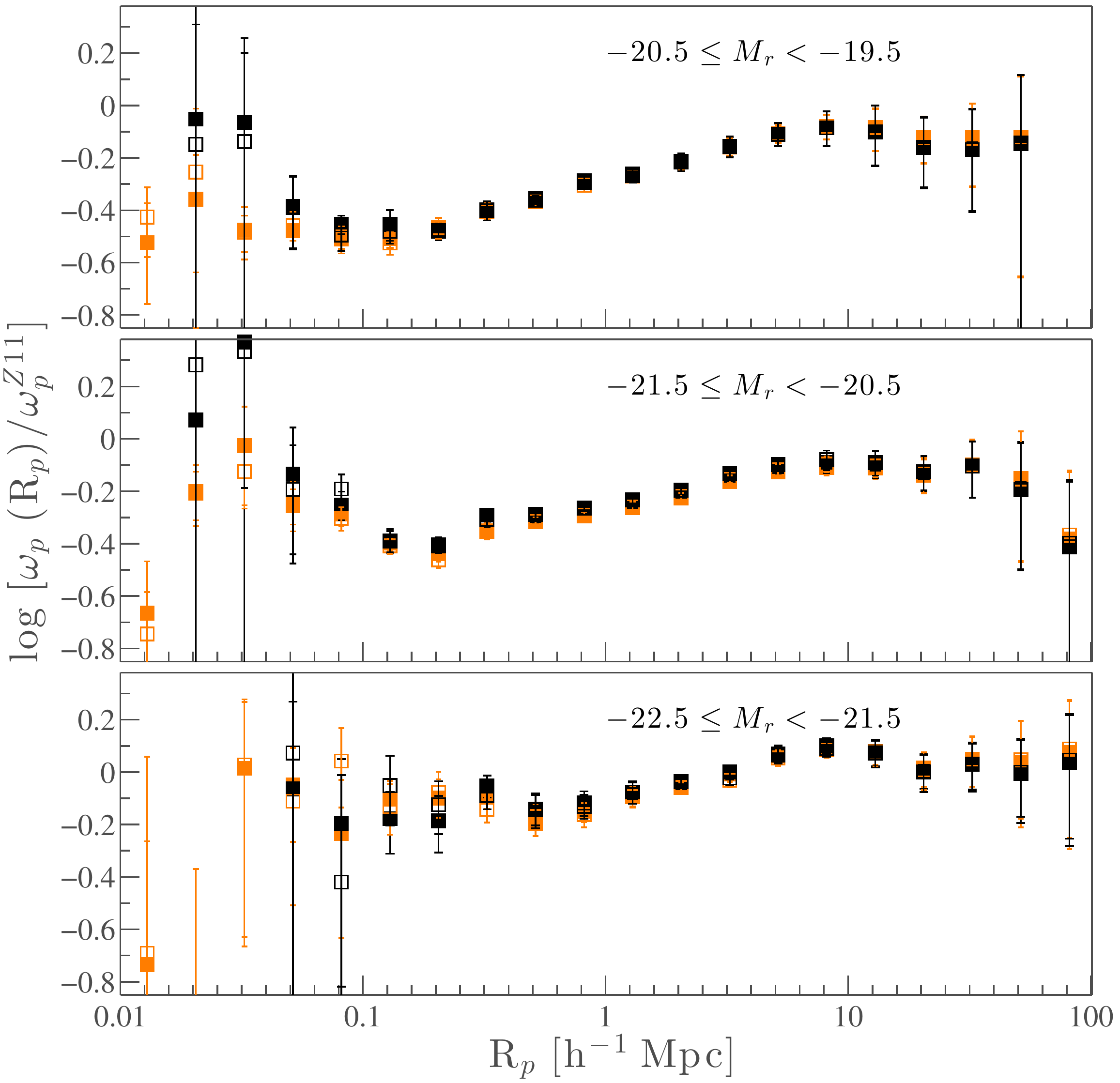}
\caption{\color{black}A comparison of ACFs computed from the \citet{LS1993} and \citet{Hamilton1993} estimators (orange and black symbols, respectively) with and without N$_{\rm{weight}}$ selections (open and filled, respectively). Left panels show the ACFs computed using random galaxies weighted as described in \S\,\ref{app:modelling}, and the right panels show the ACFs computed as described above}.
\label{fig:Auto_comparisons}
\end{center}
\end{figure*}
\begin{figure*}
\begin{center}
\includegraphics[width=0.98\textwidth]{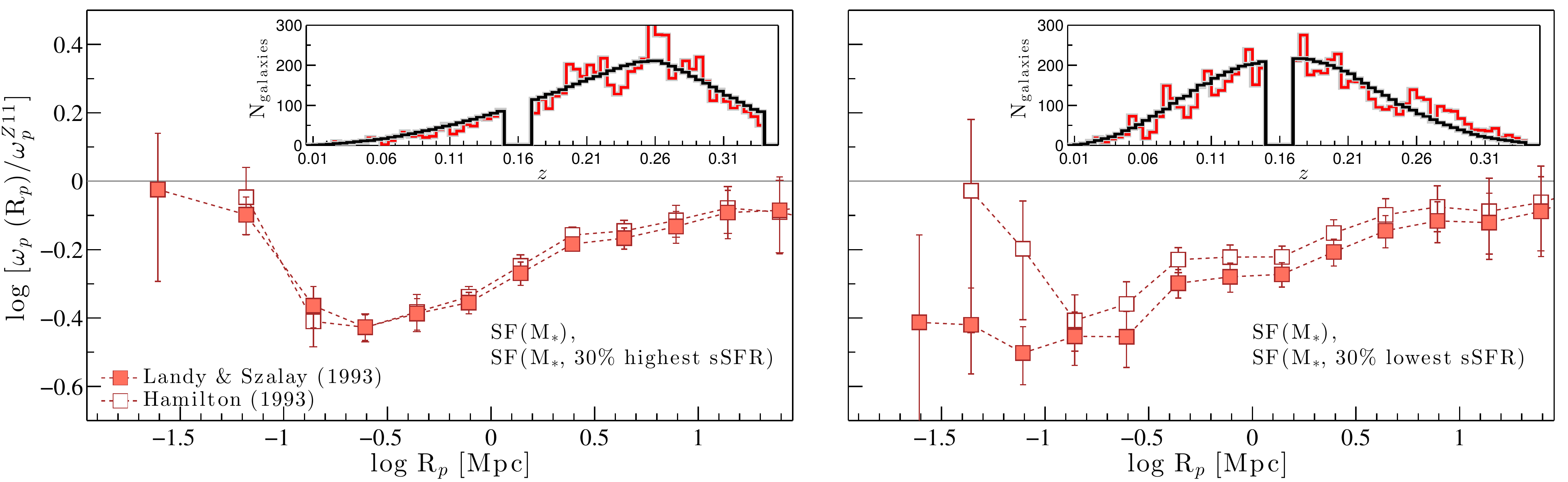}
\caption{A comparison between \citet[][filled symbols]{LS1993} and \citet[][open symbols]{Hamilton1993} cross correlation estimators. Left panel: The projected CCFs of M$_*$ SF galaxies with respect to $\omega_p^{Z11}$ (i.e.\,the reference function introduced in Figure\,\ref{fig:auto_corr_proj_main}), cross correlating all M$_*$ SF galaxies with the $30\%$ highest sSFR galaxies of M$_*$ SF sub-sample. Left panel: The projected cross CF corresponding to the cross correlation between all M$_*$ SF galaxies with the $30\%$ lowest sSFR galaxies of M$_*$ SF sub-sample. All galaxy samples used for this figure cover the full redshift range (i.e.\,$z<0.35$) over which Balmer H$\alpha$ feature is visible. As such the weights for the random clones are calculated assuming the maximum redshift of $z\approx0.35$. The insets show the redshift distributions of the highest (left) and lowest (right) sSFR galaxies of M$_*$ SF sub-sample (red line), and their respective random clones (black line). While the redshift distribution of the random clones of the highest sSFR galaxies match the respective distribution of GAMA galaxies, there is a discrepancy between the two lowest sSFR distributions at higher redshift (i.e.\,$z\gtrsim0.24$), which in turn give rise to the systematic discrepancy evident between the cross CFs calculated from the \citet[][filled symbols]{LS1993} and \citet[][open symbols]{Hamilton1993} estimators.}
\label{fig:Cross_comparisons}
\end{center}
\end{figure*}
\begin{figure}
\begin{center}
\includegraphics[width=0.45\textwidth]{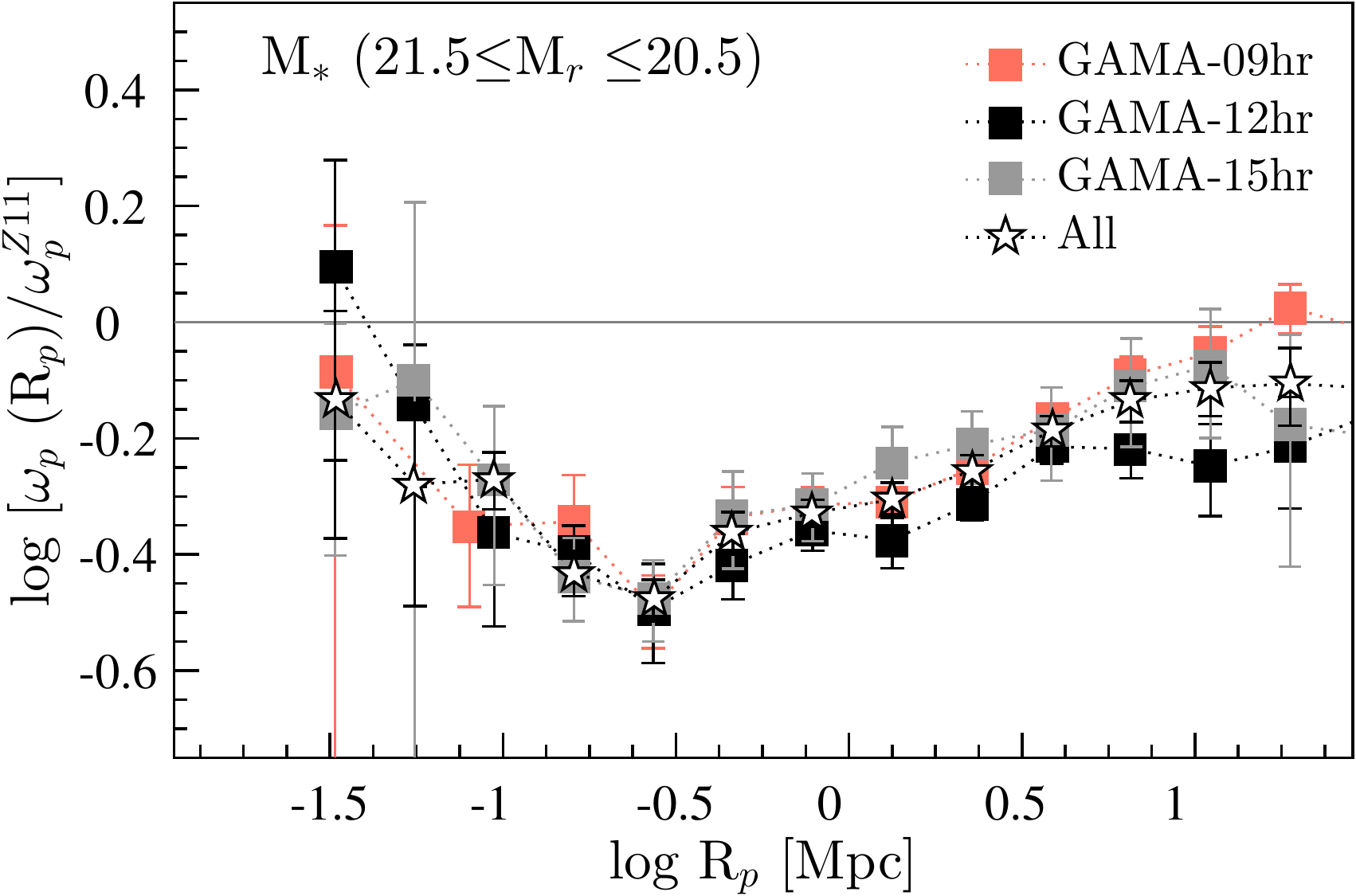}
\caption{The projected ACFs of M$_*$ galaxies relative to $\omega_p^{Z11}$. The squares denote the CFs corresponding to the three equatorial fields, and stars to all SF M$_*$ galaxies in our sample. }
\label{fig:2dfgrs}
\end{center}
\end{figure}
\begin{figure}
\begin{center}
\includegraphics[width=0.45\textwidth]{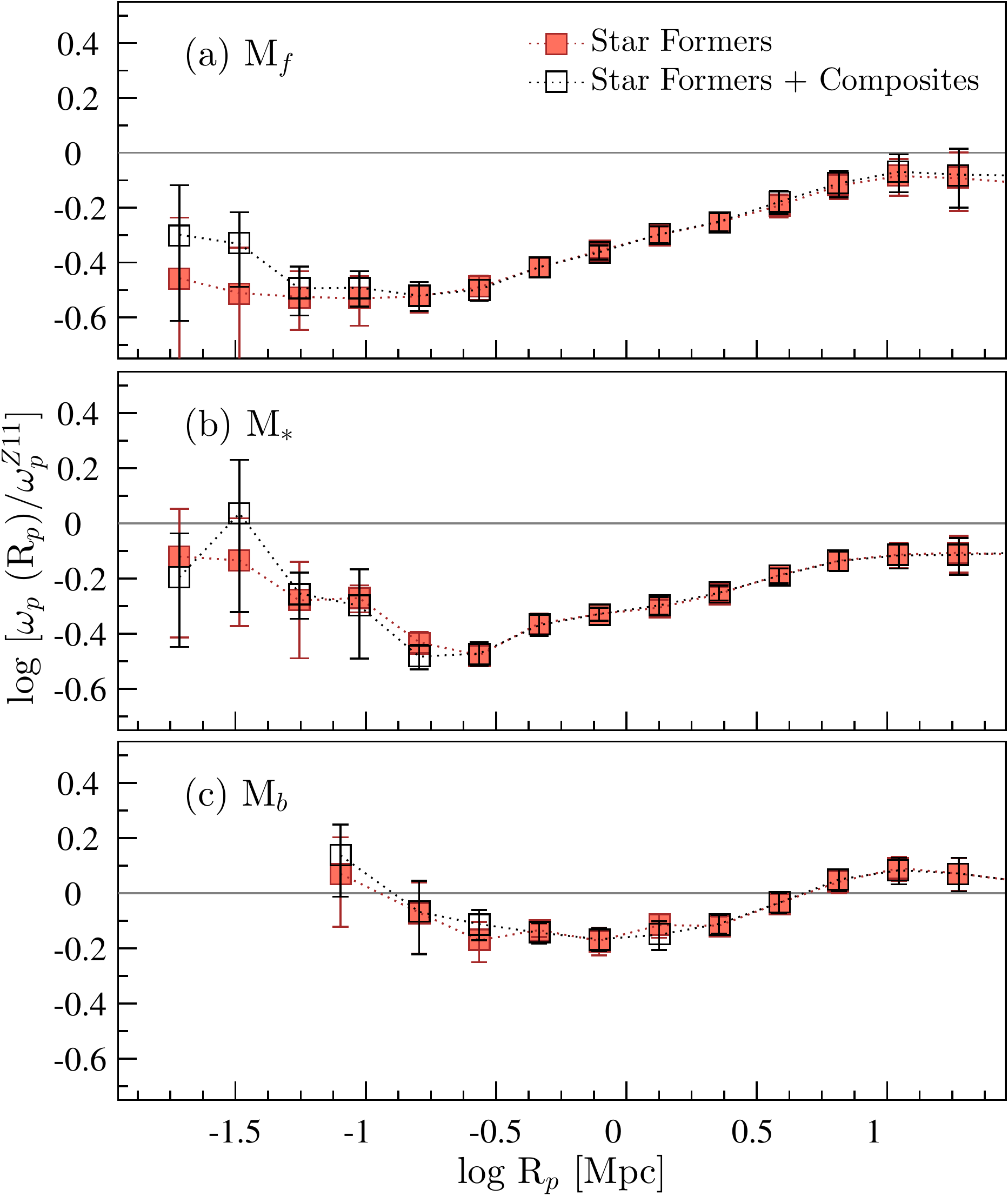}
\caption{The projected ACFs relative to $\omega_p^{Z11}$ of M$_f$, M$_*$ and M$_b$ pure star forming galaxies (filled squares) in comparison to pure star forming and composites (open squares). }
\label{fig:agns}
\end{center}
\end{figure}

Figure\,\ref{fig:randoms_weights_appendix} shows the mean distribution of $N_{\rm{weights}}$ with respect to both SFR and optical luminosity. As discussed in the main paper, the V$_{\rm{max}}$ of each galaxy is used in the computation of its $N_{\rm{weight}}$, and V$_{\rm{max}}$ can either be limited by the galaxy's SFR (i.e.\,H$\alpha$ flux), or by its $r$-band magnitude or by the upper limit of the relevant redshift slice. The solid red lines show approximate regions in SFR and M$_r$ plane where a galaxy with a given $N_{\rm{weight}}$ lie if the V$_{\rm{max}}$ of that galaxy is limited by its SFR (lower regions), by $r$-band magnitude (upper region) or by the upper redshift limits of the relevant redshift bin (left most region). The main caveat of the calculation of $N_{\rm{weights}}$ in redshift slices is that due to the relatively narrow range in redshift sampled a large fraction of galaxies are assigned $N_{\rm{weights}}>10$. 

Figure\,\ref{fig:Auto_comparisons} presents a comparison of the ACFs computed using $N_{\rm{weight}}$ estimated as described in \S\,\ref{subsec:modelling_randoms} (right panels) with that computed using $N_{\rm{weight}}$ estimated in redshift bins (left panels) as described above. In each panel, we compare the ACFs computed from the \cite[][open and filled orange squares]{LS1993} estimator with (open orange squares) and without (filled orange squares) with the respective ACFs obtained from the \cite[][open and filled black squares]{Hamilton1993} estimator,  
\begin{equation}
\xi_a(r_p, \pi)_{_{\rm{H}}} = \frac{DD(r_p, \pi)\times RR(r_p, \pi)}{DR(r_p, \pi)^2} - 1. 
\label{eq:auto_hamilton}
\end{equation}
The comparison shows that both clustering estimators as well as both methods of estimating $N_{\rm{weight}}$ yield similar auto correlation results. Given the outcome of this comparison and the caveats associated with estimating $N_{\rm{weight}}$ in redshift bins (see above), we choose to use a larger redshift range for the analysis presented in the main paper.  

Finally, we find that our method of modelling the selection function over the full redshift range over which the H$\alpha$ spectral feature is visible fails to model the low sSFR galaxy population over the $0.24<z<0.34$ range. This is demonstrated in Figure\,\ref{fig:Cross_comparisons}. While the \citet[][]{LS1993} and \citet[][]{Hamilton1993} clustering estimators produce consistent CF results for the high sSFR M$_*$ galaxies, the low sSFR M$_*$ CFs show a systematic offset. The high versus low sSFR galaxy redshift distributions shown in the insets highlight this issue; modelling the selection function over the full redshift range fails to model the redshift distribution of low sSFR M$_*$ galaxies over the $0.24<z<0.34$ range (i.e.\,random galaxy redshift distribution of low sSFR M$_*$ galaxies is under-predicted). However, limiting the redshift from 0.34 to 0.24 produces consistent results. Therefore for the cross correlation analysis presented in the main paper, we only use galaxies with redshifts in the $0.01\leq z\leq0.24$ range, and re-model the selection function to match this redshift range. 

\subsection{The sample selection \& systematics}\label{subsec:systematics}

The selection of the reference (i.e.\,REF) and SF complete samples is described in detail in \S\,\ref{subsec:selection}. Here we investigate how spectroscopic incompleteness of the star forming sample as well as our definition of star forming galaxies impact our results.

\subsubsection{The lack of 2dFGRS data}

One of the main issues discussed in \S\,\ref{subsec:selection}, in relation to the selection of the star forming galaxy sample is the incompleteness introduced by the exclusion of 2dFGRS data. Figure\,\ref{fig:completeness} demonstrates that our clustering sample is incomplete approximately between $17.7$ and $18.8$ in apparent $r$-band magnitude, and between $\sim1-0.3$ in ($g-r$)$_{\rm{app}}$. The fact that this incompleteness is not randomly distributed over the optical colour and apparent magnitude plane can be problematic for a clustering analysis.  

Figure\,\ref{fig:completeness} shows the completeness as a function of colour and magnitude for the three GAMA fields individually, as well as the total completeness. The overlap between the 2dFGRS and GAMA surveys is largest in GAMA-12hr field, followed by GAMA-15hr field. The GAMA-09hr field, on the other hand, lies completely outside of the sky regions surveyed by the 2dFGRS survey. Consequently, the spectroscopic incompleteness is significant in GAMA-12hr and relatively insignificant in GAMA-09hr. Therefore to investigate the impact of this incompleteness, we construct the ACFs of M$_*$ SF complete galaxies in GAMA-09hr, GAMA-12hr, and GAMA-15hr (Figure\,\ref{fig:2dfgrs}). Also shown is the ACF of all M$_*$ SF complete galaxies. As expected, the ACFs$_{\omega_p^{Z11}}$ of M$_*$ SF complete galaxies in GAMA-09hr and GAMA-12hr show the largest differences. Despite these differences, however, the two ACFs are in agreement with the ACF of all M$_*$ SF complete galaxies to within uncertainties. Note that the differences between the ACFs of individual GAMA fields are not only a result of the differences in spectroscopic completeness between the fields but also reflect sample variance. 

Additionally, we have also quantified the impact of excluding the 2dFGRS data on the correlation results by modifying the GAMA redshift completeness to account for the missing 2dFGRS galaxies. The comparison between the correlation functions computed using this modified GAMA redshift completeness mask with that computed using the standard redshift completeness mask (i.e.\,GAMA main sample of galaxies described in \S\,\ref{subsec:selection}) shows that the differences are minimal, and are within the measurement uncertainties. 

\subsubsection{The AGN selections}

For the analysis presented in the main paper, we selected the star forming galaxies based on the prescription of \cite{Kauffmann2003}. The \cite{Kewley2002} prescription is another popular SF/AGN discriminator widely used in the literature. Generally, the \cite{Kauffmann2003} prescription is used to select `purely' star forming galaxies, while that of \cite{Kewley2002} discriminates between galaxies with line emission likely significantly contaminated by the emission from AGNs, and galaxies with line emission likely mostly dominated by massive star formation. The latter class can include objects with some contamination from AGNs (i.e.\,composites).

In order to understand the impact of the inclusion of composites, we compare the ACFs$_{\omega_p^{Z11}}$ of star forming M$_f$, M$_*$ and M$_b$ galaxies selected using the prescription of \cite[][, SF + Composite]{Kewley2002} with the respective corresponding to star forming galaxies selected using the prescription of \cite{Kauffmann2003}. The results of this comparison shown in Figure\,\ref{fig:agns} are qualitatively and quantitatively in agreement with each other. This implies that the composites are galaxies dominated by on-going massive star formation as the AGNs have been observed to have lower clustering amplitudes than star forming galaxies \citep{Li2006, Li2008b}.  

\section{Volume limited clustering analysis} \label{app:volume_limited_analysis}
\begin{table*}
\caption{The volume limited sample definitions corresponding to the three independent magnitude and stellar mass limited samples described in Tables\,\ref{table:stats1} and \ref{table:stats3}. We note that the SF complete volume limited samples are at least $95\%$ volume limited.}
\begin{minipage}{1\textwidth}
\begin{center}
\begin{tabular}{cccc}
\hline
subset  & subset definition & N$_{\rm{galaxies}}$ & $z$ coverage  \\
\hline
\hline
\multicolumn{4}{c}{{\vspace{0.1cm}\textbf{SF complete}} } \\
\multicolumn{4}{c}{{At least $95\%$ complete with respect to both $r$-band magnitude and H$\alpha$ flux selection} } \\
\hline  
M$_{b, v_1}$ & $-23.5\leq$ M$_r$<$-21.5$, \hspace{0.2cm} SFR/M$_{\odot}$yr$^{-1}$ $\geq0.28$  &   1\,491  & $0.01-0.15$  \\
M$_{*, v_1}$ & $-21.5\leq$ M$_r$<$-20.5$, \hspace{0.2cm} SFR/M$_{\odot}$yr$^{-1}$ $\geq0.25$  & 4\,188 & $0.01-0.15$  \\
M$_{f, v_1}$ & $-20.5\leq$ M$_r$<$-19.5$, \hspace{0.2cm} SFR/M$_{\odot}$yr$^{-1}$ $\geq0.33$  & 5\,298  & $0.01-0.14$  \\
 		    &  																	&            &   \\
M$_{b, v_2}$ & $-23.5\leq$ M$_r$<$-21.5$, \hspace{0.2cm} SFR/M$_{\odot}$yr$^{-1}$ $\geq0.90$  &   1\,514  & $0.17-0.23$  \\
M$_{*, v_2}$ & $-21.5\leq$ M$_r$<$-20.5$, \hspace{0.2cm} SFR/M$_{\odot}$yr$^{-1}$ $\geq1.00$  & 4\,914 & $0.17-0.23$  \\
\hline 
\hline
$\mathcal{M}_{\mathcal{H}, v}$ & $10.5\leq$ $\log \mathcal{M}$/M$_{\odot}$<$11.0$, \hspace{0.2cm} SFR/M$_{\odot}$yr$^{-1}$ $\geq0.29$   &   1\,991  & $0.01 - 0.15$  \\
$\mathcal{M}_{\mathcal{I}, v}$  & $10.0\leq$ $\log \mathcal{M}$/M$_{\odot}$<$10.5$, \hspace{0.2cm} SFR/M$_{\odot}$yr$^{-1}$ $\geq0.35$  & 4\,163 & $0.01 - 0.15$  \\
$\mathcal{M}_{\mathcal{L}, v}$  & $9.5\leq$ $\log \mathcal{M}$/M$_{\odot}$<$10.0$, \hspace{0.2cm} SFR/M$_{\odot}$yr$^{-1}$ $\geq0.38$  & 2\,906  & $0.01 - 0.126$  \\
\hline 
\multicolumn{4}{c}{{\vspace{0.1cm}\textbf{REF}} } \\
\multicolumn{4}{c}{{$100\%$ complete with respect to $r$-band magnitude selection of the survey} } \\
\hline
M$_{b, v_1}$ & $-23.5\leq$ M$_r$<$-21.5$  &   4\,064  & $0.01-0.15$  \\
M$_{*, v_1}$ & $-21.5\leq$ M$_r$<$-20.5$  & 10\,244 & $0.01-0.15$  \\
M$_{f, v_1}$ & $-20.5\leq$ M$_r$<$-19.5$  & 12\,751  & $0.01-0.14$  \\
 		    &  						   &            &   \\
M$_{b, v_2}$ & $-23.5\leq$ M$_r$<$-21.5$  &   6\,971  & $0.17-0.23$  \\
M$_{*, v_2}$ & $-21.5\leq$ M$_r$<$-20.5$  & 17\,737 & $0.17-0.23$  \\
\hline 
\hline
$\mathcal{M}_{\mathcal{H}, v}$ & $10.5\leq$ $\log \mathcal{M}$/M$_{\odot}$<$11.0$   &   7\,338  & $0.01 - 0.15$  \\
$\mathcal{M}_{\mathcal{I}, v}$  & $10.0\leq$ $\log \mathcal{M}$/M$_{\odot}$<$10.5$  & 11\,812 & $0.01 - 0.15$  \\
$\mathcal{M}_{\mathcal{L}, v}$  & $9.5\leq$ $\log \mathcal{M}$/M$_{\odot}$<$10.0$  & 8\,014  & $0.01 - 0.126$  \\
\hline 
\end{tabular}
\end{center}
\end{minipage}
\label{table:stats2}
\end{table*}%
\begin{figure*}
\begin{center}
\includegraphics[width=1\textwidth]{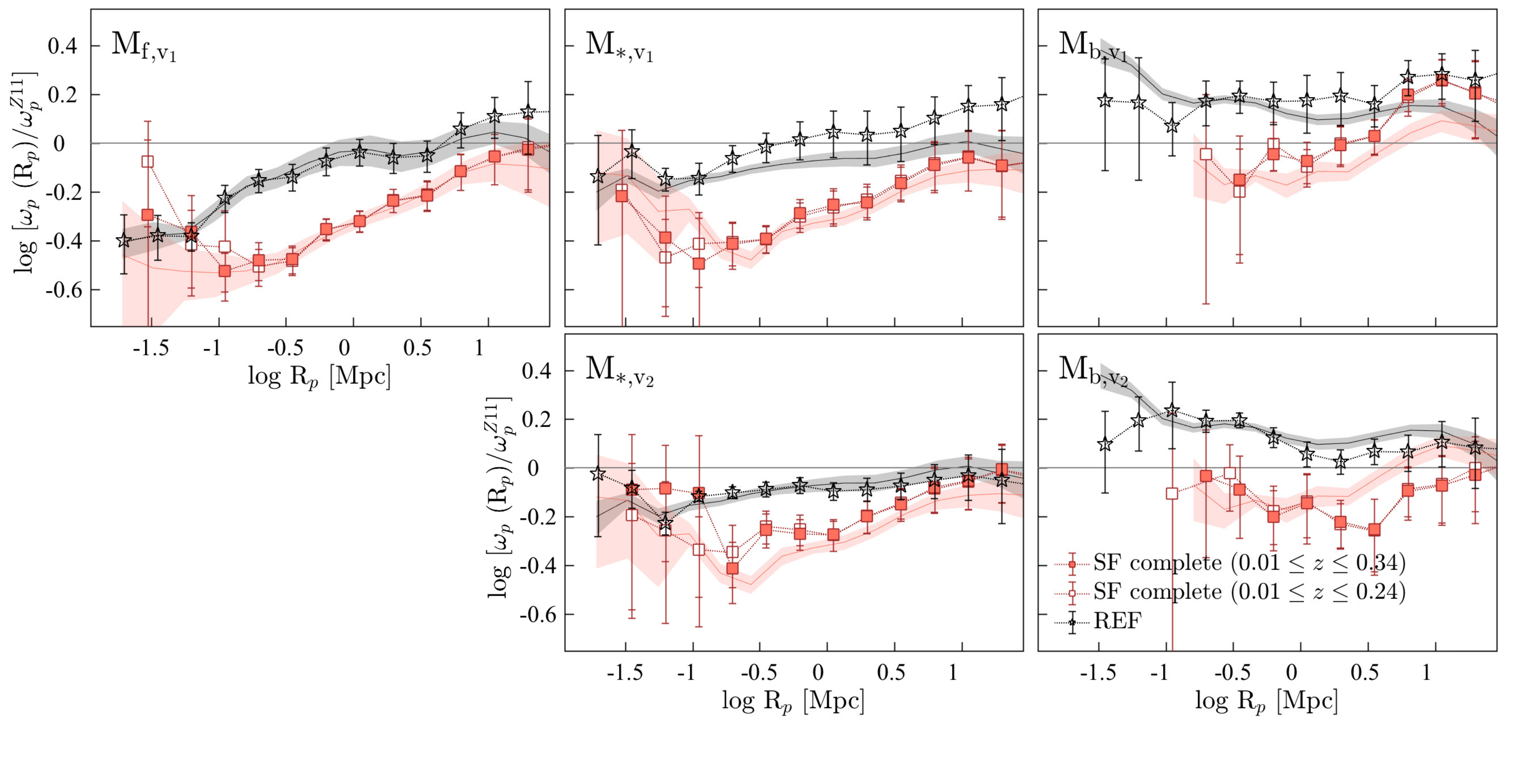}
\caption{The GAMA projected ACFs of luminosity selected volume limited samples (symbols) compared to the projected ACFs of luminosity selected samples (shaded regions, corresponding to the ACFs shown in Figure\,\ref{fig:auto_corr_proj_main}), all relative to $\omega_p^{Z11}$. The black symbols denote the REF ACFs of luminosity selected volume limited samples (Table\,\ref{table:stats2}), and the solid and open orange symbols denote the SF complete ACFs of luminosity selected volume limited samples, where N$_{\rm{weight}}$ (\S\,\ref{subsec:modelling_randoms}) is computed based on $0.001<z\leq0.34$ and $0.001<z\leq0.24$ galaxy samples, respectively. The black and dark orange shaded regions denote the REF and SF complete ACFs of luminosity selected samples (Table\,\ref{table:stats1}) presented in Figure\,\ref{fig:auto_corr_proj_main}. The ACFs of volume limited samples are in qualitative, and in most cases quantitative, agreement with the respective ACFs of luminosity selected samples.  } 
\label{fig:auto_corr_proj_vol}
\end{center}
\end{figure*}

As mentioned in \S\,\ref{subsec:clustering_samples}, we define several volume limited SF complete samples that are $\sim95\%$ complete with respect to the bivariate $r-$band magnitude and H$\alpha$ flux selections. In order to achieve this completeness without significantly limiting the redshift coverage of each volume limited SF sample, we impose a low-SFR cut, which excludes very low-SFR galaxies from the sample. The volume limited magnitude samples are, by definition, $95\%$ volume limited, however, the same cannot be said about the volume limited stellar mass samples. In order for the stellar mass samples to be $95\%$ volume limited, we need to consider the maximum volume out which a galaxy of a given stellar mass would be detectable, which has not been taken into account in this analysis. However, given the correlation between stellar mass and optical brightness, the volume limited stellar mass selected samples are likely close to $95\%$ volume limited. Furthermore, we also define several volume limited REF samples have the same redshift coverage as their SF counterpart, which are, therefore, $100\%$ complete with respect the $r-$band magnitude selection of the GAMA survey. Table\,\ref{table:stats2} presents the SF complete and REF volume limited samples used for the clustering analyses. 

We present and discuss the ACFs and CCFs constructed using magnitude (Table\,\ref{table:stats1}) and stellar mass selected (Table\,\ref{table:stats3}) non-volume limited SF complete samples in \S\,\ref{subsec:auto_corr} and \S\,\ref{subsec:cross_corr} of the main paper, respectively. In the subsequent sections, we present and discuss the respective ACFs and CCFs computed using the volume limited samples described in Table\,\ref{table:stats2}.

\subsection{Auto correlation functions of volume limited star forming and REF samples} \label{app:vanalysis_auto}

Figure\,\ref{fig:auto_corr_proj_vol} presents the ACFs$_{\omega_p^{z11}}$ of luminosity selected volume limited samples (Table\,\ref{table:stats2}), with the top (bottom) panels showing the results for the low (high) redshift volume samples. The same colour code as in Figure\,\ref{fig:auto_corr_proj_main} is used, and the shaded black and dark orange regions denote the ACFs of luminosity selected REF and SF complete samples presented in Figure\,\ref{fig:auto_corr_proj_main}. 

On small scales, all ACF$_{\omega_p^{z11}}$ of volume limited luminosity selected samples are in quantitative agreement with the respective luminosity selected functions. Compared to the ACFs$_{\omega_p^{z11}}$ of luminosity selected samples, however, the uncertainties associated with the clustering amplitudes of volume limited functions are relatively large, driven by the small number statistics of the volume samples. Given both the agreement between volume limited and non-volume limited ACF results, as well as the importance of sample statistics for studies, such as ours, that aim to investigate small scale clustering properties of star formers, we base the conclusions of this study on the analyses performed using luminosity and stellar mass selected samples.  

On large separations, however, the respective ACFs$_{\omega_p^{z11}}$ of volume limited and non-volume limited luminosity samples differ from each other. These disagreements can largely be attributed to the discrepancies between the redshift coverages of the respective volume limited and non-volume limited sample. The redshift coverage of M$_{\rm{f}}$ galaxies, for example, is similar to that of M$_{\rm{f,v1}}$, and consequently leads to a good agreement between the ACFs based on M$_{\rm{f}}$ and M$_{\rm{f,v1}}$ samples. The M$_*$ sample, on the other hand, encompasses both M$_{\rm{*, v1}}$ and M$_{\rm{*, v2}}$ galaxies. Therefore the ACF of M$_*$ galaxies can be thought of as the average of the ACFs of its respective volume samples. {\color{black} The same trends evident in the ACFs$_{\omega_p^{z11}}$ of volume limited luminosity selected samples are also evident in the ACFs$_{\omega_p^{z11}}$ of volume limited stellar mass selected samples presented in Appendix\,\ref{app:vanalysis_auto}}. 

\subsection{Cross correlation functions of volume limited star forming samples}\label{app:vanalysis}

\begin{figure*}
\begin{center}
\includegraphics[width=0.4\textwidth, trim={0 0.5cm 0 0},clip]{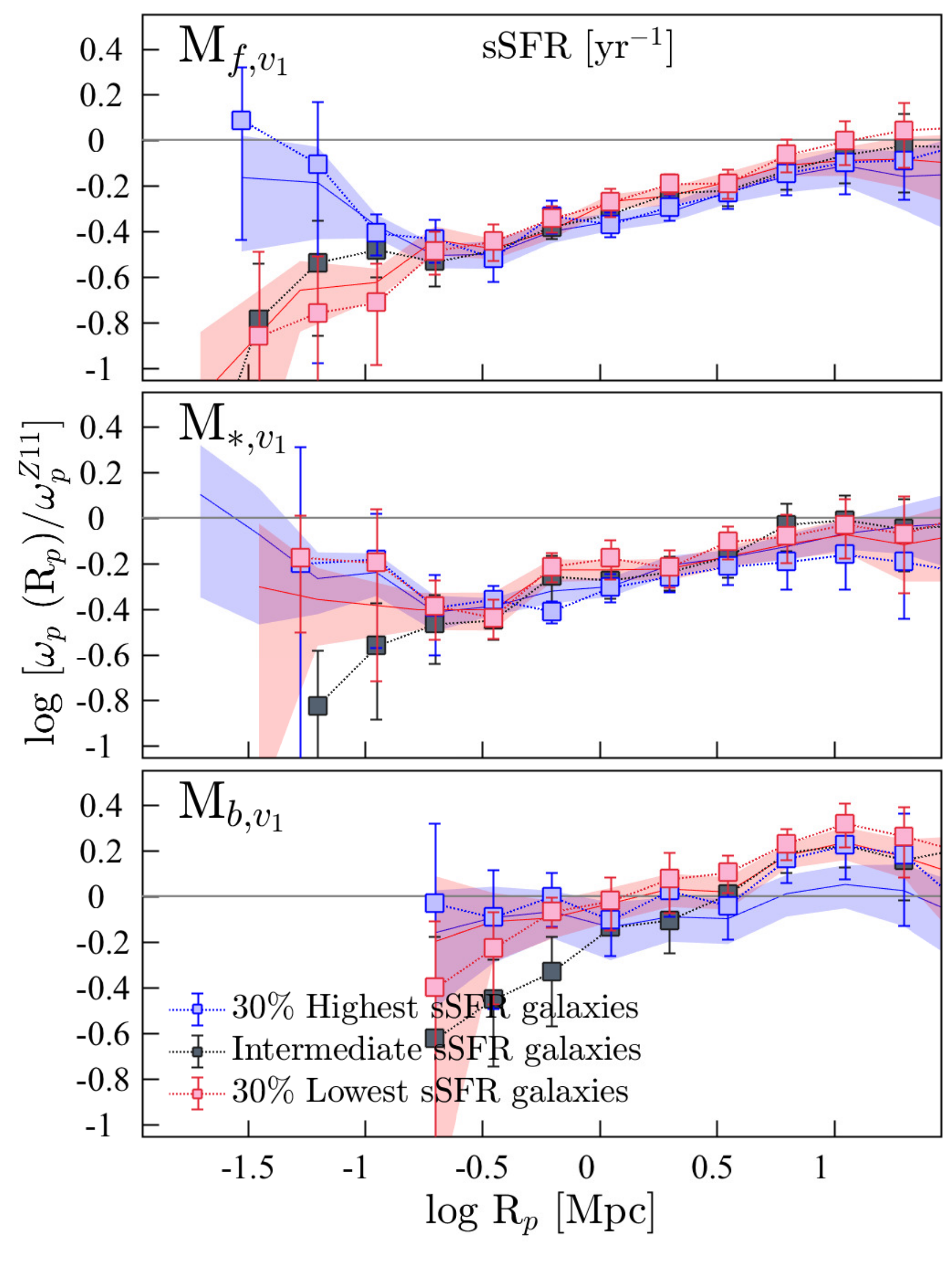}
\includegraphics[width=0.407\textwidth]{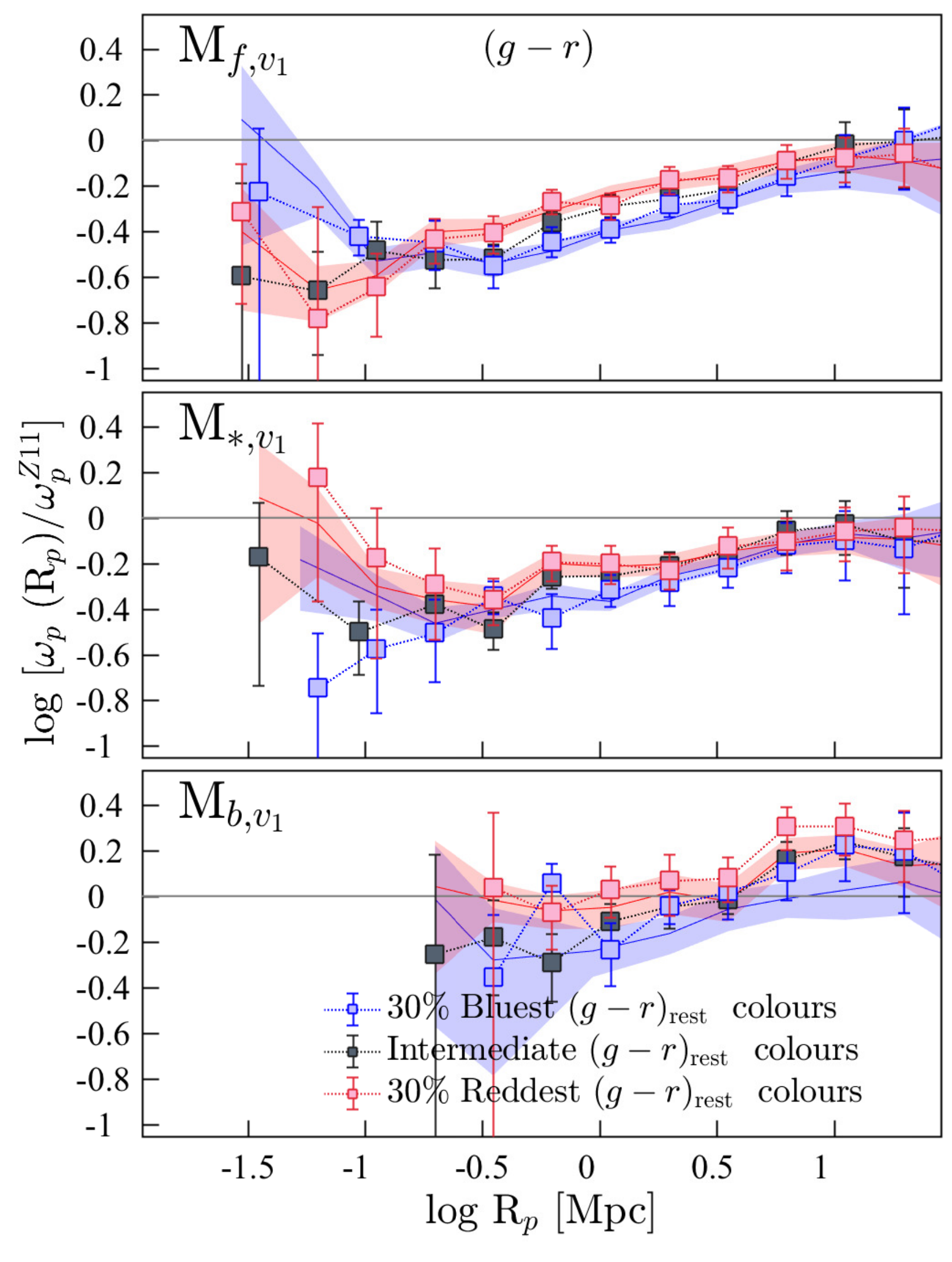}
\caption{The projected CCFs of volume limited luminosity selected SF complete samples (low redshift samples described in Table\,\ref{table:stats2}) relative to $\omega_p^{Z11}$ (luminosity increases down). Right panels: the CCFs of galaxies with optically blue (the $30\%$ bluest in ($g-r$)$_{\rm{rest}}$; blue squares), red (the $30\%$ reddest in ($g-r$)$_{\rm{rest}}$; red squares) and intermediate (the $40\%$ with intermediate ($g-r$)$_{\rm{rest}}$ measures; black squares) colours. Left panels: the CCFs of high (the $30\%$ highest in sSFR; blue squares), low (the $30\%$ lowest in sSFR; red squares) and intermediate (the $40\%$ with intermediate sSFRs; black squares) sSFR galaxies. The blue and red shaded regions show the respective CCFs of optically blue and red (right panels), and high and low sSFR (left panels) galaxies of magnitude selected SF complete samples described in Table\,\ref{table:stats1} of the main paper. }
\label{fig:xcrr_volume_mag_z1}
\end{center}
\end{figure*}

\begin{figure*}
\begin{center}
\includegraphics[width=0.407\textwidth]{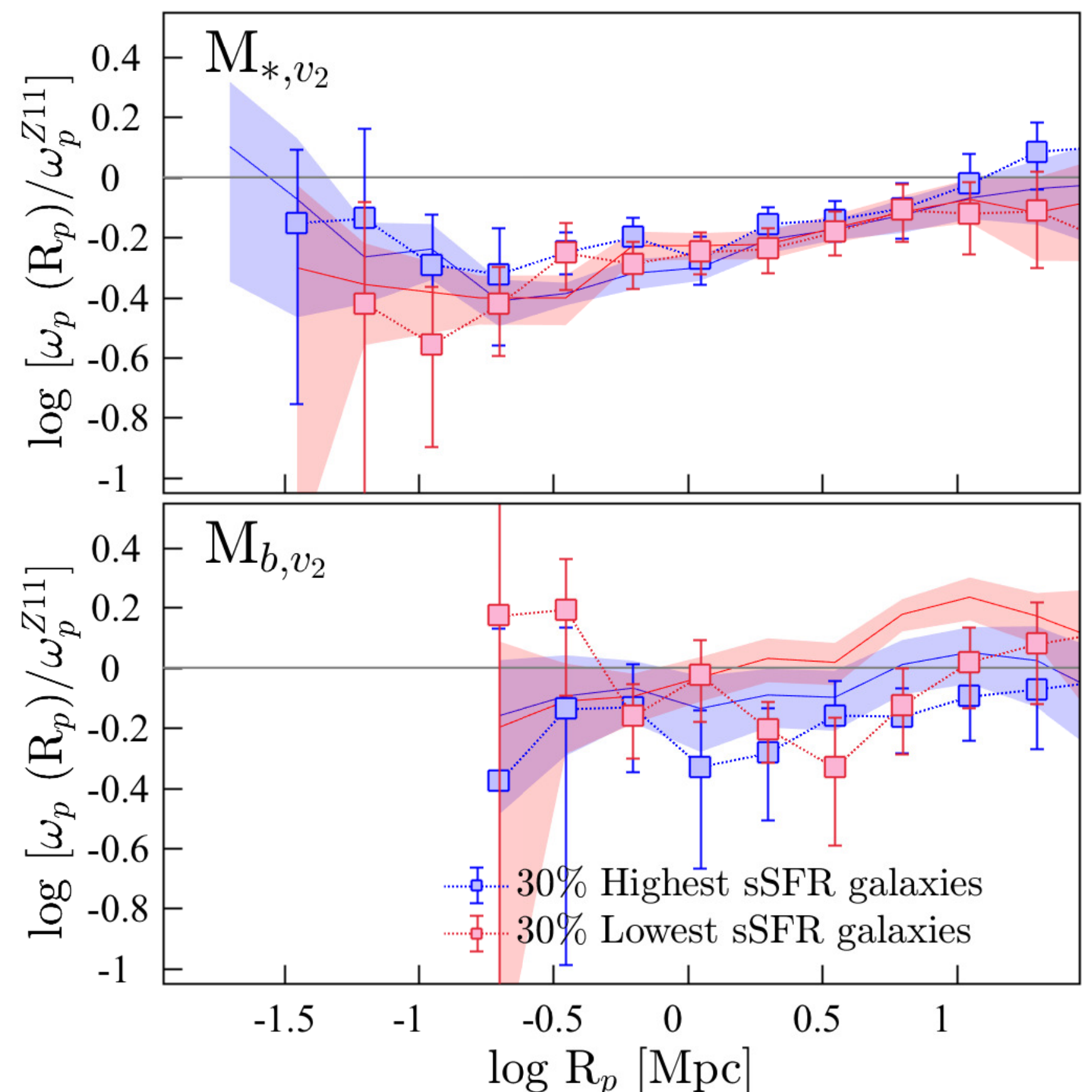}
\includegraphics[width=0.4\textwidth, trim={0 0.cm 0 0},clip]{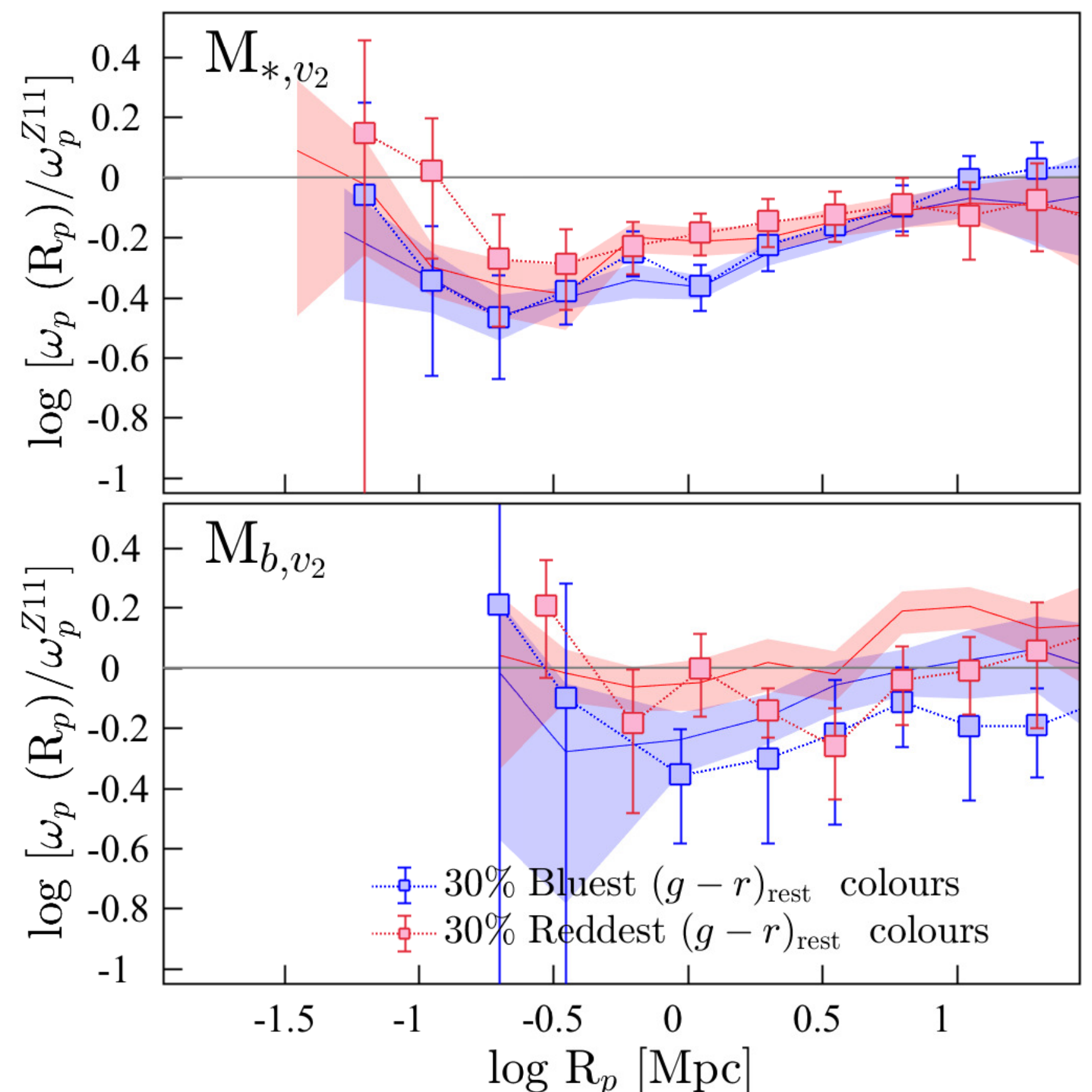}
\caption{The projected CCFs of volume limited luminosity selected SF complete samples (high redshift samples described in Table\,\ref{table:stats2}) relative to $\omega_p^{Z11}$ (luminosity increases down). Right panels: the CCFs of galaxies with optically blue (the $30\%$ bluest in ($g-r$)$_{\rm{rest}}$; blue squares) and red (the $30\%$ reddest in ($g-r$)$_{\rm{rest}}$; red squares) colours. Left panels: the CCFs of high (the $30\%$ highest in sSFR; blue squares) and low (the $30\%$ lowest in sSFR; red squares) sSFR galaxies. The blue and red shaded regions show the respective CCFs of optically blue and red (right panels), and high and low sSFR (left panels) galaxies of magnitude selected samples described in Table\,\ref{table:stats1} of the main paper. }
\label{fig:xcrr_volume_mag_z2}
\end{center}
\end{figure*}

\begin{figure*}
\begin{center}
\includegraphics[width=0.4\textwidth]{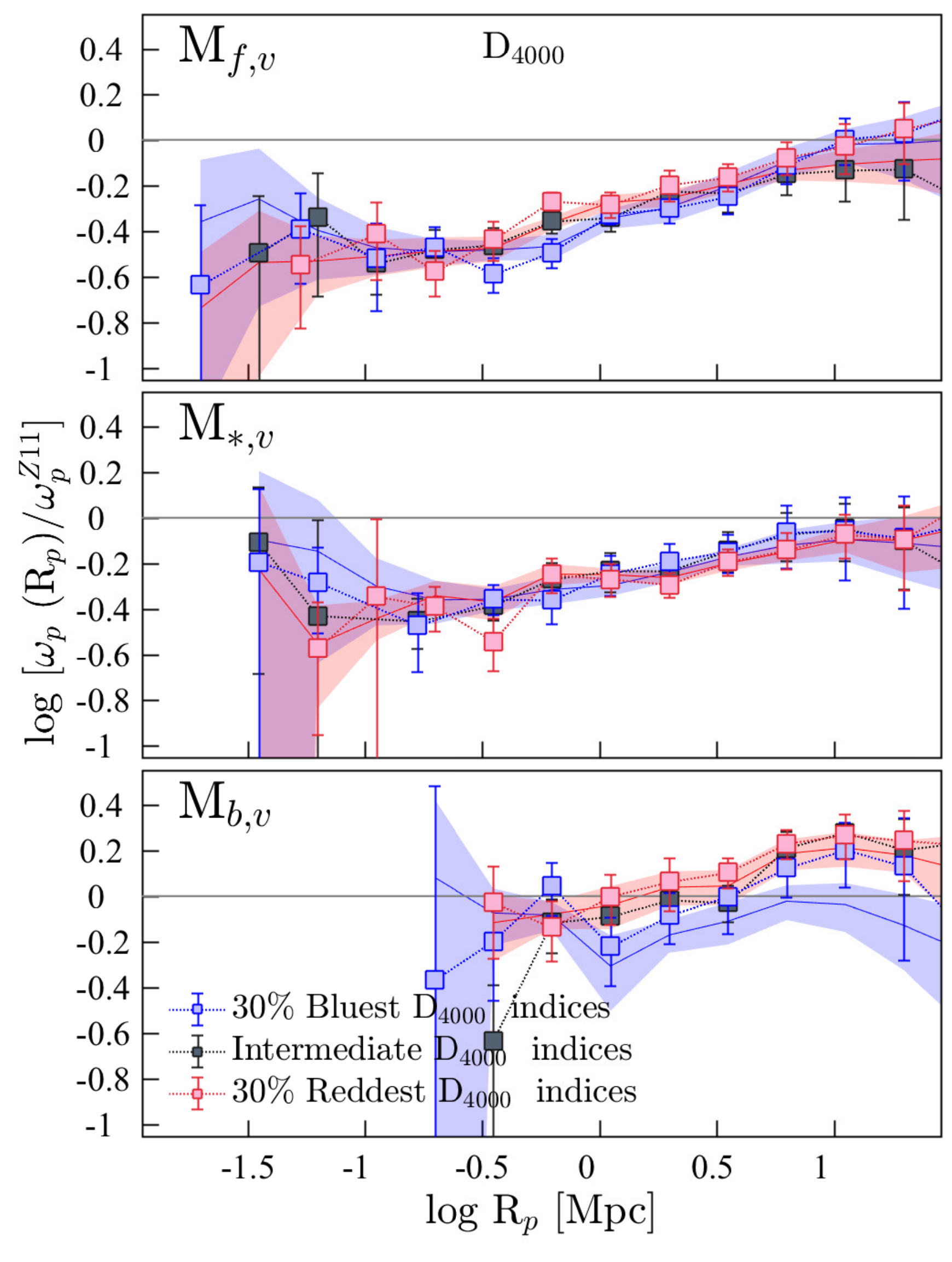}
\includegraphics[width=0.4\textwidth, trim={0 0 0 0},clip]{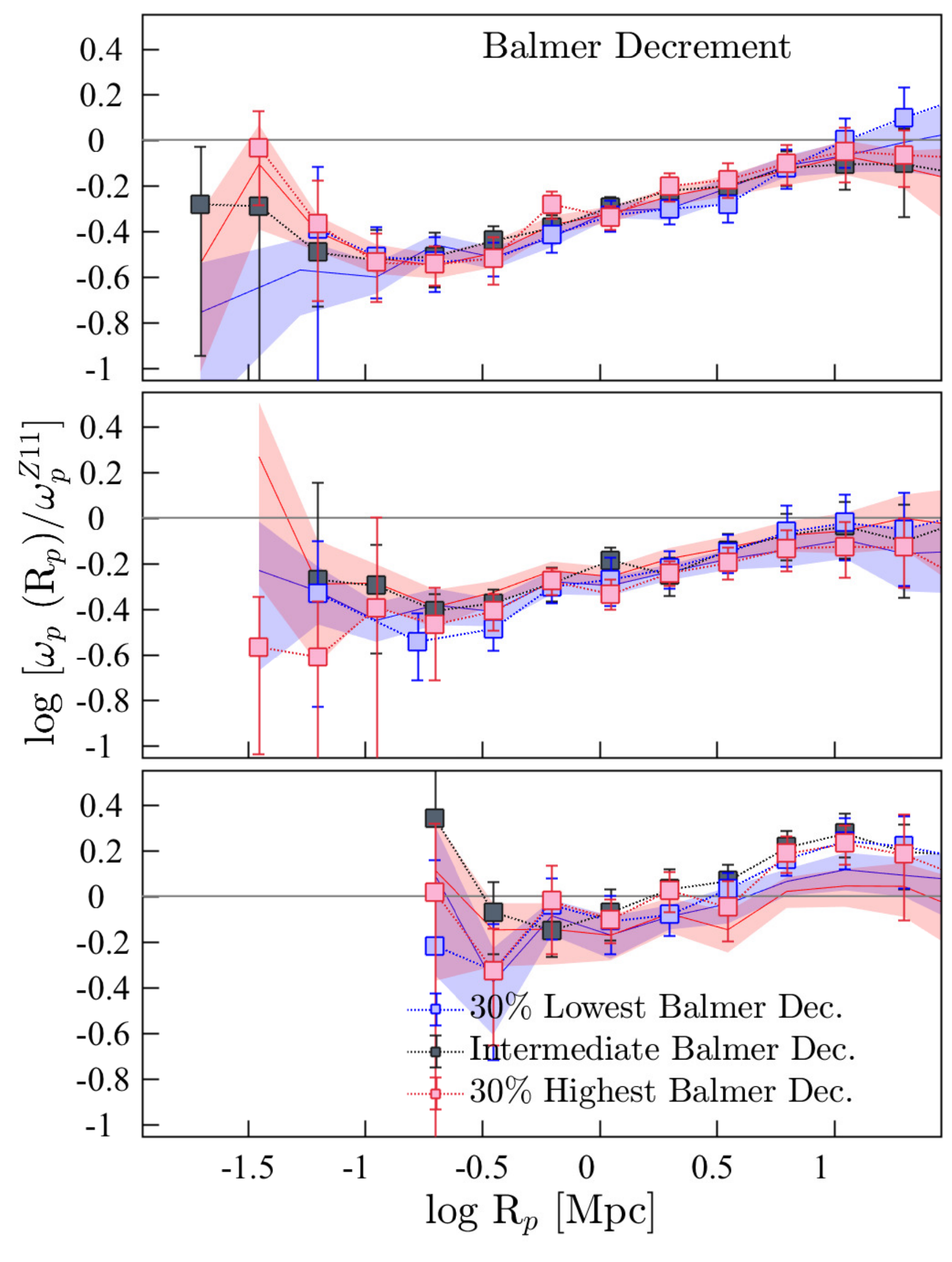}
\caption{The projected CCFs of volume limited luminosity selected SF complete samples (low redshift samples described in Table\,\ref{table:stats2}) relative to $\omega_p^{Z11}$ (luminosity increases down). Right panels: the CCFs of galaxies with low (the $30\%$ lowest in Balmer decrement; blue squares), high (the $30\%$ highest in Balmer decrement; red squares) and intermediate (the $40\%$ with intermediate Balmer decrements; black squares) dust obscurations. Left panels: the CCFs of galaxies with spectroscopically blue (the $30\%$ lowest in D$_{\rm{4000}}$ indices; blue squares), red (the $30\%$ highest in D$_{\rm{4000}}$ indices; red squares) and intermediate (the $40\%$ with intermediate D$_{\rm{4000}}$ indices; black squares) colours. The blue and red shaded regions show the respective CCFs of magnitude selected samples; galaxies with spectroscopically blue and red colours (right panels), and high and low Balmer decrements (left panels). }
\label{fig:xcrr_volume_mag_bd_d4000}
\end{center}
\end{figure*}

The cross correlation results of the volume limited magnitude selected samples are presented in Figures\,\ref{fig:xcrr_volume_mag_z1}-\ref{fig:xcrr_volume_mag_bd_d4000}. The left panels of Figure\,\ref{fig:xcrr_volume_mag_z1} show the projected CCFs relative to $\omega_p^{Z11}$ of high, low and intermediate sSFR galaxies (blue, red and black symbols, respectively), and the right panels show the CCFs of optically blue, red and intermediate colour (blue, red and black symbols, respectively) galaxies. Also shown as shaded regions are the ACFs of respective magnitude selected samples (non-volume limited) with blue and red on left (right) panels denoting high (optically blue) and low (optically red) sSFR galaxies in each magnitude sample, respectively.  


The CCFs$_{\omega_p^{Z11}}$ of volume limited M$_f$ galaxies of SF complete sample show a strong agreement with their non-volume limited counterparts on all scales. Recall that there is a large overlap in redshift between volume limited and non-volume limited magnitude selected M$_f$ galaxies. Therefore the respective  CCFs$_{\omega_p^{Z11}}$ likely mostly probe the clustering properties of similar galaxy populations (as is the case with the respective ACF$_{\omega_p^{Z11}}$ of M$_{\rm{f}}$ galaxies of SF complete sample). In comparison, the CCFs$_{\omega_p^{Z11}}$ of M$_*$ and M$_b$ galaxies of volume limited luminosity selected samples noticeably differ from their respective non-volume limited counterparts. 

On R$_p\lesssim0.3$ Mpc, the CCF$_{\omega_p^{Z11}}$ of optically blue M$_*$ galaxies with $0.01\lesssim z\lesssim0.15$ (i.e.\,low redshift volume limited M$_*$ sample; Figure\,\ref{fig:xcrr_volume_mag_z1}) show a steady decline in clustering with decreasing R$_p$. The CCF of optically blue M$_*$ galaxies with $0.17\lesssim z\lesssim0.24$ (i.e.\,high redshift volume limited M$_*$ sample; Figure\,\ref{fig:xcrr_volume_mag_z2}), on the other hand, show a steady incline in clustering amplitude with decreasing R$_p$, in agreement with that seen in the CCFs$_{\omega_p^{Z11}}$ of M$_*$ galaxies over the $0.01\lesssim z\lesssim0.24$ range.  On larger separations, the respective CCFs$_{\omega_p^{Z11}}$ optically blue galaxies are in agreement with each other to within their uncertainties. Finally, the respective volume limited and non-volume limited CCFs$_{\omega_p^{Z11}}$ of optically red and low sSFR M$_*$ galaxies show similar clustering behaviour to within their uncertainties over all separations. 

On R$_p\lesssim0.3$ Mpc, the M$_b$ CCFs$_{\omega_p^{Z11}}$ of volume limited magnitude selected low redshift galaxies with high (low) sSFRs and optically blue (red) colours show on average higher clustering strengths than (similar clustering strength to) their respective magnitude limited CCFs$_{\omega_p^{Z11}}$. The clustering of the respective high redshift volume limited samples, on the other hand, show the opposite. These differences could be driven by the differences in the redshift distributions of low and high redshift volume samples.  

Figure\,\ref{fig:xcrr_volume_mag_bd_d4000} shows the CCFs$_{\omega_p^{Z11}}$ of spectroscopically blue and red (left panels) galaxies, and of galaxies with low and high dust obscuration measures (right panels) of volume limited (blue and red symbols, respectively) and non-volume limited (blue and red shaded regions, respectively). Overall, there is a good agreement between the respective CCFs$_{\omega_p^{Z11}}$. Interestingly, the clustering signal on R$_p\lesssim0.3$ Mpc of spectroscopically blue M$_*$ galaxies does not appear to mirror the decline in amplitude with decreasing R$_p$ evident in the respective CCF$_{\omega_p^{Z11}}$ of optically blue M$_*$ galaxies, despite being a proxy for optical colour. This is perhaps a result of D$_{\rm{4000}}$ index being more sensitive to the colour changes in the central regions of galaxies than optical colours (see \S\,\ref{subsec:cross_corr} for detailed discussion).

\begin{figure*}
\begin{center}
\includegraphics[width=1\textwidth]{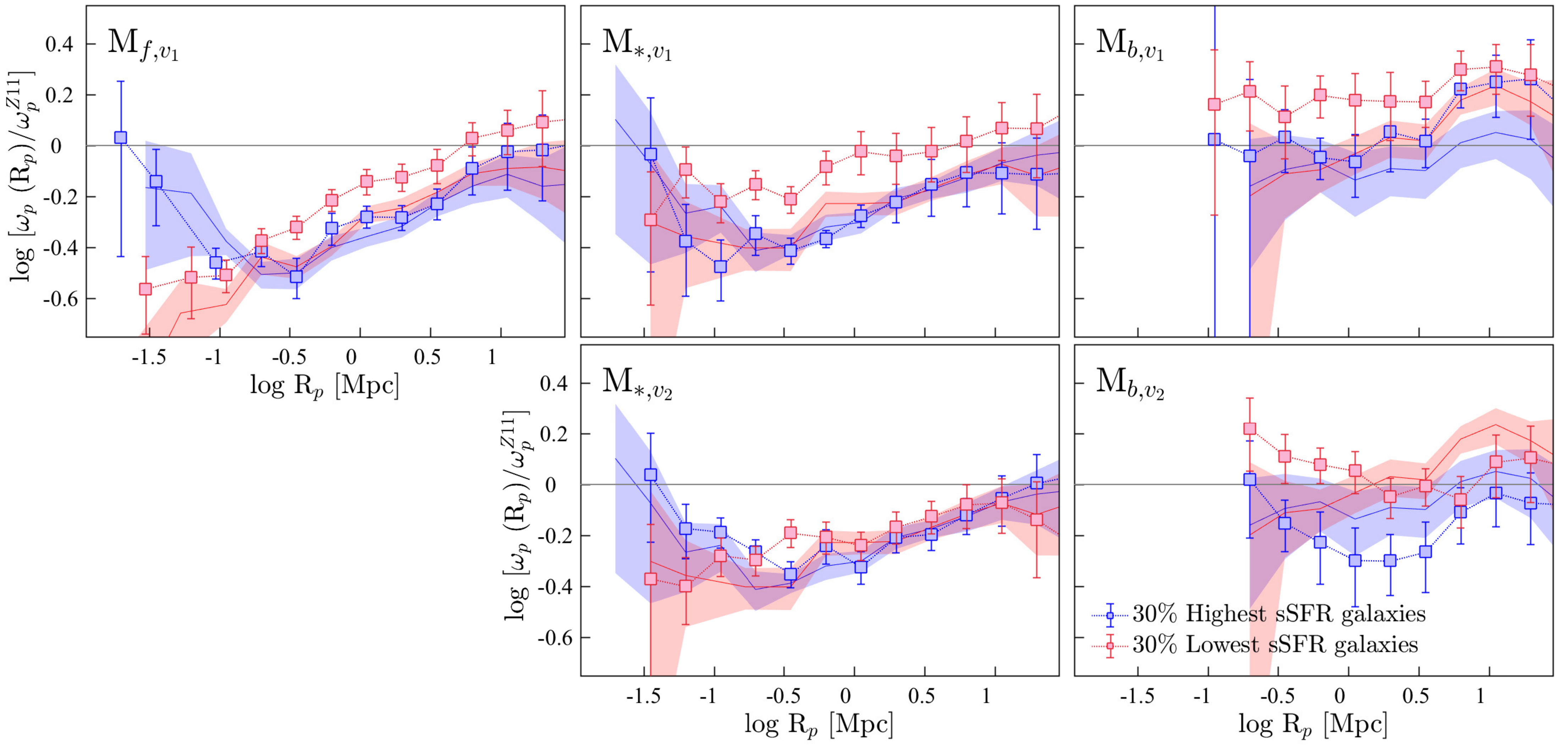}
\caption{The projected CCFs relative to $\omega_p^{Z11}$ computed from cross correlating luminosity selected volume limited SF complete samples (i.e.\,high and low sSFR galaxies drawn from volume limited SF complete samples) with the respective volume limited REF samples (optical luminosity increases across). Top panels: the CCFs of low redshift high (the $30\%$ highest in sSFR; blue squares) and low (the $30\%$ lowest in SFR; red squares) sSFR galaxies. Bottom panels: the CCFs of high redshift high (the $30\%$ highest in sSFR; blue squares) and low (the $30\%$ lowest in SFR; red squares) sSFR galaxies. The blue and red shaded regions show the CCFs of high and low sSFR galaxies relative to $\omega_p^{Z11}$ of magnitude selected volume limited SF complete samples discussed in \S\,\ref{app:vanalysis} and \S\,\ref{subsec:cross_corr}.}
\label{fig:xcrr_Mr1_color_ssfr_func_ref2}
\end{center}
\end{figure*}
\begin{figure*}
\begin{center}
\includegraphics[width=1\textwidth]{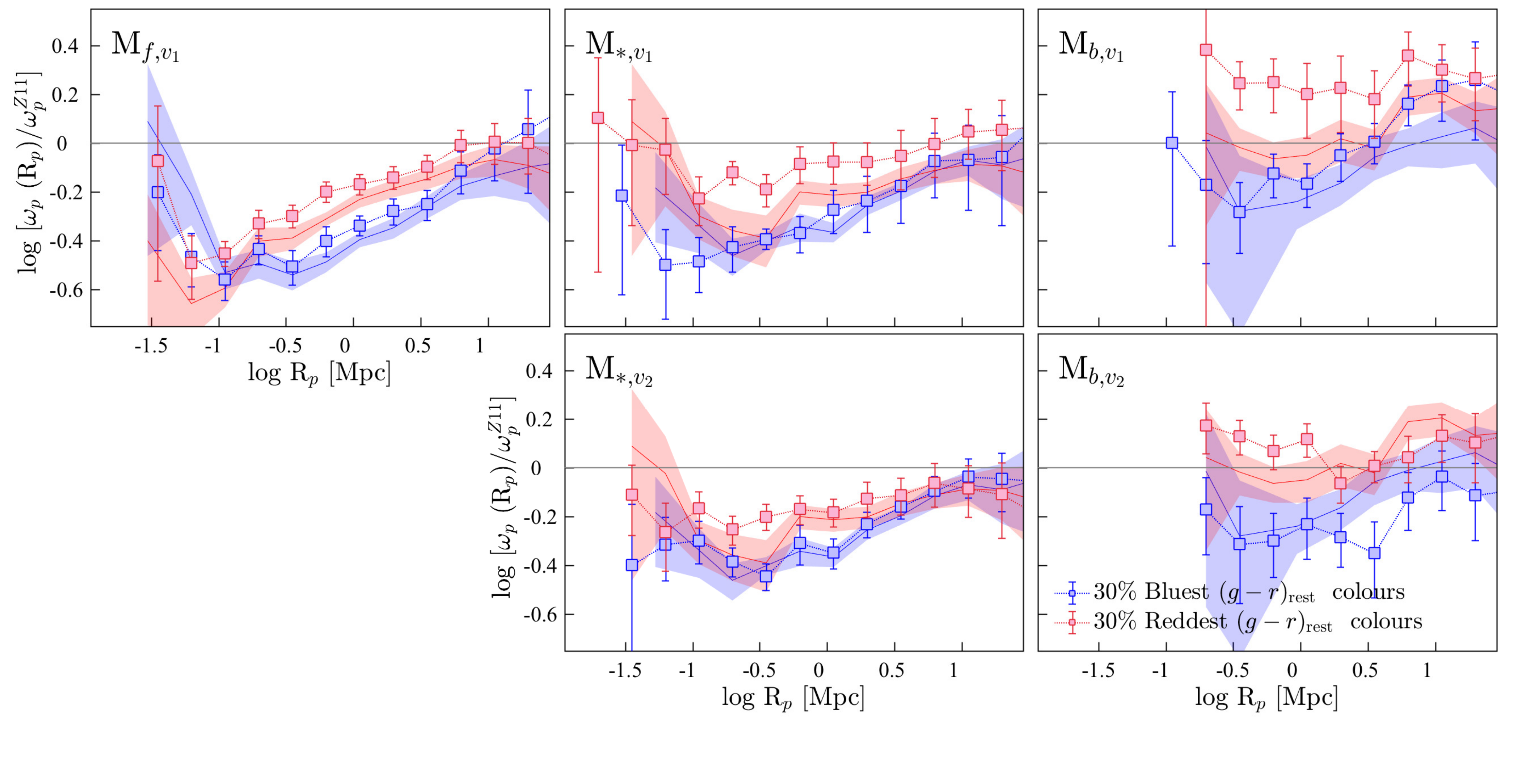}
\caption{The projected CCFs relative to $\omega_p^{Z11}$ computed from cross correlating luminosity selected volume limited SF complete samples (i.e.\,optically blue and red galaxies drawn from volume limited SF complete samples) with the respective volume limited REF samples (optical luminosity increases across). Top panels: the CCFs of low redshift galaxies with optically blue (the $30\%$ bluest in ($g-r$)$_{\rm{rest}}$; blue squares) and red (the $30\%$ reddest in ($g-r$)$_{\rm{rest}}$; red squares) colours. Bottom panels:  the CCFs of high redshift galaxies with optically blue (the $30\%$ bluest in ($g-r$)$_{\rm{rest}}$; blue squares) and red (the $30\%$ reddest in ($g-r$)$_{\rm{rest}}$; red squares) colours. The blue and red shaded regions show the CCFs of high and low sSFR galaxies relative to $\omega_p^{Z11}$ of magnitude selected volume limited SF complete samples discussed in \S\,\ref{app:vanalysis} and \S\,\ref{subsec:cross_corr}.}
\label{fig:xcrr_Mr1_color_ssfr_func_ref3}
\end{center}
\end{figure*}

Finally, the CCFs$_{\omega_p^{Z11}}$ of high, low and intermediate SFR galaxies, and of galaxies with optically blue, red and intermediate colours of low redshift volume limited stellar mass selected samples (not shown) indicate a clustering behaviour similar to that observed in Figure\,\ref{fig:xcrr_volume_mag_z1}.

\section{Cross correlating star forming and REF samples}\label{App:XC_ref}

The clustering results presented in the main paper are computed from cross correlating different star forming samples. We have also investigated the clustering of different galaxy populations by cross correlating star forming and reference (REF) galaxy samples. This approach significantly reduces the uncertainties arising from small number statistics as REF samples contain a higher number of galaxies than SF complete samples, however, with the caveat that by mixing star forming and non-star forming populations it becomes difficult to interpret and understand the clustering properties of star forming galaxies. For this reason, we present and discuss the results of this analysis here. Also, in order to differentiate the results of the cross correlation between different SF complete samples of galaxies (i.e.\,the CCFs presented in \S\,\ref{subsec:cross_corr}) from the results of the cross correlations between SF complete and REF samples, we use the labels SF CCFs$_{\omega_p^{Z11}}$ and REF CCFs$_{\omega_p^{Z11}}$, respectively. 

Figures\,\ref{fig:xcrr_Mr1_color_ssfr_func_ref2} and \ref{fig:xcrr_Mr1_color_ssfr_func_ref3} present the REF (symbols) and SF (coloured regions) CCFs$_{\omega_p^{Z11}}$ of volume limited magnitude selected samples. The red and blue colours in Figure\,\ref{fig:xcrr_Mr1_color_ssfr_func_ref2} denote low and high sSFR galaxies, and in Figure\,\ref{fig:xcrr_Mr1_color_ssfr_func_ref3}, optically red and blue galaxies. The REF CCFs$_{\omega_p^{Z11}}$ are in qualitative agreement with the respective SF CCFs$_{\omega_p^{Z11}}$ on all scales. The CCFs$_{\omega_p^{Z11}}$ of low and high redshift volume limited samples of M$_*$ and M$_b$ galaxies show the evolution of the clustering of optically bright star forming galaxies. These evolutionary effects are present to varying degrees over all R$_p$ probed, however, are particularly notable over the R$_p>0.5$ Mpc. For instance, the CCFs$_{\omega_p^{Z11}}$ of M$_*$ and M$_b$ galaxies show higher clustering amplitudes at low redshift than at high redshift. This result is in agreement with previous studies, in particular with those that investigate the dependence of galaxy clustering on optical luminosity \citep[e.g.][]{Adelberger2005, Marulli2013}, that report an increase in clustering strength with decreasing redshift.    

To summarise, our results based on cross correlating star forming galaxies with all galaxies regardless of star formation (i.e.\,REF) show that, on most scales, the redder, low-sSFR galaxies at a fixed M$_r$ are clustered more strongly than bluer, high-sSFR systems, in agreement with previous photometric studies of clustering in the local Universe \citep[e.g.][]{Norberg2001, Marulli2013, Guo2014}.  The environmental effects are likely to be largely responsible for the differences observed in the clustering strengths between the REF CCFs$_{\omega_p^{Z11}}$ of low-sSFR and high-sSFR galaxies. Both the lower clustering strengths exhibited by high-sSFR galaxies and the observational evidence of higher SFR enhancements in galaxy pairs of equal mass \citep{Ellison2008} suggest that, at a fixed M$_r$, it is more likely for an interacting high-sSFR galaxy to have a SF companion than a non-SF companion, and reside in relatively low density environment. The higher clustering of low-sSFR and redder systems at a fixed M$_r$, on the other hand, suggest that an interacting companion of a low-sSFR galaxy can either be a SF or a non-SF galaxy, both preferentially inhabiting a higher-density environment, and rather than triggering SF in interacting members, their environment has triggered quenching of star formation. %

\section{The standard mark correlation functions of star forming galaxies}{\label{app:mark_CFs}}
\begin{figure*}
\begin{center}
\includegraphics[width=1.01\textwidth]{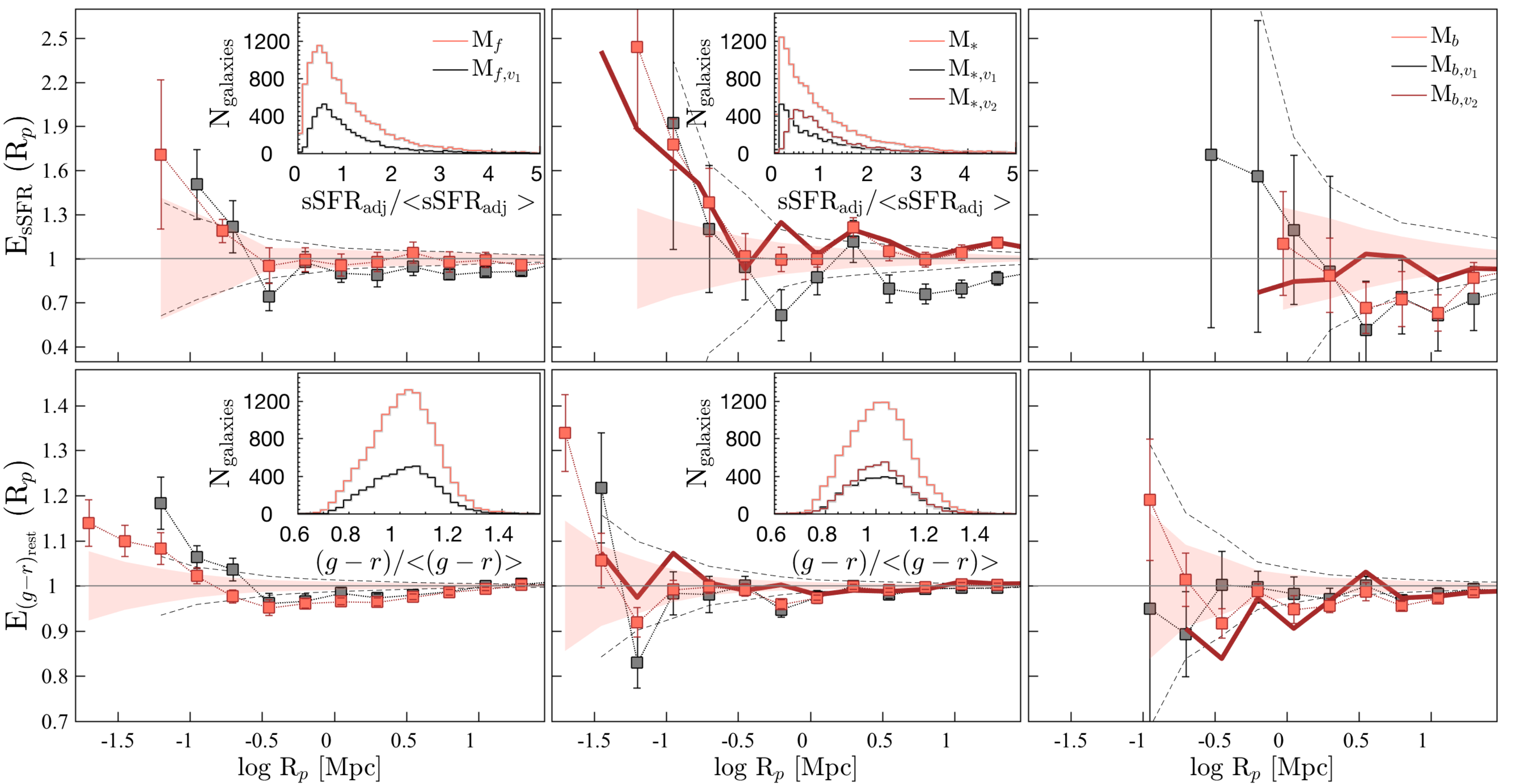}
\caption{The MCFs of non-volume limited and volume limited luminosity selected samples (with increasing luminosity from left-to-right), with sSFR$_{\rm{adj}}$ (top panels) and ($g-r$)$_{\rm{rest}}$ colour (bottom panels) as marks. From left-to-right: The filled symbols show the MCFs of SF M$_f$, M$_*$ and M$_b$ galaxies, and their respective low-$z$ volume samples. We show the MCFs of higher-$z$ volume samples as thick solid lines for clarity, noting that they only exist for M$_*$ and M$_b$ samples (centre and right panels). The shaded regions and the regions enclosed by black dashed lines denote the $1\sigma$ scatter from scrambling the marks of luminosity selected and corresponding volume limited samples. For clarity, we do not show the scatter on the MCFs of higher-$z$ M$_*$ and M$_b$ volume samples. The insets on left and middle panels show the distribution of the marks. Again, for clarity, we do not show the distributions of M$_b$ galaxies.}
\label{fig:mcrr_Mr1_color_ssfr_func}
\end{center}
\end{figure*}
\begin{figure*}
\begin{center}
\includegraphics[width=1.01\textwidth]{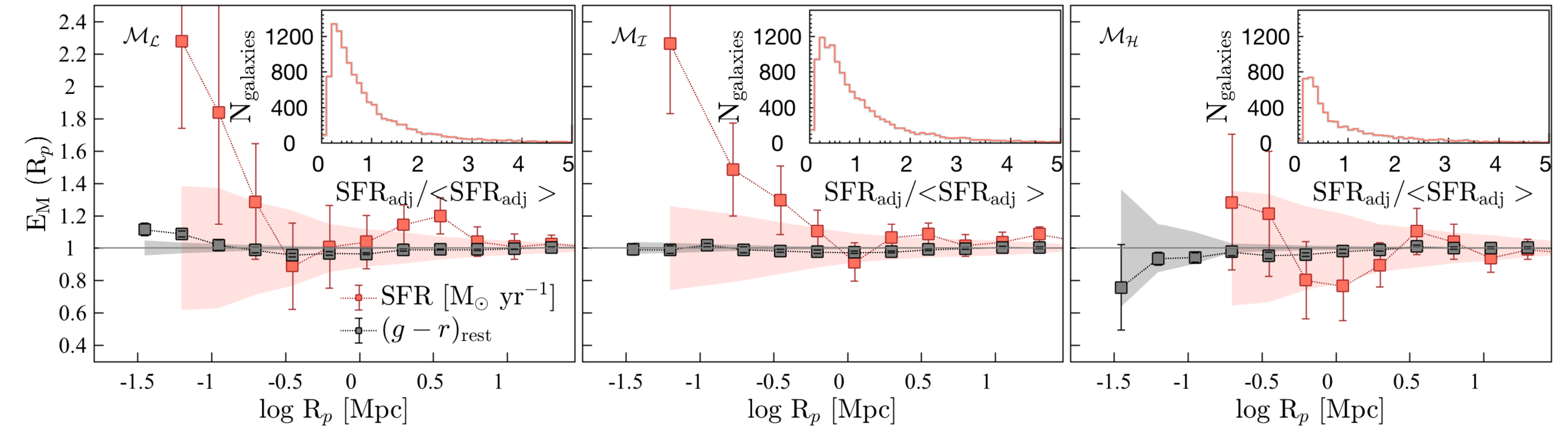}
\caption{The MCFs of stellar mass selected SF complete samples (with increasing stellar mass from left-to-right), with SFR$_{\rm{adj}}$ and ($g-r$)$_{\rm{rest}}$ colour as marks (orange and grey symbols, respectively). The shaded regions denote the $1\sigma$ scatter from scrambling the marks. The insets show the distributions of the SFR$_{\rm{adj}}$ marks. The distributions of colour marks are similar to that shown in Figure\,\ref{fig:mcrr_Mr1_color_ssfr_func}. }
\label{fig:mcrr_Mass_color_ssfr_func}
\end{center}
\end{figure*}

The MCFs of luminosity selected and their equivalent volume limited samples, with sSFR (top panels) and ($g-r$)$_{\rm{rest}}$ colour (bottom panels) as marks, are presented in Figure\,\ref{fig:mcrr_Mr1_color_ssfr_func}.  The measurement uncertainties of sSFRs are relatively large compared to those of ($g-r$)$_{\rm{rest}}$ colours, and our SF sample likely contains some overestimated sSFR measures. Therefore to limit the impact of sSFR outliers on the MCFs while not removing true starbursts from the clustering samples, we re-adjust sSFRs as follows, 
\begin{equation}
 \rm{sSFR_{{adj}}} = \frac{sSFR \times sSFR_{max}}{sSFR + sSFR_{max}},
\label{eq:mark_ssfr_adjust}
\end{equation}
where sSFR$_{\rm{max}}=10^{-9.0}$ [yr$^{-1}$]. The ($g-r$)$_{\rm{rest}}$ and sSFR$_{\rm{adj}}$ distributions of different clustering samples are shown in the insets of Figure\,\ref{fig:mcrr_Mr1_color_ssfr_func}. 

The sSFR$_{\rm{adj}}$ MCFs (top panels of Figure\,\ref{fig:mcrr_Mr1_color_ssfr_func}) not only show a clear dependence of sSFR on the environment, but also show a small-scale dependence of enhancement in the spatial distribution of sSFR (i.e.\,E$_{\rm{sSFR}}$) on optical brightness. This E$_{\rm{sSFR}}$-optical brightness dependence is in the sense that E$_{\rm{sSFR}}$ shows a strengthening in magnitude at a given R$_p$ on R$_p\lesssim0.35$ Mpc scales with increasing optical brightness, a behaviour similar to that seen in the SF  ACFs of magnitude-limited samples (\S\,\ref{subsec:auto_corr}). The ($g-r$)$_{\rm{rest}}$ MCFs (bottom panels of Figure\,\ref{fig:mcrr_Mr1_color_ssfr_func}) also show an enhancement in E$_{(g-r)_{\rm{rest}}}$, in particular in M$_f$ galaxies, however, the strength of E$_{(g-r)_{\rm{rest}}}$ enhancement does not appear to depend on optical brightness of galaxies.  

The MCFs of stellar mass selected SF complete samples are presented in Figure\,\ref{fig:mcrr_Mass_color_ssfr_func}. For the same reason mentioned above, we adjust the SFRs as follows, 
\begin{equation}
 \rm{SFR_{{adj}}} = \frac{SFR \times SFR_{max}}{SFR + SFR_{max}},
\label{eq:mark_sfr_adjust}
\end{equation}
where SFR$_{\rm{max}}=50$ [M$_{\odot}$ yr$^{-1}$], and SFR$_{\rm{adj}}$ distributions of different clustering samples are shown in the insets of Figure\,\ref{fig:mcrr_Mass_color_ssfr_func}.

The SFR MCFs show an enhancement in E$_{\rm{SFR}}$, similar to that observed in sSFR MCFs of luminosity selected samples. The ($g-r$)$_{\rm{rest}}$ colour MCFs, other than that of $\mathcal{M}_{\mathcal{L}}$ galaxies, show no enhancement in the signal over small scales.  

\label{lastpage}

\end{document}